# The Pierre Auger Observatory V: Enhancements

THE PIERRE AUGER COLLABORATION

*Observatorio Pierre Auger, Av. San Martín Norte 304, 5613 Malargüe, Argentina*

# The Pierre Auger Collaboration

P. Abreu[74], M. Aglietta[57], E.J. Ahn[93], I.F.M. Albuquerque[19], D. Allard[33], I. Allekotte[1], J. Allen[96], P. Allison[98], J. Alvarez Castillo[67], J. Alvarez-Muñiz[84], M. Ambrosio[50], A. Aminaei[68], L. Anchordoqui[109], S. Andringa[74], T. Antičić[27], A. Anzalone[56], C. Aramo[50], E. Arganda[81], F. Arqueros[81], H. Asorey[1], P. Assis[74], J. Aublin[35], M. Ave[41], M. Avenier[36], G. Avila[12], T. Bäcker[45], M. Balzer[40], K.B. Barber[13], A.F. Barbosa[16], R. Bardenet[34], S.L.C. Barroso[22], B. Baughman[98], J. Bäuml[39], J.J. Beatty[98], B.R. Becker[106], K.H. Becker[38], A. Bellétoile[37], J.A. Bellido[13], S. BenZvi[108], C. Berat[36], X. Bertou[1], P.L. Biermann[42], P. Billoir[35], F. Blanco[81], M. Blanco[82], C. Bleve[38], H. Blümer[41, 39], M. Boháčová[29, 101], D. Boncioli[51], C. Bonifazi[25, 35], R. Bonino[57], N. Borodai[72], J. Brack[91], P. Brogueira[74], W.C. Brown[92], R. Bruijn[87], P. Buchholz[45], A. Bueno[83], R.E. Burton[89], K.S. Caballero-Mora[99], L. Caramete[42], R. Caruso[52], A. Castellina[57], O. Catalano[56], G. Cataldi[49], L. Cazon[74], R. Cester[53], J. Chauvin[36], S.H. Cheng[99], A. Chiavassa[57], J.A. Chinellato[20], A. Chou[93, 96], J. Chudoba[29], R.W. Clay[13], M.R. Coluccia[49], R. Conceição[74], F. Contreras[11], H. Cook[87], M.J. Cooper[13], J. Coppens[68, 70], A. Cordier[34], U. Cotti[66], S. Coutu[99], C.E. Covault[89], A. Creusot[33, 79], A. Criss[99], J. Cronin[101], A. Curutiu[42], S. Dagoret-Campagne[34], R. Dallier[37], S. Dasso[8, 4], K. Daumiller[39], B.R. Dawson[13], R.M. de Almeida[26], M. De Domenico[52], C. De Donato[67, 48], S.J. de Jong[68, 70], G. De La Vega[10], W.J.M. de Mello Junior[20], J.R.T. de Mello Neto[25], I. De Mitri[49], V. de Souza[18], K.D. de Vries[69], G. Decerprit[33], L. del Peral[82], O. Deligny[32], H. Dembinski[41], N. Dhital[95], C. Di Giulio[47, 51], J.C. Diaz[95], M.L. Díaz Castro[17], P.N. Diep[110], C. Dobrigkeit [20], W. Docters[69], J.C. D'Olivo[67], P.N. Dong[110, 32], A. Dorofeev[91], J.C. dos Anjos[16], M.T. Dova[7], D. D'Urso[50], I. Dutan[42], J. Ebr[29], R. Engel[39], M. Erdmann[43], C.O. Escobar[20], A. Etchegoyen[2], P. Facal San Luis[101], I. Fajardo Tapia[67], H. Falcke[68, 71], G. Farrar[96], A.C. Fauth[20], N. Fazzini[93], A.P. Ferguson[89], A. Ferrero[2], B. Fick[95], A. Filevich[2], A. Filipčič[78, 79], S. Fliescher[43], C.E. Fracchiolla[91], E.D. Fraenkel[69], U. Fröhlich[45], B. Fuchs[16], R. Gaior[35], R.F. Gamarra[2], S. Gambetta[46], B. García[10], D. García Gámez[83], D. Garcia-Pinto[81], A. Gascon[83], H. Gemmeke[40], K. Gesterling[106], P.L. Ghia[35, 57], U. Giaccari[49], M. Giller[73], H. Glass[93], M.S. Gold[106], G. Golup[1], F. Gomez Albarracin[7], M. Gómez Berisso[1], P. Gonçalves[74], D. Gonzalez[41], J.G. Gonzalez[41], B. Gookin[91], D. Góra[41, 72], A. Gorgi[57], P. Gouffon[19], S.R. Gozzini[87], E. Grashorn[98], S. Grebe[68, 70], N. Griffith[98], M. Grigat[43], A.F. Grillo[58], Y. Guardincerri[4], F. Guarino[50], G.P. Guedes[21], A. Guzman[67], J.D. Hague[106], P. Hansen[7], D. Harari[1], S. Harmsma[69, 70], J.L. Harton[91], A. Haungs[39], T. Hebbeker[43], D. Heck[39], A.E. Herve[13], C. Hojvat[93], N. Hollon[101], V.C. Holmes[13], P. Homola[72], J.R. Hörandel[68], A. Horneffer[68], M. Hrabovský[30, 29], T. Huege[39], A. Insolia[52], F. Ionita[101], A. Italiano[52], C. Jarne[7], S. Jiraskova[68], M. Josebachuili[2], K. Kadija[27], K.-H. Kampert[38], P. Karhan[28], P. Kasper[93], B. Kégl[34], B. Keilhauer[39], A. Keivani[94], J.L. Kelley[68], E. Kemp[20], R.M. Kieckhafer[95], H.O. Klages[39], M. Kleifges[40], J. Kleinfeller[39], J. Knapp[87], D.-H. Koang[36], K. Kotera[101], N. Krohm[38], O. Krömer[40], D. Kruppke-Hansen[38], F. Kuehn[93], D. Kuempel[38], J.K. Kulbartz[44], N. Kunka[40], G. La Rosa[56], C. Lachaud[33], P. Lautridou[37], M.S.A.B. Leão[24], D. Lebrun[36], P. Lebrun[93], M.A. Leigui de Oliveira[24], A. Lemiere[32], A. Letessier-Selvon[35], I. Lhenry-Yvon[32], K. Link[41], R. López[63], A. Lopez Agüera[84], K. Louedec[34], J. Lozano Bahilo[83], A. Lucero[2, 57], M. Ludwig[41], H. Lyberis[32], M.C. Maccarone[56], C. Macolino[35], S. Maldera[57], D. Mandat[29], P. Mantsch[93], A.G. Mariazzi[7], J. Marin[11, 57], V. Marin[37], I.C. Maris[35], H.R. Marquez Falcon[66], G. Marsella[54], D. Martello[49], L. Martin[37], H. Martinez[64], O. Martínez Bravo[63],

H.J. Mathes[39], J. Matthews[94, 100], J.A.J. Matthews[106], G. Matthiae[51], D. Maurizio[53], P.O. Mazur[93], G. Medina-Tanco[67], M. Melissas[41], D. Melo[2, 53], E. Menichetti[53], A. Menshikov[40], P. Mertsch[85], C. Meurer[43], S. Mićanović[27], M.I. Micheletti[9], W. Miller[106], L. Miramonti[48], S. Mollerach[1], M. Monasor[101], D. Monnier Ragaigne[34], F. Montanet[36], B. Morales[67], C. Morello[57], E. Moreno[63], J.C. Moreno[7], C. Morris[98], M. Mostafá[91], C.A. Moura[24, 50], S. Mueller[39], M.A. Muller[20], G. Müller[43], M. Münchmeyer[35], R. Mussa[53], G. Navarra[57 †], J.L. Navarro[83], S. Navas[83], P. Necesal[29], L. Nellen[67], A. Nelles[68, 70], J. Neuser[38], P.T. Nhung[110], L. Niemietz[38], N. Nierstenhoefer[38], D. Nitz[95], D. Nosek[28], L. Nožka[29], M. Nyklicek[29], J. Oehlschläger[39], A. Olinto[101], V.M. Olmos-Gilbaja[84], M. Ortiz[81], N. Pacheco[82], D. Pakk Selmi-Dei[20], M. Palatka[29], J. Pallotta[3], N. Palmieri[41], G. Parente[84], E. Parizot[33], A. Parra[84], R.D. Parsons[87], S. Pastor[80], T. Paul[97], M. Pech[29], J. Pękala[72], R. Pelayo[84], I.M. Pepe[23], L. Perrone[54], R. Pesce[46], E. Petermann[105], S. Petrera[47], P. Petrinca[51], A. Petrolini[46], Y. Petrov[91], J. Petrovic[70], C. Pfendner[108], N. Phan[106], R. Piegaia[4], T. Pierog[39], P. Pieroni[4], M. Pimenta[74], V. Pirronello[52], M. Platino[2], V.H. Ponce[1], M. Pontz[45], P. Privitera[101], M. Prouza[29], E.J. Quel[3], S. Querchfeld[38], J. Rautenberg[38], O. Ravel[37], D. Ravignani[2], B. Revenu[37], J. Ridky[29], S. Riggi[84, 52], M. Risse[45], P. Ristori[3], H. Rivera[48], V. Rizi[47], J. Roberts[96], C. Robledo[63], W. Rodrigues de Carvalho[84, 19], G. Rodriguez[84], J. Rodriguez Martino[11, 52], J. Rodriguez Rojo[11], I. Rodriguez-Cabo[84], M.D. Rodríguez-Frías[82], G. Ros[82], J. Rosado[81], T. Rossler[30], M. Roth[39], B. Rouillé-d'Orfeuil[101], E. Roulet[1], A.C. Rovero[8], C. Rühle[40], F. Salamida[47, 39], H. Salazar[63], G. Salina[51], F. Sánchez[2], M. Santander[11], C.E. Santo[74], E. Santos[74], E.M. Santos[25], F. Sarazin[90], B. Sarkar[38], S. Sarkar[85], R. Sato[11], N. Scharf[43], V. Scherini[48], H. Schieler[39], P. Schiffer[43], A. Schmidt[40], F. Schmidt[101], O. Scholten[69], H. Schoorlemmer[68, 70], J. Schovancova[29], P. Schovánek[29], F. Schröder[39], S. Schulte[43], D. Schuster[90], S.J. Sciutto[7], M. Scuderi[52], A. Segreto[56], M. Settimo[45], A. Shadkam[94], R.C. Shellard[16, 17], I. Sidelnik[2], G. Sigl[44], H.H. Silva Lopez[67], A. Śmiałkowski[73], R. Šmída[39, 29], G.R. Snow[105], P. Sommers[99], J. Sorokin[13], H. Spinka[88, 93], R. Squartini[11], S. Stanic[79], J. Stapleton[98], J. Stasielak[72], M. Stephan[43], E. Strazzeri[56], A. Stutz[36], F. Suarez[2], T. Suomijärvi[32], A.D. Supanitsky[8, 67], T. Šuša[27], M.S. Sutherland[94, 98], J. Swain[97], Z. Szadkowski[73], M. Szuba[39], A. Tamashiro[8], A. Tapia[2], M. Tartare[36], O. Taşcău[38], C.G. Tavera Ruiz[67], R. Tcaciuc[45], D. Tegolo[52, 61], N.T. Thao[110], D. Thomas[91], J. Tiffenberg[4], C. Timmermans[70, 68], D.K. Tiwari[66], W. Tkaczyk[73], C.J. Todero Peixoto[18, 24], B. Tomé[74], A. Tonachini[53], P. Travnicek[29], D.B. Tridapalli[19], G. Tristram[33], E. Trovato[52], M. Tueros[84, 4], R. Ulrich[99, 39], M. Unger[39], M. Urban[34], J.F. Valdés Galicia[67], I. Valiño[84, 39], L. Valore[50], A.M. van den Berg[69], E. Varela[63], B. Vargas Cárdenas[67], J.R. Vázquez[81], R.A. Vázquez[84], D. Veberič[79, 78], V. Verzi[51], J. Vicha[29], M. Videla[10], L. Villaseñor[66], H. Wahlberg[7], P. Wahrlich[13], O. Wainberg[2], D. Walz[43], D. Warner[91], A.A. Watson[87], M. Weber[40], K. Weidenhaupt[43], A. Weindl[39], S. Westerhoff[108], B.J. Whelan[13], G. Wieczorek[73], L. Wiencke[90], B. Wilczyńska[72], H. Wilczyński[72], M. Will[39], C. Williams[101], T. Winchen[43], L. Winders[109], M.G. Winnick[13], M. Wommer[39], B. Wundheiler[2], T. Yamamoto[101 a], T. Yapici[95], P. Younk[45], G. Yuan[94], A. Yushkov[84, 50], B. Zamorano[83], E. Zas[84], D. Zavrtanik[79, 78], M. Zavrtanik[78, 79], I. Zaw[96], A. Zepeda[64], M. Zimbres-Silva[20, 38] M. Ziolkowski[45]

[1] *Centro Atómico Bariloche and Instituto Balseiro (CNEA- UNCuyo-CONICET), San Carlos de Bariloche, Argentina*
[2] *Centro Atómico Constituyentes (Comisión Nacional de Energía Atómica/CONICET/UTN-FRBA), Buenos Aires, Argentina*
[3] *Centro de Investigaciones en Láseres y Aplicaciones, CITEFA and CONICET, Argentina*
[4] *Departamento de Física, FCEyN, Universidad de Buenos Aires y CONICET, Argentina*
[7] *IFLP, Universidad Nacional de La Plata and CONICET, La Plata, Argentina*
[8] *Instituto de Astronomía y Física del Espacio (CONICET- UBA), Buenos Aires, Argentina*
[9] *Instituto de Física de Rosario (IFIR) - CONICET/U.N.R. and Facultad de Ciencias Bioquímicas y Farmacéuticas U.N.R., Rosario, Argentina*
[10] *National Technological University, Faculty Mendoza (CONICET/CNEA), Mendoza, Argentina*
[11] *Observatorio Pierre Auger, Malargüe, Argentina*
[12] *Observatorio Pierre Auger and Comisión Nacional de Energía Atómica, Malargüe, Argentina*
[13] *University of Adelaide, Adelaide, S.A., Australia*
[16] *Centro Brasileiro de Pesquisas Fisicas, Rio de Janeiro, RJ, Brazil*
[17] *Pontifícia Universidade Católica, Rio de Janeiro, RJ, Brazil*

[18] *Universidade de São Paulo, Instituto de Física, São Carlos, SP, Brazil*
[19] *Universidade de São Paulo, Instituto de Física, São Paulo, SP, Brazil*
[20] *Universidade Estadual de Campinas, IFGW, Campinas, SP, Brazil*
[21] *Universidade Estadual de Feira de Santana, Brazil*
[22] *Universidade Estadual do Sudoeste da Bahia, Vitoria da Conquista, BA, Brazil*
[23] *Universidade Federal da Bahia, Salvador, BA, Brazil*
[24] *Universidade Federal do ABC, Santo André, SP, Brazil*
[25] *Universidade Federal do Rio de Janeiro, Instituto de Física, Rio de Janeiro, RJ, Brazil*
[26] *Universidade Federal Fluminense, EEIMVR, Volta Redonda, RJ, Brazil*
[27] *Rudjer Bošković Institute, 10000 Zagreb, Croatia*
[28] *Charles University, Faculty of Mathematics and Physics, Institute of Particle and Nuclear Physics, Prague, Czech Republic*
[29] *Institute of Physics of the Academy of Sciences of the Czech Republic, Prague, Czech Republic*
[30] *Palacky University, RCATM, Olomouc, Czech Republic*
[32] *Institut de Physique Nucléaire d'Orsay (IPNO), Université Paris 11, CNRS-IN2P3, Orsay, France*
[33] *Laboratoire AstroParticule et Cosmologie (APC), Université Paris 7, CNRS-IN2P3, Paris, France*
[34] *Laboratoire de l'Accélérateur Linéaire (LAL), Université Paris 11, CNRS-IN2P3, Orsay, France*
[35] *Laboratoire de Physique Nucléaire et de Hautes Energies (LPNHE), Universités Paris 6 et Paris 7, CNRS-IN2P3, Paris, France*
[36] *Laboratoire de Physique Subatomique et de Cosmologie (LPSC), Université Joseph Fourier, INPG, CNRS-IN2P3, Grenoble, France*
[37] *SUBATECH, École des Mines de Nantes, CNRS-IN2P3, Université de Nantes, Nantes, France*
[38] *Bergische Universität Wuppertal, Wuppertal, Germany*
[39] *Karlsruhe Institute of Technology - Campus North - Institut für Kernphysik, Karlsruhe, Germany*
[40] *Karlsruhe Institute of Technology - Campus North - Institut für Prozessdatenverarbeitung und Elektronik, Karlsruhe, Germany*
[41] *Karlsruhe Institute of Technology - Campus South - Institut für Experimentelle Kernphysik (IEKP), Karlsruhe, Germany*
[42] *Max-Planck-Institut für Radioastronomie, Bonn, Germany*
[43] *RWTH Aachen University, III. Physikalisches Institut A, Aachen, Germany*
[44] *Universität Hamburg, Hamburg, Germany*
[45] *Universität Siegen, Siegen, Germany*
[46] *Dipartimento di Fisica dell'Università and INFN, Genova, Italy*
[47] *Università dell'Aquila and INFN, L'Aquila, Italy*
[48] *Università di Milano and Sezione INFN, Milan, Italy*
[49] *Dipartimento di Fisica dell'Università del Salento and Sezione INFN, Lecce, Italy*
[50] *Università di Napoli "Federico II" and Sezione INFN, Napoli, Italy*
[51] *Università di Roma II "Tor Vergata" and Sezione INFN, Roma, Italy*
[52] *Università di Catania and Sezione INFN, Catania, Italy*
[53] *Università di Torino and Sezione INFN, Torino, Italy*
[54] *Dipartimento di Ingegneria dell'Innovazione dell'Università del Salento and Sezione INFN, Lecce, Italy*
[56] *Istituto di Astrofisica Spaziale e Fisica Cosmica di Palermo (INAF), Palermo, Italy*
[57] *Istituto di Fisica dello Spazio Interplanetario (INAF), Università di Torino and Sezione INFN, Torino, Italy*
[58] *INFN, Laboratori Nazionali del Gran Sasso, Assergi (L'Aquila), Italy*
[61] *Università di Palermo and Sezione INFN, Catania, Italy*
[63] *Benemérita Universidad Autónoma de Puebla, Puebla, Mexico*
[64] *Centro de Investigación y de Estudios Avanzados del IPN (CINVESTAV), México, D.F., Mexico*
[66] *Universidad Michoacana de San Nicolas de Hidalgo, Morelia, Michoacan, Mexico*
[67] *Universidad Nacional Autonoma de Mexico, Mexico, D.F., Mexico*
[68] *IMAPP, Radboud University Nijmegen, Netherlands*
[69] *Kernfysisch Versneller Instituut, University of Groningen, Groningen, Netherlands*
[70] *Nikhef, Science Park, Amsterdam, Netherlands*
[71] *ASTRON, Dwingeloo, Netherlands*
[72] *Institute of Nuclear Physics PAN, Krakow, Poland*

[73] University of Łódź, Łódź, Poland
[74] LIP and Instituto Superior Técnico, Lisboa, Portugal
[78] J. Stefan Institute, Ljubljana, Slovenia
[79] Laboratory for Astroparticle Physics, University of Nova Gorica, Slovenia
[80] Instituto de Física Corpuscular, CSIC-Universitat de València, Valencia, Spain
[81] Universidad Complutense de Madrid, Madrid, Spain
[82] Universidad de Alcalá, Alcalá de Henares (Madrid), Spain
[83] Universidad de Granada & C.A.F.P.E., Granada, Spain
[84] Universidad de Santiago de Compostela, Spain
[85] Rudolf Peierls Centre for Theoretical Physics, University of Oxford, Oxford, United Kingdom
[87] School of Physics and Astronomy, University of Leeds, United Kingdom
[88] Argonne National Laboratory, Argonne, IL, USA
[89] Case Western Reserve University, Cleveland, OH, USA
[90] Colorado School of Mines, Golden, CO, USA
[91] Colorado State University, Fort Collins, CO, USA
[92] Colorado State University, Pueblo, CO, USA
[93] Fermilab, Batavia, IL, USA
[94] Louisiana State University, Baton Rouge, LA, USA
[95] Michigan Technological University, Houghton, MI, USA
[96] New York University, New York, NY, USA
[97] Northeastern University, Boston, MA, USA
[98] Ohio State University, Columbus, OH, USA
[99] Pennsylvania State University, University Park, PA, USA
[100] Southern University, Baton Rouge, LA, USA
[101] University of Chicago, Enrico Fermi Institute, Chicago, IL, USA
[105] University of Nebraska, Lincoln, NE, USA
[106] University of New Mexico, Albuquerque, NM, USA
[108] University of Wisconsin, Madison, WI, USA
[109] University of Wisconsin, Milwaukee, WI, USA
[110] Institute for Nuclear Science and Technology (INST), Hanoi, Vietnam
[†] Deceased
[a] at Konan University, Kobe, Japan

# The HEAT Telescopes of the Pierre Auger Observatory Status and First Data

T. HERMANN-JOSEF MATHES[1], FOR THE PIERRE AUGER COLLABORATION[2]
[1] *Karlsruhe Institute of Technology, D-76021 Karlsruhe, Germany*
[2] *Observatorio Pierre Auger, (5613) Malargüe, Av. San Martín Norte 304, Argentina*
*(Full author list: http://www.auger.org/archive/authors_2011_05.html)*
*auger_spokespersons@fnal.gov*

**Abstract:** The southern Pierre Auger Observatory was designed to detect ultrahigh energy cosmic rays above $10^{18}$ eV with high accuracy exploiting a hybrid detection technique. A surface array of 1660 water Cherenkov detectors on a 1500 m triangular grid covers an area of 3000 km$^2$. The atmosphere above the array is viewed by 24 wide angle telescopes. These telescopes observe the faint fluorescence light of the extensive air showers at dark moonless nights, i.e. with a duty cycle of about 15 %. As an enhancement to this baseline design of the Auger Observatory three additional telescopes with elevated field of view were built and now constitute HEAT, the high-elevation Auger telescopes. These telescopes are similar to the 24 other fluorescence telescopes (FD) but can be tilted by 29 degrees upward. They cover an elevation range from 30 to 58 degrees above horizon to enable the unbiased detection of nearby low energy air showers. Especially in combination with the detector information from an infill array of water tanks on a 750 m grid close to the HEAT site the energy range of high quality hybrid air shower measurements is extended down to below $10^{17}$ eV. HEAT is fully commissioned and is taking data continuously since September 2009. The status and prospects of HEAT are discussed and first (preliminary) data are presented.

**Keywords:** Pierre Auger Observatory, cosmic rays, HEAT, high-elevation fluorescence telescope

## 1 Introduction

The Pierre Auger Observatory was designed to measure the energy, arrival direction and composition of cosmic rays from about $10^{18}$ eV to the highest energies with high precision and statistical significance. The construction of the southern site near Malargüe, Province of Mendoza, Argentina has been completed since mid 2008 and the analysis of the recorded data has already provided important results with respect to, for example, the energy spectrum of cosmic rays [1], their distribution of arrival directions [2], their composition [3], and upper limits on the gamma ray and neutrino flux [4, 5]. The measured cosmic ray observables at the highest energies are suitable to tackle open questions like flux suppression due to the GZK effect, to discriminate between bottom-up and top-down models and to locate possible extragalactic point sources.

However, for the best discrimination between astrophysical models, the knowledge of the evolution of the cosmic ray composition in the expected transition region from galactic to extragalactic cosmic rays in the range $10^{17}$ eV to $10^{19}$ eV is required. Tests of models for the acceleration and transport of galactic and extragalactic cosmic rays are sensitive to the composition and its energy dependence in the transition region where the observatory with the original design had a low detection efficiency.

The fluorescence technique is best suited to determine the cosmic ray composition by a measurement of the depth of shower maximum. As the fluorescence light signal is roughly proportional to the primary particle energy, low energy showers can be detected only at short distances from the telescopes. In addition, as these showers develop earlier in the atmosphere, their shower maximum lies higher in the atmosphere and thus is not accessible to the standard Auger telescopes due to their limited field of view in elevation (30°). Furthermore, the geometric orientation of the shower axis with respect to the telescope imposes a bias on the shower selection [6]. This was the motivation to build HEAT, the high-elevation Auger telescopes. The three HEAT telescopes are similar to the 24 standard ones and can be tilted, providing the extension of the field of view to larger elevation angles. From the data collected with the Pierre Auger Observatory we know that the quality of the reconstruction is improved considerably if the showers are recorded by a hybrid trigger. Hybrid events provide information about the shower profile from the data of the fluorescence detector (FD) and of at least one surface detector (SD). The time and location of the shower impact point on ground set further constraints which improves the reconstruction of the shower energy significantly [7, 8]. The Pierre Auger Collaboration has built an additional infill surface detector array located at an optimal distance in the field of view of HEAT [9].

## 2 Design and properties of HEAT

In 2006, the Auger Collaboration decided to extend the original fluorescence detector, a system consisting of 24 telescopes located at four sites at the periphery of the surface detector array, by three High Elevation Auger Telescopes (HEAT). These telescopes have now been constructed, and they are located $180\,\mathrm{m}$ north-east of the Coihueco FD building. At the same time, the collaboration deployed extra surface detector stations as an infill array of $24\,\mathrm{km}^2$ close to and in the field of view of HEAT. The layout of HEAT and the infill array is shown in Fig. 1.

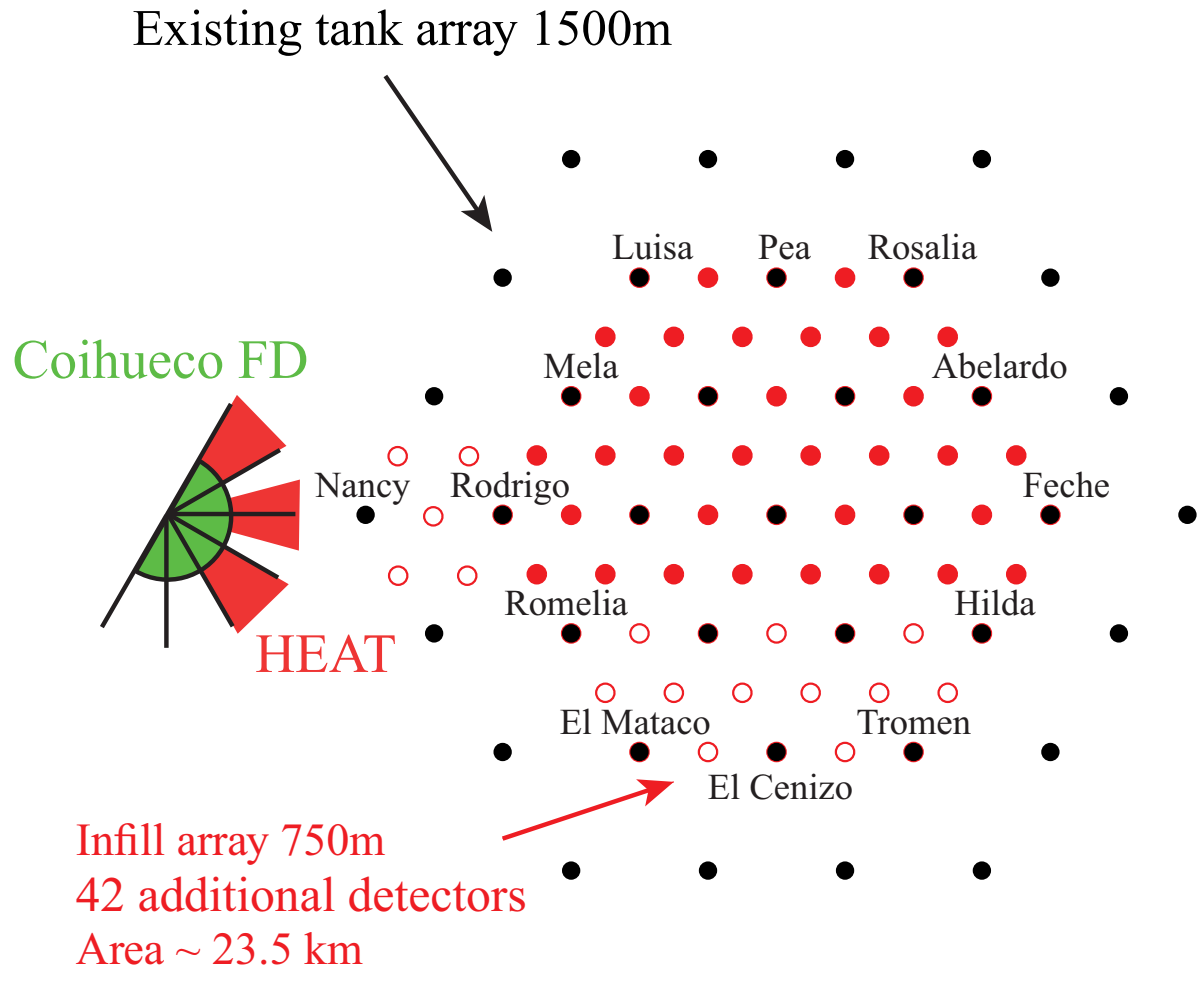

Figure 1: Layout of the infill array and the Coihueco and HEAT fluorescence telescopes. The additional infill stations are marked red (solid: already integrated in data taking).

The design of HEAT is very similar to the original FD system [10], except for the ability to tilt the telescopes upwards by 29°. In both cases a large field of view of about $30° \times 30°$ is obtained using Schmidt optics (approx. $30° \times 40°$ when tilted). Fluorescence light entering the aperture is focused by a spherical mirror onto a camera containing 440 hexagonal PMTs. A UV transmitting filter mounted at the entrance window reduces background light from stars effectively. An annular corrector ring assures a spot size of about $0.6°$ despite the large effective aperture of about $3\,\mathrm{m}^2$. The high sensitivity of the Auger FD telescopes enables the detection of high energy showers up to $40\,\mathrm{km}$ distance. A slow control system for remote operation from Malargüe allows safe handling.

Differences between the conventional FD telescopes and HEAT are caused by the tilting mechanism. While the original 24 FD telescopes are housed in four solid concrete buildings, the three HEAT telescopes are installed in individual, pivot-mounted enclosures. Each telescope shelter is made out of lightweight insulated walls coupled to a steel structure. It rests on a strong steel frame filled with concrete. An electrically driven hydraulic system can tilt this heavy platform by 29° within two minutes. The whole design is very rigid and can stand large wind and snow loads as required by legal regulations. All optical components are connected to the heavy and stiff ground plate to avoid wind induced vibrations and to keep the geometry fixed. Mirror and camera were initially adjusted in the horizontal position (service position). To ensure sufficient mechanical stability of camera body and mirror support system, additional steel rods and overall improved support structures are employed. The mechanical stability is monitored by means of distance and inclination sensors. The principle of tilting is illustrated for one bay in Fig. 2.

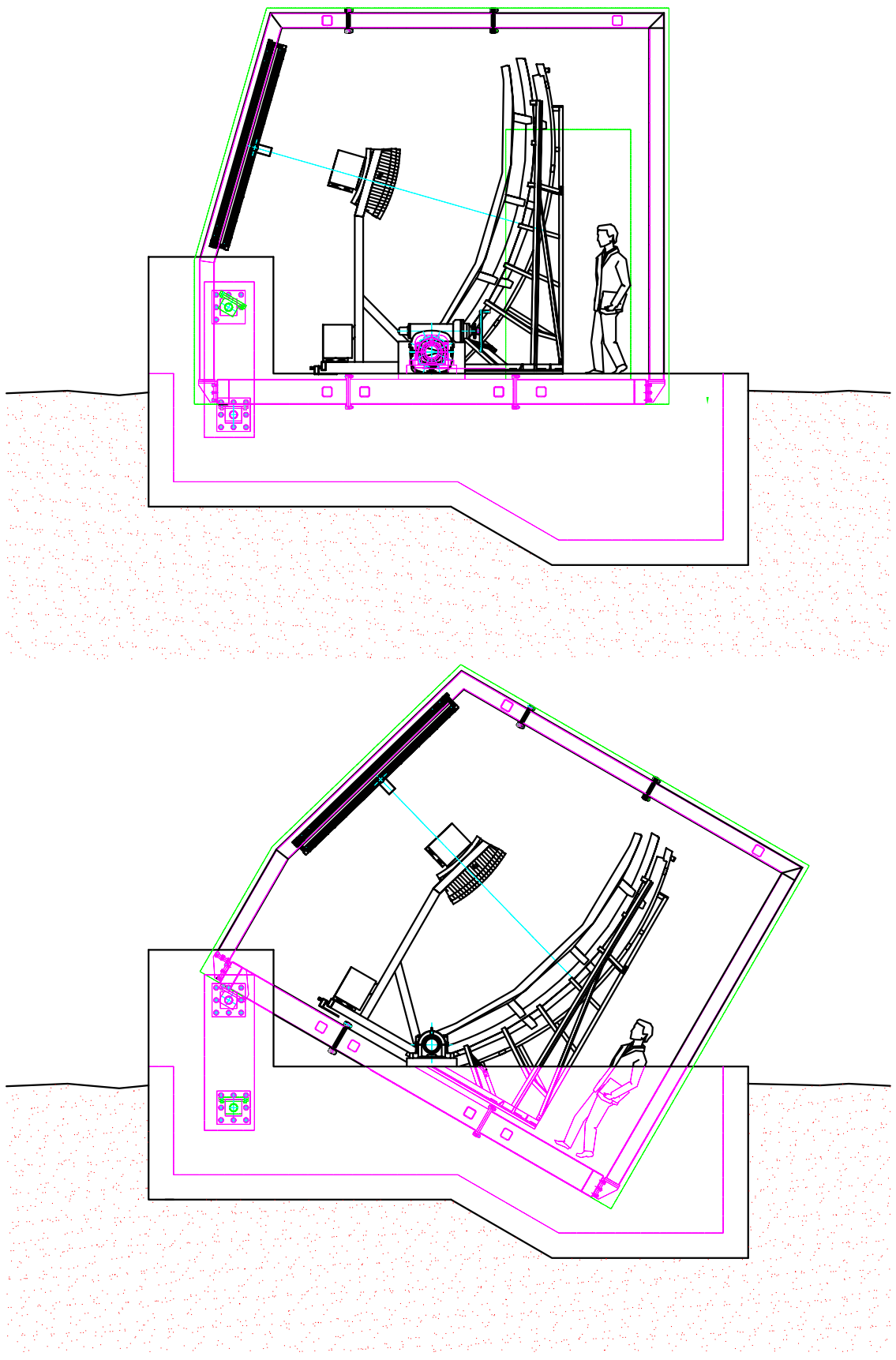

Figure 2: Schematic view of the cross-section of one of the HEAT telescopes. Top: Horizontal mode for service and cross-calibration, bottom: data-taking mode in tilted orientation.

Another design change for HEAT is the use of an improved data acquisition electronics (DAQ) whose concept and partitioning is, in principle, the same as with the previous version. The new design of the electronic is modernized and updated with larger and faster FPGAs. Along with that, the sampling rate of the digitizing system was increased from $10\,\mathrm{MHz}$ to $20\,\mathrm{MHz}$ and the overall readout speed can be potentially increased. From the point of view of the data taking and operation, HEAT acts as an independent fifth telescope site.

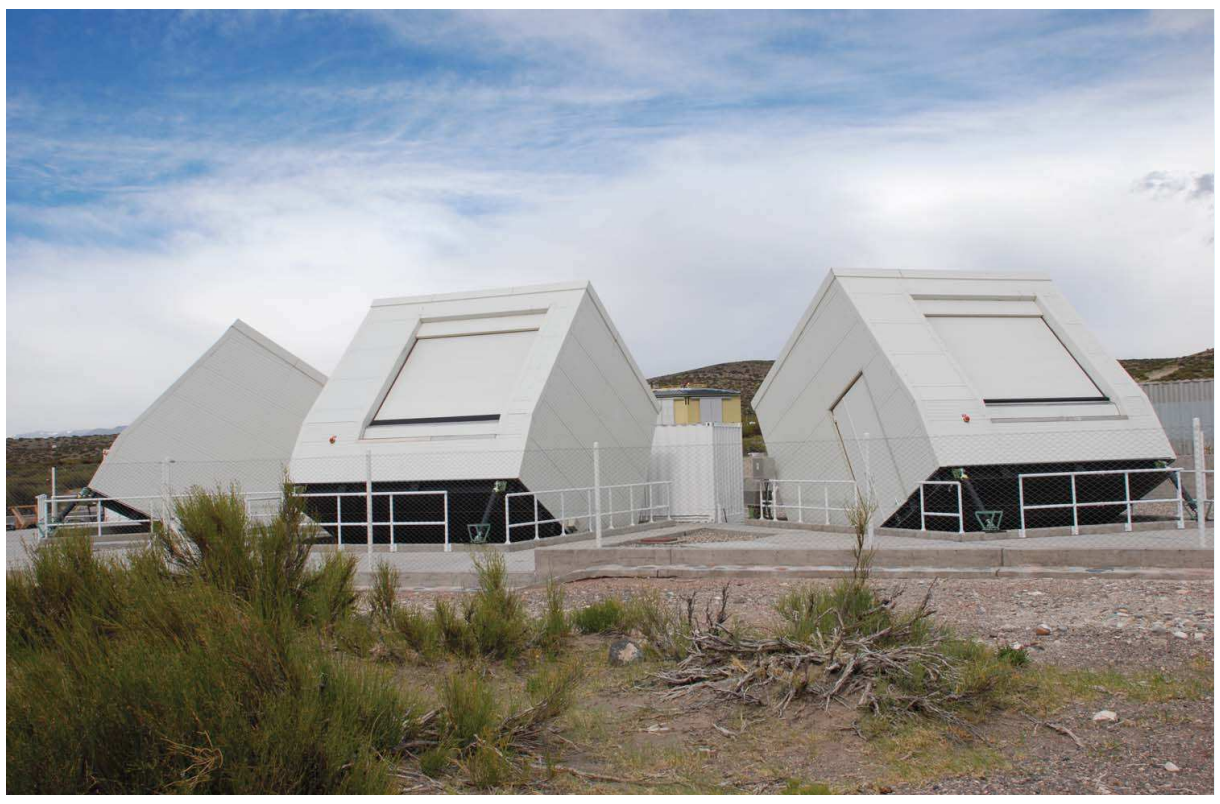

Figure 3: Photograph of HEAT in tilted mode with closed shutters.

The aforementioned distance monitoring system has been used to prove that the tilting of the telescope enclosures does not modify the optical parameters of the telescopes significantly. In addition, reference calibration measurements at different tilting angles have shown that the influence of the Earth's magnetic field on the performance of the PMTs is only marginal and can thus be neglected.

A photograph of HEAT in tilted mode is shown in Fig. 3.

## 3 Measurements

Data taking with the new telescopes of HEAT is possible in horizontal ('down') position as well as in the tilted ('up') position. The horizontal position of the HEAT telescopes, which is used for installation, commissioning and maintenance of the hardware, is also the position in which the absolute calibration of the telescopes takes place. In this position the field of view of the HEAT telescopes overlaps with those of the Coihueco telescopes. This offers the possibility of doing special analyses of events recorded simultaneously at both sites. In addition, these events can be used to check the alignment of the new telescopes and provide a cross-check of their calibration constants.

With the HEAT enclosures in the tilted position, the combined HEAT-Coihueco telescopes cover an elevation range from the horizon to 58°. This extended field of view enables the reconstruction of low energy showers for close-by shower events and resolve ambiguities in the $X_{\rm max}$ determination. The improved resolution in energy and $X_{\rm max}$ determination is especially visible in the low energy regime.

The first measurements with a single HEAT telescope started in January 2009 whereas measurements using the new DAQ electronics with all three telescopes commenced in September 2009. An example of one of the first low-energy showers recorded with HEAT and the Coihueco station is shown in Fig. 4.

The initial data taking period served as commissioning and learning period. Since June 2010, the data taking and data quality reached a satisfactory performance level and all results presented here are based on the latter data taking period. During this period, an absolute calibration campaign with a uniformly lit drum [11] and a roving laser for these new telescopes has been performed successfully.

The alignment of the regular fluorescence telescopes is obtained from star tracking. In addition to this method, a new method was introduced to determine the alignment of the HEAT telescopes. Given a reference geometry from any number of sources (SD, hybrid, reconstruction from other sites, laser shots) and the observation of the corresponding light traces in a HEAT telescope, the developed algorithm determines the optimal pointing direction for the telescope. The accuracy of the method increases when applied to many events. This method results in a statistical accuracy of $0.3°$ or better for elevation and azimuth. For the HEAT telescope 1, which has the Central Laser Facility (CLF) [12] in its field of view, accuracies of better than $0.1°$ can be achieved.

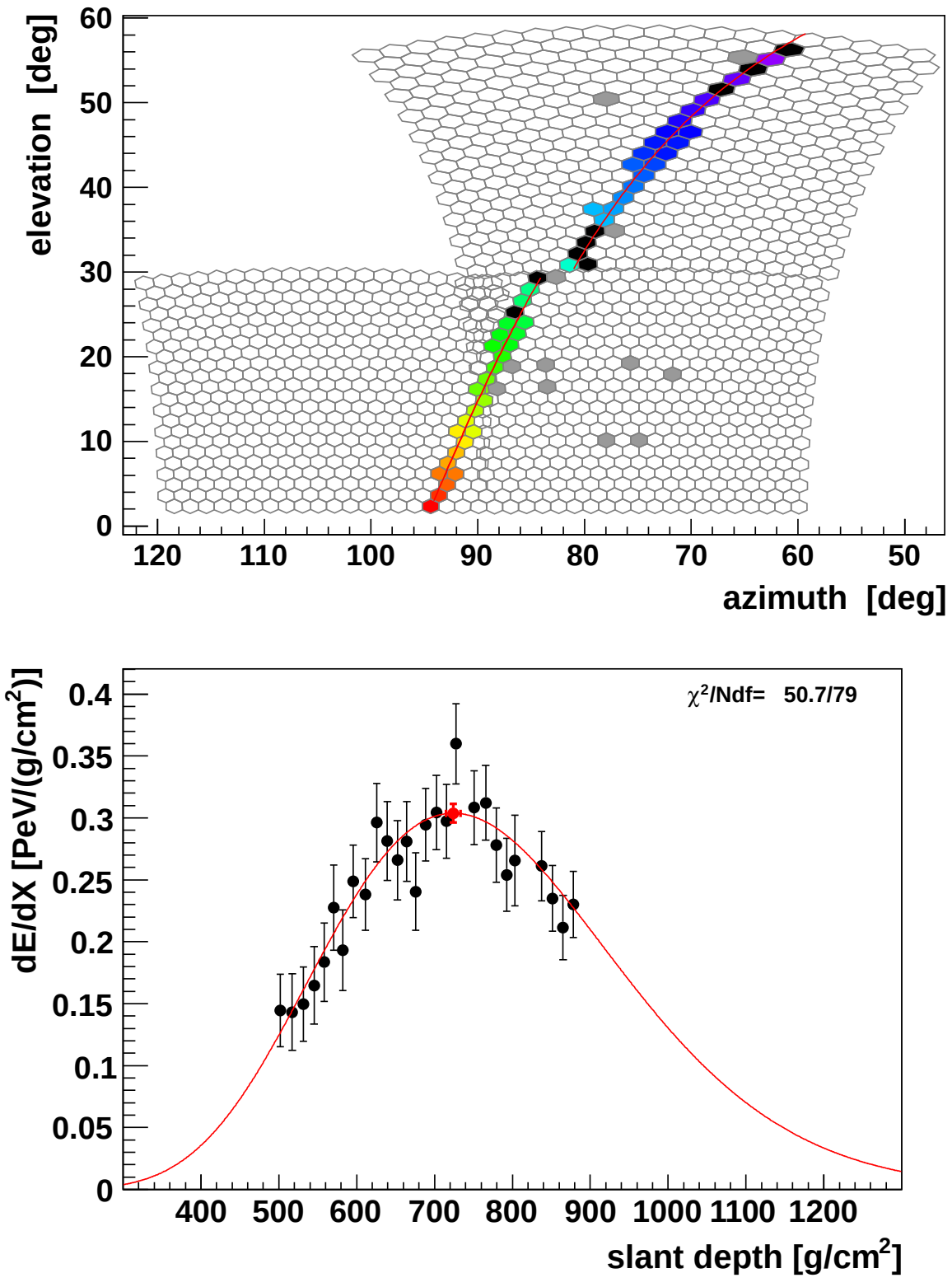

Figure 4: Example of a low-energy event recorded in coincidence with HEAT and two Coihueco telescopes. Top: Camera image of the recorded signal. The arrival time of the light is color-coded (blue early, late red). Bottom: Reconstructed energy deposit profile. This nearly vertical event with a zenith angle of 19° has a reconstructed energy of about $1.7 \times 10^{17}$ eV.

The different FD sites and the SD are operated independently. Their data are merged offline using the GPS pulse-per-second timestamps. It is thus vital for the reconstruction based on event times to measure and control the time-offsets between the components. Instead of minimizing the angular difference, a similar approach is also used to minimize the time difference between SD and HEAT. A correctly determined SD-HEAT time offset is also characterized by the fact, that in the combined SD-HEAT (time) geometry fit, the least number of pixels is rejected in the event reconstruction.

## 4 Data analysis

For the shower reconstruction it is desired to combine the data from HEAT and Coihueco. However, in the standard shower reconstruction chain the data from each building is used separately. Thus, the analysis software package Offline [13] has been generalized from a building-based to a telescope-based reconstruction. The different telescopes can then be combined in any desired form to build a virtual site.

In the same manner, the module for the hybrid geometry finding was extended in a way that shower detector plane (SDP) times and the time determined by the analysis of the SD data are combined in a kind of global fit procedure. In each fitting step the parameters describing the SDP and the shower core are calculated together.

Initially, when the HEAT telescopes went into operation, no absolute calibration for the HEAT telescopes was available, the gain of the camera was only flat-fielded. Using showers measured in coincidence with HEAT (downward mode) and Coihueco which fulfill certain quality criteria, a set of preliminary calibration constants could be achieved. The cross-check with the updated calibration constants (after an absolute calibration was performed) resulted in differences of 2 % or less (depending on the telescope). With this method also the energy resolution of the telescopes could be determined to be of the order of 10 %, see Fig. 5.

## 5 Conclusions

The telescopes of the HEAT site have operated since September 2009 and are producing high quality data in a stable manner. The first data from this site were used to produce rough calibration and alignment constants for this newly built part of the observatory.

Even from a short period of data taking so far it is clear that these new telescopes improve the quality of data for energy and mass composition analyses significantly at low energies. Their extended field of view allows for an unbiased measurement of $\langle X_{\max} \rangle$ down to much lower energies than has been possible with the standard Auger fluorescence detectors.

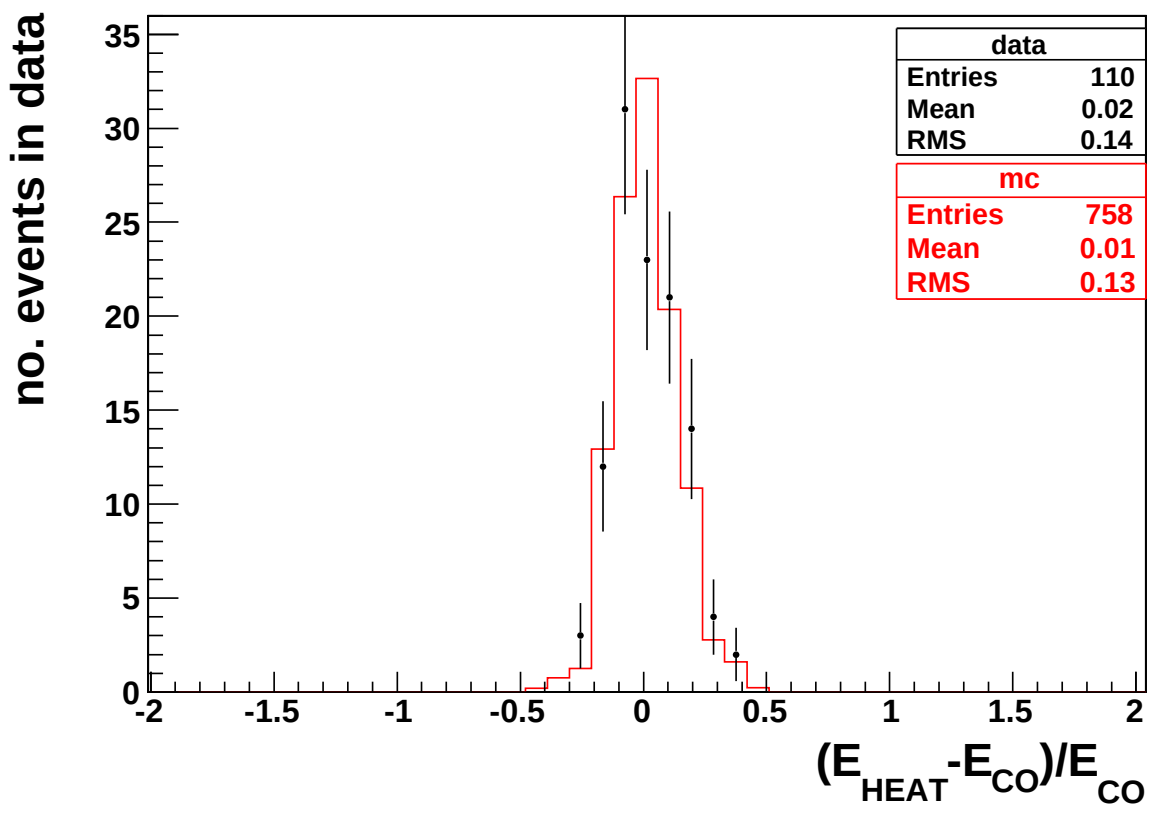

Figure 5: Determination of the energy resolution of the HEAT and Coihueco telescopes. Shown is the difference between the energy reconstructed with HEAT in downward mode with respect to that of Coihueco. Assuming that all telescopes have the same energy resolution one has to divide the RMS by $\sqrt{2}$ to obtain the resolution of one telescope. The histogram also shows simulated showers reconstructed the same way as data.

The improvements to the combined field of view of old and new detectors required major changes to the Offline analysis framework and the development of new reconstruction components. This is currently still work in progress.

# The AMIGA detector of the Pierre Auger Observatory: an overview

FEDERICO SÁNCHEZ[1] FOR THE PIERRE AUGER COLLABORATION[2]
[1]*Instituto de Tecnologías en Detección y Astropartículas (CNEA-CONICET-UNSAM), Av. Gral Paz 1499, (1650) Buenos Aires, Argentina*
[2]*Observatorio Pierre Auger, Av. San Martín Norte 304, 5613 Malargüe, Argentina*
*(Full author list: http://www.auger.org/archive/authors_2011_05.html)*
*auger_spokespersons@fnal.gov*

**Abstract:** The Pierre Auger Observatory is currently being enhanced with the AMIGA detector (**A**uger **M**uons and **I**nfill for the **G**round **A**rray) to bring the energy threshold down to $10^{17}$ eV and to enable the muon content of air showers to be determined. Its baseline layout consists of a 23.5 km$^2$ infilled area within the Pierre Auger Observatory array deployed with synchronised pairs of water-Cherenkov surface stations and buried scintillator counters that sample simultaneously the particles of air showers at ground level and at a depth of 2.3 m respectively. At present, both detectors are placed on a triangular grid of 750 m, half the spacing of the main array. In this work we present the status of AMIGA, the performance of the surface array and the analysis of the first data of the scintillator detectors.

**Keywords:** Pierre Auger Observatory, Low Energy Extensions, AMIGA

## 1 Introduction

The energy region from $\sim\!10^{17}$ eV to $\sim\!10^{18}$ eV is of the outmost importance to understand the origin of the high-energy cosmic-rays: it is the range where the transition from a galactic to an extragalactic dominated flux may occur. Measurements of the energy spectrum and the mass composition within that range are expected to enable discrimination between different astrophysical models [1, 2, 3]. The Pierre Auger Observatory has the unique characteristic of combining the observation of the fluorescence light induced by extensive air showers with the measurement of their secondary particles that reach the ground level. This hybrid approach allows cosmic ray observables to be interpreted with unprecedented precision.

The southern site of the Observatory, located in the Province of Mendoza, Argentina, spans an area of 3000 km$^2$ covered with over 1600 surface detectors (SDs) deployed on a 1500 m triangular grid. The SD array is overlooked by 24 fluorescence detector (FD) telescopes grouped in units of 6 at four sites on the array periphery. Each telescope has a $30^\circ \times 30^\circ$ elevation and azimuth field of view. The regular array of the Observatory is fully efficient above $3 \times 10^{18}$ eV [4] and in the hybrid mode this range is extended to $\sim\!10^{18}$ eV [5] which does not suffice to study the transition region.

The Auger Collaboration has already measured the energy spectrum of cosmic rays from $10^{18}$ eV to above $10^{20}$ eV [6]. The first enhancements of the Auger Observatory, AMIGA and HEAT (High Elevation Auger Telescopes [7]) aim at measuring the cosmic ray spectrum and its chemical composition components down to $10^{17}$ eV. Both extensions started in 2008 after the construction of the Observatory was completed. HEAT complements the Auger FD with three additional telescopes that are tilted upwards to extend the range of vertical viewing angles from $30^\circ$ up to $60^\circ$. In turn AMIGA consists of an array of water-Cherenkov detectors (WDs) set out on a hexagonal spacing with sides of 750 m and 433 m (named *graded infill* array or simply *infill*) and an associated set of muon detectors (MDs) each of 30 m$^2$ buried at a depth of 2.3 m corresponding to 540 g cm$^{-2}$. Any impinging muon with energy $\geq$1 GeV propagates in the soil and is capable of reaching the buried muon detectors that are located near to the WDs as shown in Fig. 1. The new MD is the core of AMIGA that further develops the Auger Observatory as a multi-detector facility. Since the cosmic ray flux increases rapidly with the decreasing energy, the 750 m infill is laid out over an area of 23.5 km$^2$ while the planned 433 m array will cover only 5.9 km$^2$ within the larger area. The infill spacings allow cosmic rays to be detected with full efficiency down to an energy of $3 \times 10^{17}$ eV and $10^{17}$ eV respectively.

In the following sections we will describe the status of the AMIGA project. A description of the detector performance is presented in [8, 9].

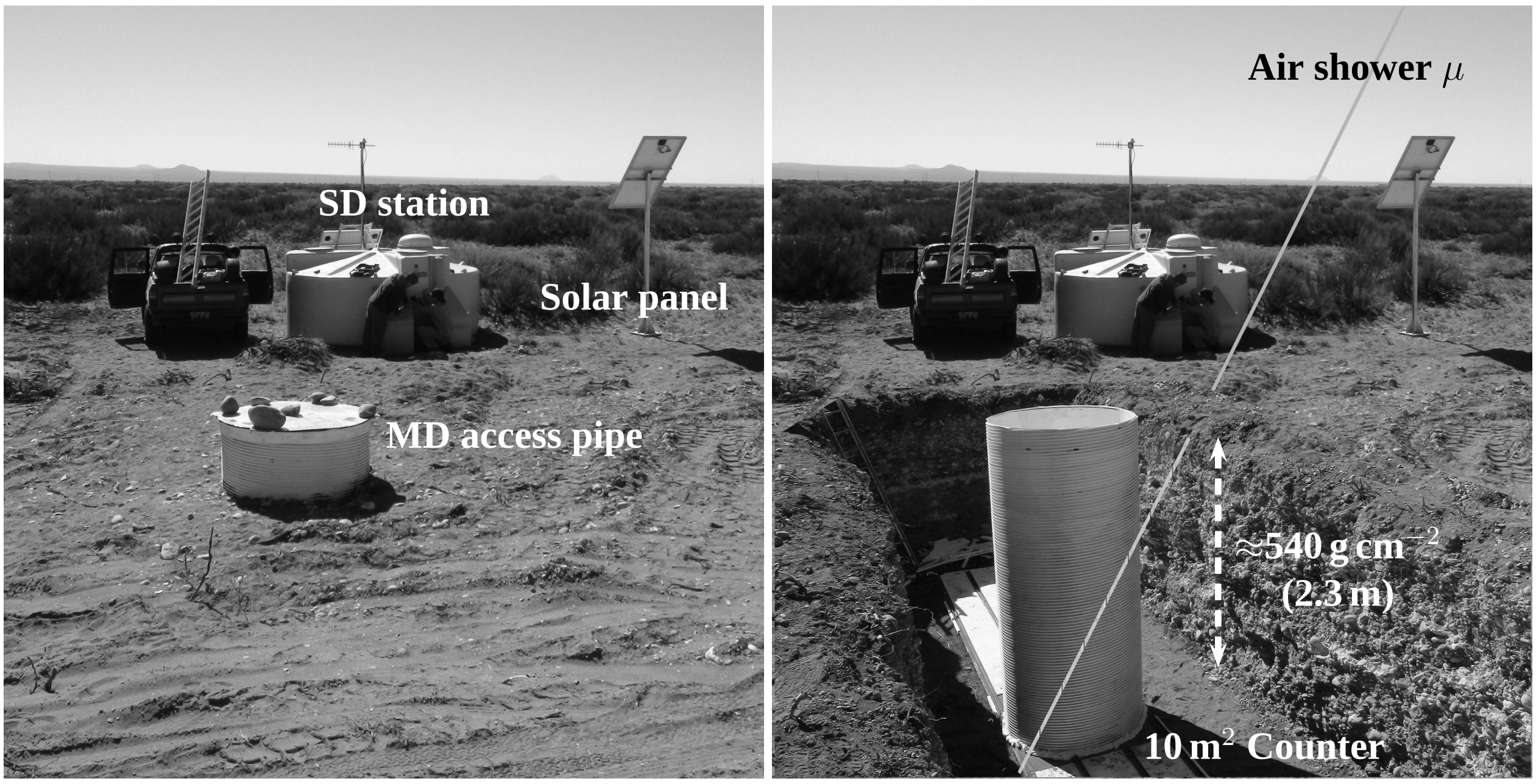

Figure 1: AMIGA concept: (Left) Surface infill SD station with its associated muon counter already buried. Once instrumented, the access pipe is filled with bags containing soil from the installation site. (Right) Photo-montage to depict the detector concept: any impinging muon with energy ≥1 GeV propagates in the soil and is capable of reaching the buried scintillator.

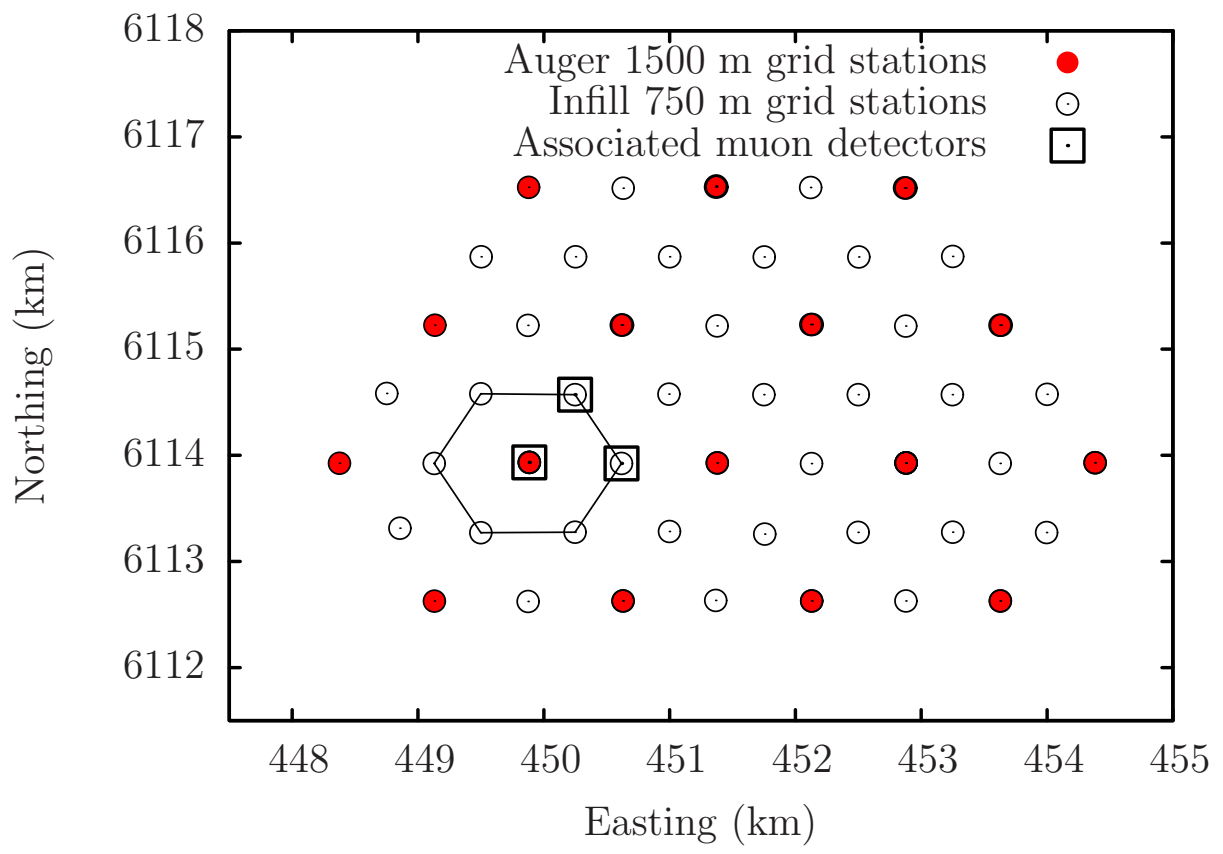

Figure 2: AMIGA array status in April 2011: the surface stations deployed on the 750 m grid are shown together with the associated muon detectors. The prototype MD will consist of 7 counters on the marked hexagon.

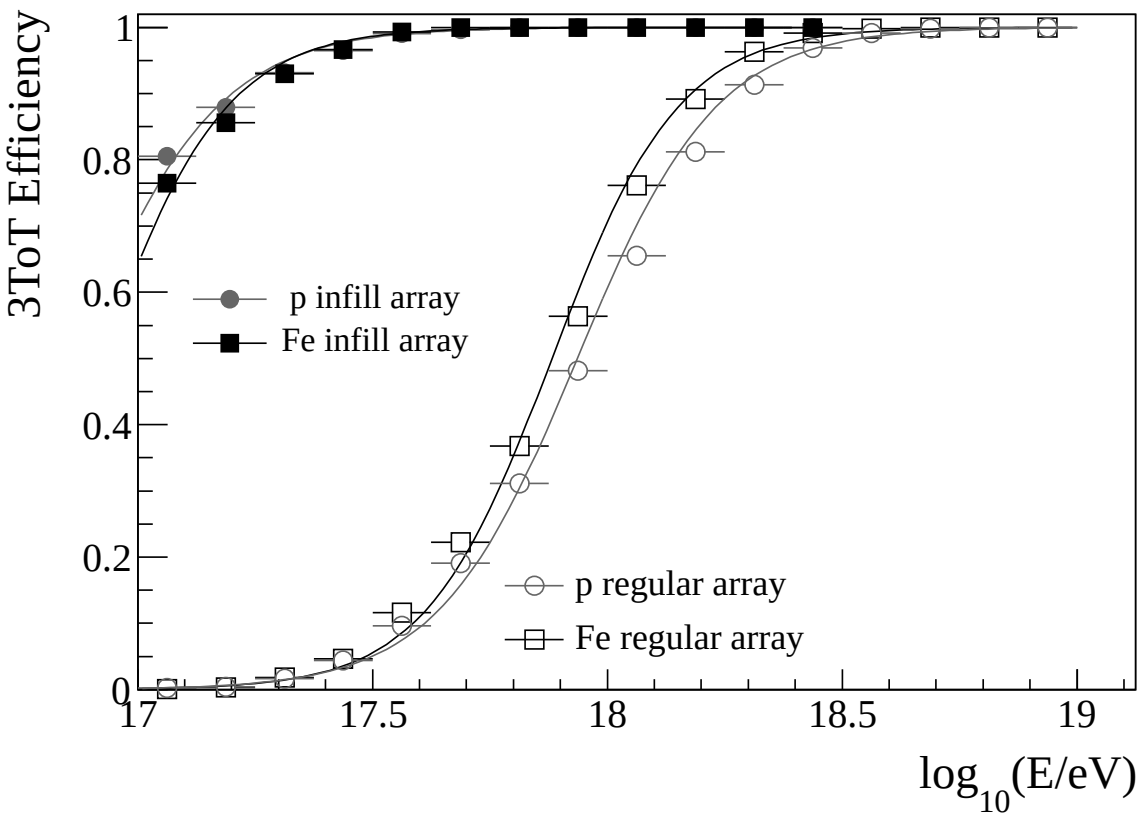

Figure 3: 3ToT trigger efficiency for the infill and regular array obtained from simulations of iron and proton primaries.

## 2 Surface Graded Infill

As of April 2011, 53 out of 61 surface stations planned for the 750 m infill have been deployed (see Fig. 2). For the smaller 433 m grid 24 new detectors will be installed after the completion of the 750 m array.

The water-Cherenkov detectors used in the infill array are identical to those used in the main array with the benefit of being a well-proven technology. Moreover, the infill is embedded in the regular array and therefore the same selection and reconstruction strategies may be employed to the observed showers. The trigger efficiency, the aperture and exposure calculation, the event selection, the geometry reconstruction and lateral distribution functions used (LDFs), as well as the energy estimator and the energy calibration, all benefit from algorithms tested successfully over the past years for the regular Auger SD (for a detailed description see [8]). Any infill event with at least 3 stations forming a triangle and satisfying a local trigger of the type Time-over-Threshold (3ToT) [4] is accepted. The trigger efficiency as a function of energy for simulated 3ToT events with zenith angles below 55° is shown in Fig. 3 for both the regular and the infill array. As can be seen, the 750 m spacing of the infill allows cosmic rays to be detected down to an en-ergy of $3 \times 10^{17}$ eV with full efficiency. Integrating the instantaneous effective infill aperture over the time when the detector was stable, the exposure between August 2008 and March 2011 amounts to (26.4±1.3) km$^2$ sr yr. With the current configuration consisting of 16 hexagons the mean rate of 3ToT events is 55 events/day/hexagon out of which ~51% satisfy the fiducial selection that allows events that fall close to the boundary of the array to be rejected.

As was done for the 1.5 km array, measurement uncertainties have been derived from data [10]. The angular resolution of the 750 m array was found to be 1.3° for events with at least 4 stations. To reconstruct the events, the distribution of SD signals on ground as function of their distance to the shower axis is fitted with a LDF, S(r). The optimum distance, $r_{\rm opt}$, where the signal fluctuations are minimised depends on the array spacing. The signal S($r_{\rm opt}$) is the ground parameter eventually used to obtain an energy estimator. For the regular array $r_{\rm opt}$=1000 m whereas for the 750 m infill was found to be 450 m. Besides the uniform treatment of data from the highest energies down to $3 \times 10^{17}$ eV, an additional advantage of having the infill within the regular array is that it is possible to make cross-checks of results in the overlap region. As an example the LDF of a well-contained infill event of $2.7 \times 10^{18}$ eV impinging with zenith angle of 27° is shown in the top panel of Fig. 4. The same event reconstructed using stations from the regular array alone is illustrated in the bottom panel. Both reconstructions are compatible. From event to event, the main statistical sources of uncertainty in the parameter S(450) are the shower-to-shower fluctuations, the finite size of the detectors and the sparse sampling of the LDF. There is also the systematic contribution due to the lack of knowledge of the LDF. The total S(450) uncertainty derived from the infill data goes from ~22% at 10 VEM to ~13% at 100 VEM. The energy of an event is estimated from the ground parameter independently of the zenith angle of the air shower by means of the *Constant Intensity Cut* (CIC) method [11]. The method allows S(450) to be evaluated at a reference angle of 35° ($S_{35}$). $S_{35}$ is calibrated using the events simultaneously observed by the SD and the FD. The infill array was found to be fully efficient from $S_{35}$~20 VEM.

## 3 The Array of Muon Detectors

The MD is currently in its prototype phase, named the *Unitary Cell* (UC). The UC will consist of 7 buried detectors to be installed on one hexagon and on its centre. Once completed, the UC will be composed of 30 m$^2$ scintillator counters. Each counter will consist of two 10 m$^2$ plus two 5 m$^2$ modules associated to a single water-Cherenkov detector. In turn each module is made up of 64 scintillator strips 4.1 cm wide × 1.0 cm high. The length of the strips is 4 m and 2 m for the 10 m$^2$ and 5 m$^2$ modules respectively [9]. At present, three 10 m$^2$ detectors have been deployed in the positions shown in Fig. 2 with prototype electronics sampling every 12.5 ns (80 MHz). At the hexagon centre, an additional 5 m$^2$ module was also installed.

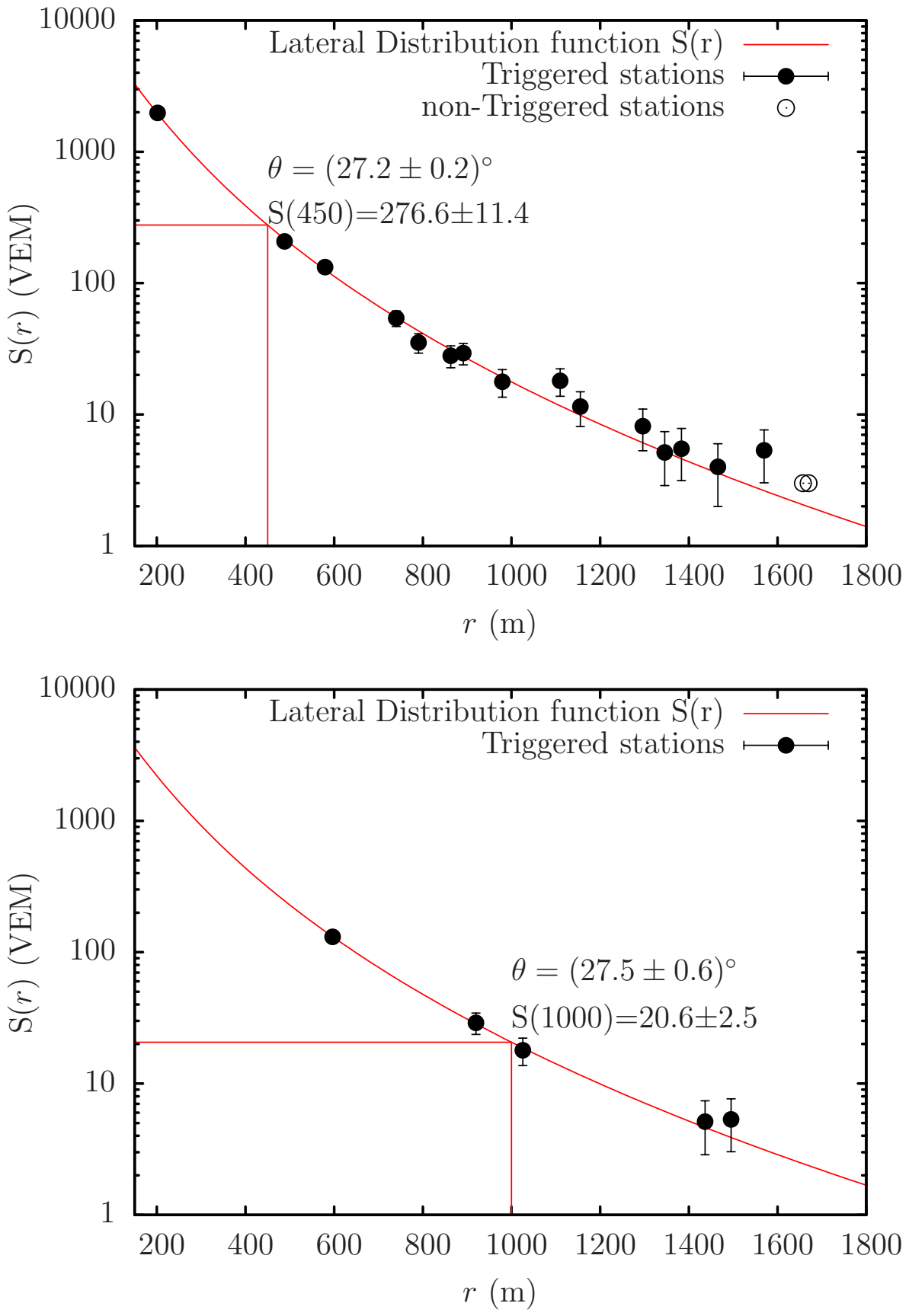

Figure 4: The same event reconstructed with the infill (top) and the regular (bottom) array. Solid and open circles are triggered and non-triggered stations respectively.

Although the MD has an internal stand-alone trigger mode of operation for monitoring and self-calibration purposes (see [9] for details), the buried scintillation detectors are triggered by signals from the associated WD station. The event geometry and the primary energy of the showers detected by AMIGA are reconstructed by means of the SD data alone as explained in the previous section. Once the shower parameters are established, the MD data are analysed [9] to provide the number of muons of the observed event. As of April 2011, the surface stations provide their lower-level trigger (T1, in single mode) to the associated scintillation detectors. The electronics required to link the MD data with the SD higher trigger levels is under development and therefore, is not yet possible to include the muon data in the shower reconstruction chain.

We will now describe the observed correlation, at T1 level, between the MD acquisitions and its associated SD station.

The surface stations have two independent T1 trigger modes [4]: a mode based on threshold discrimination (TH-T1), and a mode where the signal above threshold must be maintained over more than 325 ns (ToT-T1). The rate of triggers which satisfies the TH-T1 condition in the WD is around 100 Hz while the rate of those satisfying the ToT-T1 is ~1.5 – 2 Hz. Once the MD is triggered, it records the digital signal of each one of the 64 scintillator strips in a local memory. Each signal has 768 logical samples and up to 1024 events can be stored. Within each signal, a logical 1 is stored every time the corresponding pulse is above a certain (adjustable) discrimination threshold.

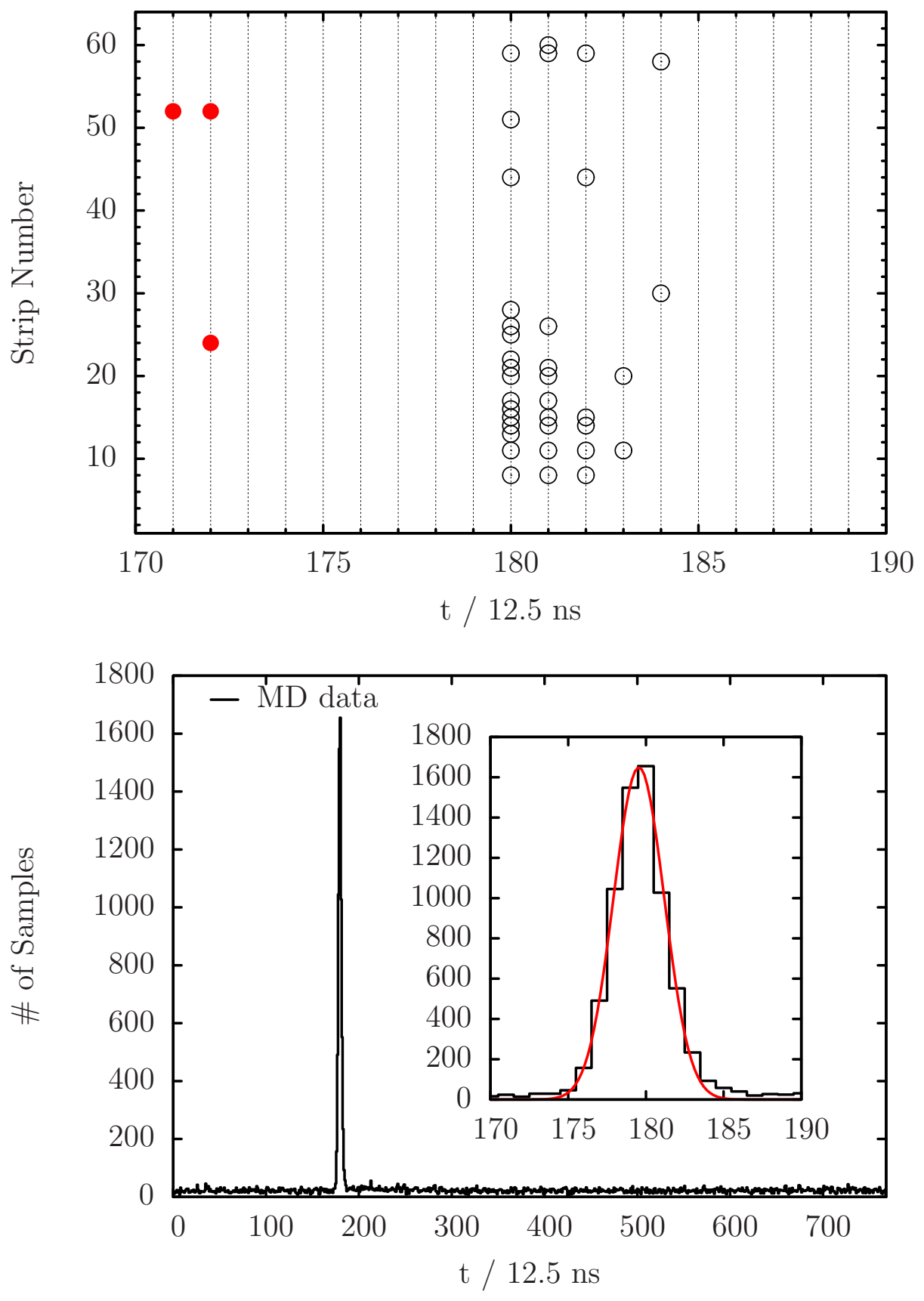

Figure 5: (Top) Two superimposed T1 events in the MD: a typical one with low-multiplicity (solid circles) and an unusual one with high-multiplicity (open circles). The low-multiplicity events account for around 95% of the T1 events. (Bottom) Projection over the time axis of the logical 1s of MD events triggered by WD T1s.

Only 1% of the WD T1 events also have data in the MD. The rate of these events is around 1.6 Hz. Furthermore, within this small fraction, in 95% of the cases fewer than 4 strips were struck. These low-multiplicity events spread in time across 1 or 2 intervals of 12.5 ns. In the top panel of Fig. 5 two T1 events, as recorded by the MD, are shown superimposed as an example. One is a typical low-multiplicity event while the other is a rather unusual event with high-multiplicity. The projection over the time axis of the logical 1s of all non-empty MD events triggered by the WD T1s is shown in the bottom panel of Fig. 5. The discrimination threshold was set to ~80% of the mean single photo-electron pulse height. Around 40% of these events contribute to the well-defined Gaussian peak illustrated in the inset, showing a time correlation with the WD T1s and therefore indicating a common physical source. The remaining 60%, which constitutes the histogram baseline, is most likely atmospheric background given that it is uniformly distributed in time. The same results were obtained with different discrimination levels.

## 4 Discussion and Conclusion

The AMIGA enhancement of the Pierre Auger Observatory is being built. It comprises both, SD stations identical to the ones used in the main array of the Observatory deployed over an infilled area, and buried scintillation detectors that are used to count muons. The deployment of the SD infill component started in 2008 and more than 85% of the 750 m array has already been deployed. The prototype phase of the MD consists of 30 $m^2$ scintillation counters to be associated to 7 SD infill stations corresponding to one 750 m hexagon and to its centre. At present, four scintillator modules have been deployed. The uncertainties of the AMIGA infill have been studied with data and simulations using well-proven algorithms developed for the regular array during the past years. The total S(450) uncertainty goes from ~22% at 10 VEM to ~13% at 100 VEM. At the highest energies, the main driver of the uncertainties are the shower-to-shower fluctuations. The angular resolution of the infill array is below 1.3° for events with at least 4 SD stations. The first data from the muon detectors being triggered by the lowest level trigger signal of the associated WD station were analysed. Although it is still preliminary, a time correlation between the T1 trigger of the water-Cherenkov stations and the muon detector events was shown indicating a common physical source of both signals. The analysis of the MD events is on-going.

# The AMIGA muon counters of the Pierre Auger Observatory: performance and first data

BRIAN WUNDHEILER[1] FOR THE PIERRE AUGER COLLABORATION[2]
[1] *Instituto de Tecnologías en Detección y Astropartículas (CNEA-CONICET-UNSAM), Av. Gral Paz 1499 (1650) Buenos Aires, Argentina.*
[2] *Observatorio Pierre Auger, Av. San Martín Norte 304 (5613) Malargüe, Argentina.*
*(Full author list: http://www.auger.org/archive/authors_2011_05.html)*
*auger_spokespersons@fnal.gov*

**Abstract:** In this paper we introduce a full simulation of the AMIGA muon counters including, the modeling of its scintillators, wavelength shifter fibers, multi-anode photomultipliers, and front-end electronics. A novel technique for muon counting for such underground detectors based on their signal-time structure is presented. The proposed counting technique is evaluated with real and simulated muon pulses. Simulations of extensive air showers and particle propagation through matter were included in an end-to-end simulation chain. Preliminary results of the first muon counter modules installed at the Pierre Auger Observatory are presented.

**Keywords:** Muon Counter, AMIGA, Pierre Auger Observatory.

## 1 Introduction

The transition from galactic to extragalactic cosmic rays is still a poorly known phenomena, and composition studies are fundamental to thoroughly understand it. The Pierre Auger Collaboration is building an upgrade named AMIGA (**A**uger **M**uons and **I**nfill for the **G**round **A**rray) to both extend the energy threshold of the ground array down to 0.1 EeV, and determine the muon component of extensive air showers aiding primary particle identification. AMIGA consists of an infilled area having 85 pairs of water Cherenkov surface detectors and scintillator muon counters buried underground to avoid the electromagnetic component of the shower to be detected [1, 2].

In this work we describe a full simulation for the AMIGA muon detectors, and introduce a novel technique for muon counting. Finally, we present some preliminary outcomes of the first counter modules installed at the Pierre Auger Observatory.

## 2 The muon counters

The AMIGA muon detector (MD) consists of $30\,\mathrm{m}^2$ scintillator counters buried 2.3 m underground. The MD counts muons of air showers observed by the Pierre Auger Observatory, which are reconstructed by its Surface and Fluorescence Detector systems. Each counter has three $10\,\mathrm{m}^2$ modules with $4.1\,\mathrm{cm}$ wide $\times$ $1.0\,\mathrm{cm}$ high $\times$ $400\,\mathrm{cm}$ long strips. Each module consists of 64 strips made of extruded polystyrene doped with fluor and co-extruded with a $TiO_2$ reflecting coating. The strips have a groove where a wavelength shifter optical fiber (WLS) is glued and covered with a reflective foil. The manifold of fibers ends in an optical connector matched to a 64 multi-anode PMT from the Hamamatsu H8804 series ($2\,\mathrm{mm} \times 2\,\mathrm{mm}$ pixel size). Scintillators are grouped in two sets of 32 strips on each side of the PMT.

The bandwidth of the front-end electronics is set to 180 MHz to limit the pulse width. Signal sampling is performed by an FPGA at 320 MHz with an external memory to store up to 6 ms of data, equivalent to 1024 showers [3, 4]. Stored samples of each strip are collections of logical 1 or 0 depending on whether the signal surpasses a given adjustable threshold, which is foreseen to be around 30% of the mean height of a single photoelectron (SPE). This method is very robust since it neither relies on deconvoluting the number of muons from an integrated signal, nor on the PMT gain or its fluctuations, nor on the muon hitting position on the scintillator strip and the corresponding light attenuation along the fiber. It also does not require a thick scintillator to control poissonian fluctuations in the number of SPEs per impinging muon. This *one-bit* electronics design relies on a fine counter segmentation in strips to prevent undercounting due to simultaneous muon arrivals [5]. One unwanted feature of multi-anode PMTs is the crosstalk (XT) between neighboring pixels. In systems in which the discrimination is set below the SPE height, this effect can lead to a considerable overcounting. An efficient technique is then needed to avoid this effect while not missing real signals in the process. This subject is addressed in section 3.2.

## 3 Simulations and counting strategies

The MD simulation chain implements phenomenological models which include several experimental parameters. In this way most of first principle processes are parametrised improving considerably the computational performance. Before considering the counting techniques, two main blocks of the simulation are described in the following section: injection of particle traces and front-end electronics signal processing.

### 3.1 Injection methods and front-end electronics

While the front-end simulation is common for every particle injection method, there are three alternatives for injecting traces: injection of real particle traces (*real injection method*), traces from simulated particle tracks (*simulated injection method*), and traces derived from simulated energy deposition (*energy deposition method*).

*Real injection method*

This method allows laboratory measured muon pulses to be incorporated as an input to the front-end electronics simulation. These pulses are measured using the same type of scintillators, fibers and PMTs employed in the construction of the MDs. In Fig. 1 a real muon pulse is shown. Each measured signal is interpolated with a spline function to feed the analog electronics stage of the simulation.

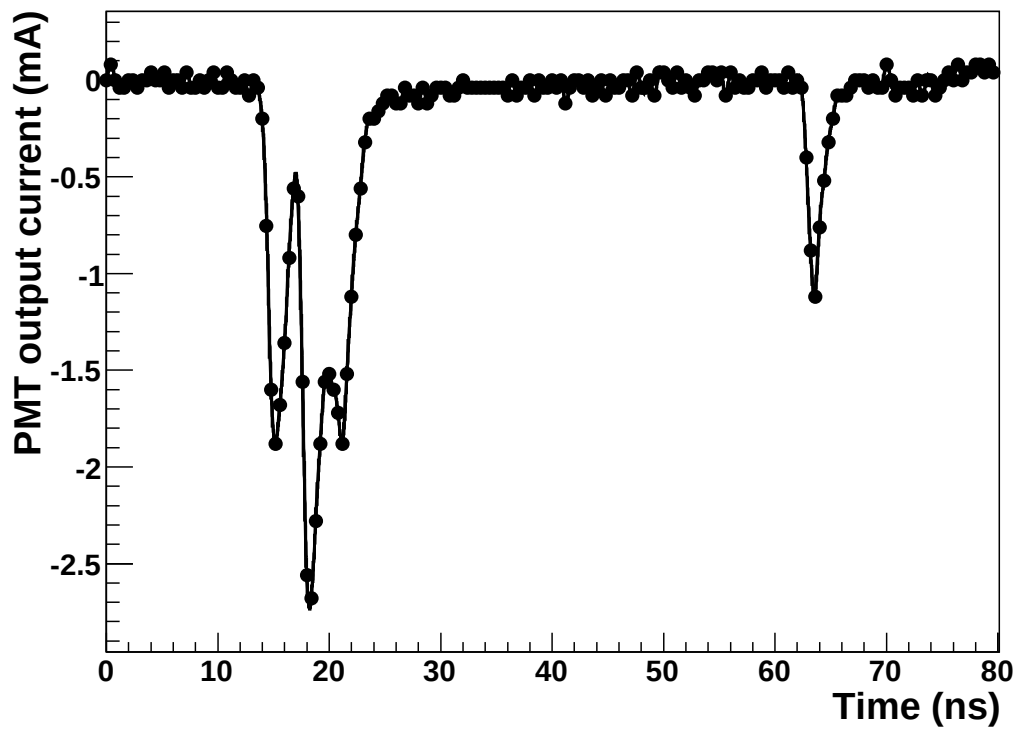

Figure 1: A laboratory measurement of a muon pulse is incorporated as coming out of the PMT in the real injection method. A spline interpolation is used as an input of the front-end electronics in the next link of the simulation chain.

*Simulated injection method*

In this method both the muon impinging position on the MD and its momentum direction can be selected. The muon signal is modeled as a superposition of SPE pulses

$$I^{\mu}(t) = \sum_{i=0}^{N_{spe}} I_i^{spe}(t - t_i). \quad (1)$$

Each $I^{spe}$ is taken to be of gaussian shape, whose amplitude and time width are fluctuated around their mean measured values. Its reference time $t_i$ has two contributions: the decay time of both scintillator and fiber, and the delay due to propagation through the WLS. While the latter is geometrically calculated, the mean value of the former is measured and then used to randomly generate exponentially distributed times. The number of SPEs ($N_{spe}$) is a poissonian random integer whose mean value depends on the impinging point according to the $\langle N_{spe} \rangle$ obtained from the measured fiber attenuation curve. This mean value is rescaled with the particle zenith angle of incidence (as $1/\cos\theta$). The XT effect is incorporated in the simulation by tossing the destination of each SPE produced: it can end up in the pixel corresponding to the impinged scintillator, or in any of its neighbors. The probability is determined according to the XT ratio which is measured for each of the 64 pixels.

*Energy deposition method*

Finally, in the third method of injection, energy deposition ($E_{dep}$) on a given point of the MD is indicated. This input can be generated with dedicated packages that simulate particle propagation through matter [6], as described in section 4. The deposited energy is normalised with the $\langle E_{dep} \rangle$ of vertical simulated muons in 1 cm thick scintillator [7] and correlated with $\langle N_{spe} \rangle$ extracted from the same attenuation curve used in the previous method.

*Front-end electronics*

The analog front-end behavior is mimicked as being an ideal inverter amplifier with a 3 dB point at 180 MHz. Each real or simulated muon pulse is convoluted with the amplifier transference function to obtain the analog voltage response. A voltage threshold is set to discriminate the amplified signal resulting in a digital pulse. The fall and rise times of the discriminator are taken to be finite at a constant 2.2 V/ns.

The simulated FPGA sampling, which is performed on the digital pulse of the previous stage, has two levels of thresholds. If the signal is higher than a given $V_{high}$ the sample is set to 1, if the signal is lower than $V_{low}$ it is set to 0. If the signal falls in between these voltages of reference, the sample remains in the same state as the previous one.

In Fig. 2 the processing of the real muon pulse of Fig. 1 by the simulated front-end electronics is shown. Since the bandwidth is set to 180 MHz, higher frequency details of the muon pulse are weakened. The discriminated signal at a threshold of 30% $\langle V_{spe} \rangle$ is also shown. The displayed

FPGA samples of this signal constitute the digital trace to be analysed in the counting process.

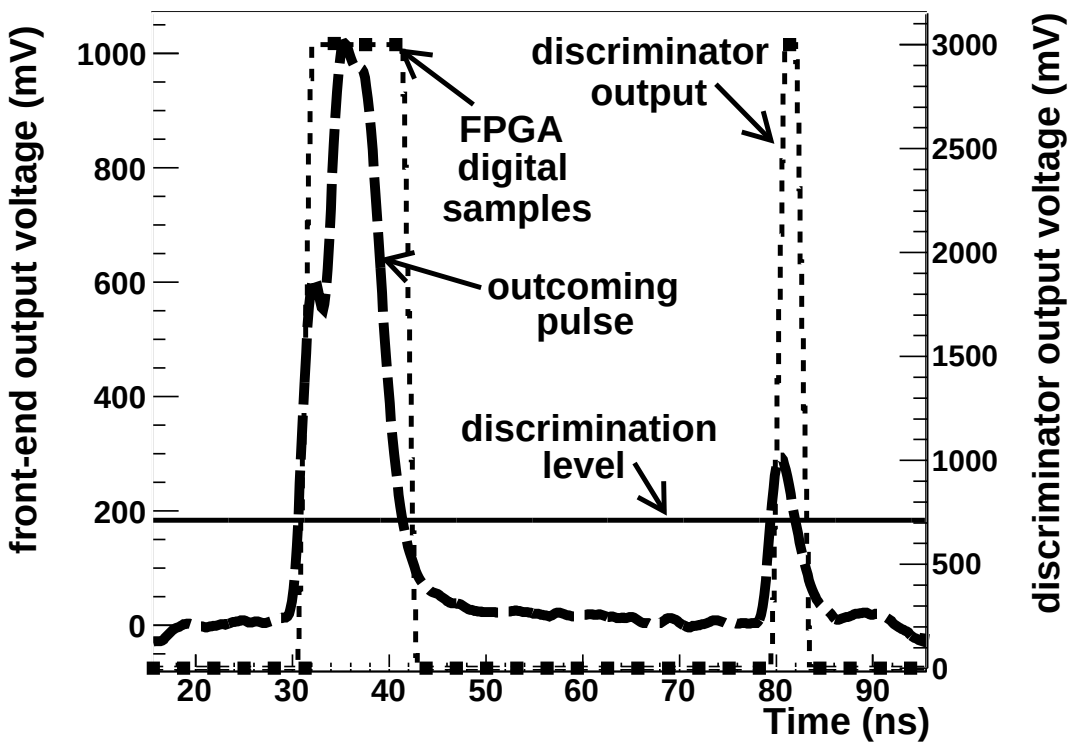

Figure 2: Simulated electronics response to a real muon pulse. The signal is inverted, amplified and the higher frequencies are filtered (thick dash line), the discrimination level is at 30% $\langle V_{spe} \rangle$ and FPGA sampling produces the digitised signal.

## 3.2 Analyses and counting techniques

The digital traces obtained by any of the previous methods, must be analysed to determine the number of muons. The counting strategies are based on a pattern recognition within a given time window $w$. Three strategies are considered: $nQ_w$, $nC_w$, and $nG_w$. They are defined as follows: a muon is counted with $nQ_w$ if $n$ *ones* are found in any position in the time window, with $nC_w$ the *ones* must be also consecutive, and with $nG_w$ if $n$ consecutive *ones* are found with one sample (0 or 1) among them. The $w$ parameter depends mainly on the fiber type and on the PMT model and operation. It can be selected from measurements of muon pulse widths at the discrimination level, or from simulated signals if they can reproduce the real time width distributions. As an example, in Fig. 3 a comparison between real muon pulses with simulated ones is made.

Simulated muon pulses were analysed with each technique, the $2G_{30\mathrm{ns}}$ strategy was found to have the best counted-to-impinging muons ratio when sampling at 320 MHz. Using the real injection method a counting ratio of 99% was found at 110 cm from the PMT, 99% at 297 cm, and 88% at 482 cm. The last point corresponds to the longest length of fiber present in the MDs, where light attenuation makes pulses more likely to be one SPE than in any other position, setting a lower limit to the counting ratio. In Fig. 4 the counting ratio of each strategy is evaluated at different thresholds for simulated muons at 2 m from the PMT. The $2G_{30\mathrm{ns}}$ is the best suited method for counting muons as it eliminates the overcounting due to XT regardless of the threshold, and it has the best counting ratio for the main pixel at 30% $\langle V_{spe} \rangle$. The measured mean width of an SPE pulse after discrimination is (3.75 $\pm$ 0.36) ns. The FPGA working at 320 MHz can sample one SPE pulse at most twice consecutively. The $2G_{30\mathrm{ns}}$ precludes it to be counted demanding an extra sample between two *ones*, irrespective of whether it is 0 or 1.

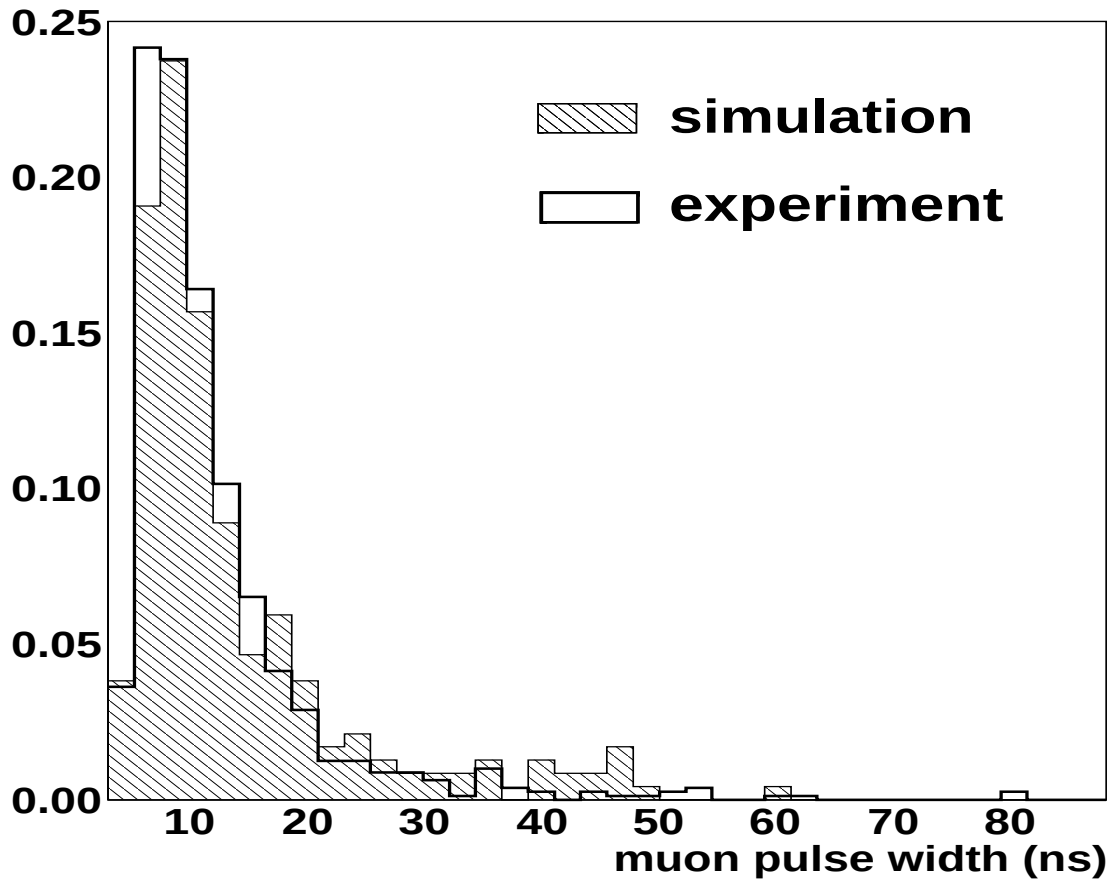

Figure 3: Normalised histograms for the width of real and simulated muon pulses at an impinging distance of 1.5 m from the PMT at 30% $\langle V_{spe} \rangle$ level.

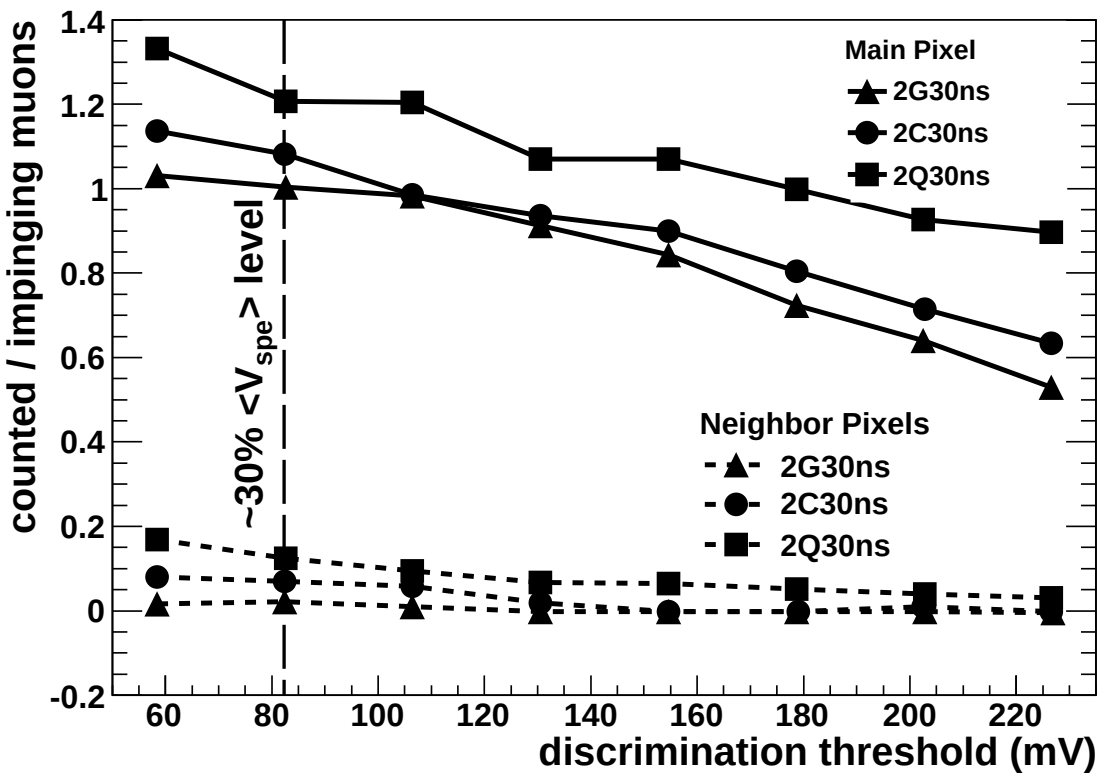

Figure 4: Counting techniques at 2 m from the PMT with simulated muons. The $2G_{30\mathrm{ns}}$ strategy has a counting ratio of 1 at 30% $\langle V_{spe} \rangle$ whereas the sum of the contributions of the neighboring pixels is negligible.

# 4 First data from the Observatory

The Pierre Auger Collaboration has installed four MD modules at the Observatory site since Nov 2009. The MD is meant to be triggered externally by its associated SD station [2]. Nevertheless an internal trigger mode is used for debugging the system and for monitoring and calibration purposes. This internal trigger is activated whenever $n$ or more channels coincidently show discriminated signals. The software development for the 320 MHz FPGA sampling is in its final stage, the preliminary measurements on site were taken at 80 MHz. A clustering structure, which shows activated strips grouped together, is very common to be found among internally triggered events with $n$ = 8 (IT8). Since multiple scintillator strips triggered by one muon are very unlikely, IT8s are thought to be generated by small showers, and the low rate of these events (∼0.1 Hz) is consistent with this idea. The clustering effect is still under study.

To have a better understanding of IT8 structure, air showers from $10^{14.5}$ to $10^{16}$ eV randomizing the core position over an area of 150 m radius and the angle of incidence of their axis ($0° < \phi < 360°$ and $0° < \theta < 30°$) were simulated [8]. The particles at the ground level were propagated towards the MD [6] and the energy deposition method was applied. In Fig. 5 an histogram of measured IT8s is shown on top of the one achieved by the application of the same triggering conditions to the simulated events.

Since most events are clustered, it is expected that the central scintillator strips on each side of the MD participate in more events than the strips of the sides. This consideration explains the double bump structure of the histogram. Each band corresponds to a side of the MD with its 32 scintillators involved. It is worth noting that, although a fine tuning is still needed, the general structure of the double bump histogram is being reproduced by the simulation.

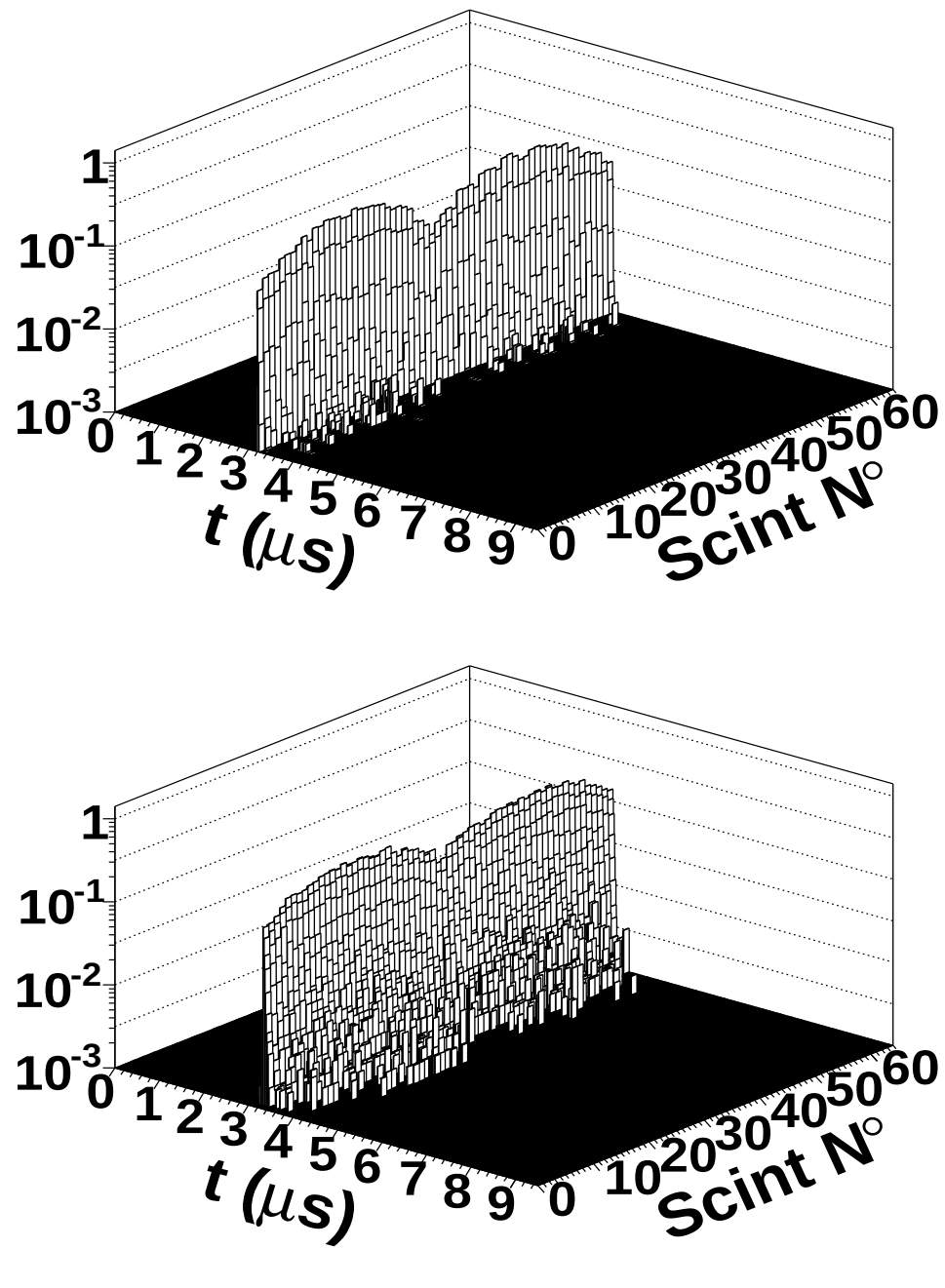

Figure 5: (Top) Normalised histogram of IT8 events from the Observatory site, the double bump structure is expected from clustered data. (Bottom) Normalised histogram of simulated events, air showers from $10^{14.5}$ to $10^{16}$ eV propagated underground were used to build it.

Field measured time structure is matched by muon signals from laboratory measurements. In Fig. 6 such a comparison is shown. The bin width of 12.5 ns corresponds to the sampling rate of 80 MHz used for MD data acquisition. Both sets show similar features.

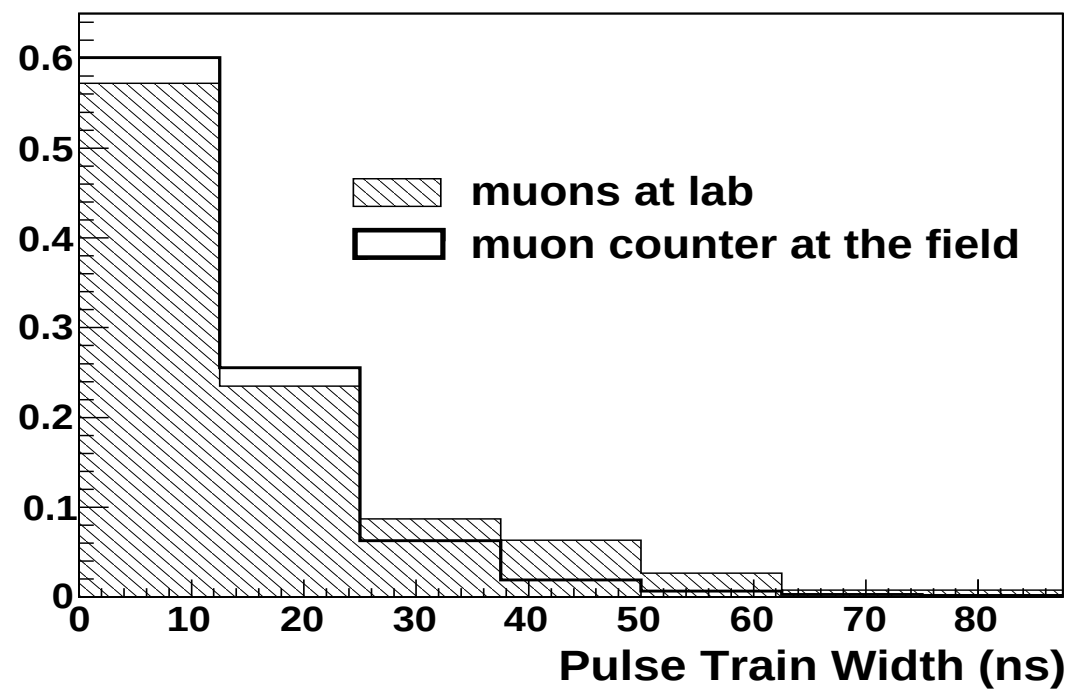

Figure 6: Normalised histograms of time width for MD data and muon signals measured at the laboratory. Both data sets show similar structures.

## 5 Summary

A full simulation of the AMIGA muon counter was presented. Its three methods of injecting traces were described. They include the possibility of injecting real muon signals from laboratory measurements, the generation of pulses modeled with experimental parameters, and a third method which builds the muon signals from deposited energy in the scintillator strips. Simulated pulses show good agreement with measured muons. The cross talk effect of the PMT is incorporated to find the right counting strategy to avoid overcounting. The $2G_{30\mathrm{ns}}$ technique shows the best counting ratio among the inspected ones. The overcounting due to cross talk is negligible with this strategy regardless of the discrimination threshold. Real muon pulses were injected and counting ratios of 99% at 110 cm from the PMT, 99% at 297 cm, and 88% at 482 cm were found. The latter constitutes a lower limit giving the fiber attenuation. Preliminary results of the first muon counter modules installed at the Observatory were presented. The performance was shown to be in line with the design specifications.

# AERA: the Auger Engineering Radio Array

JOHN L. KELLEY[1] FOR THE PIERRE AUGER COLLABORATION[2]
[1]*IMAPP / Dept. of Astrophysics, Radboud University Nijmegen, 6500GL Nijmegen, Netherlands*
[2]*Observatorio Pierre Auger, Av. San Martín Norte 304, 5613 Malargüe, Argentina*
*(Full author list: http://www.auger.org/archive/authors_2011_05.html)*
*auger_spokespersons@fnal.gov*

**Abstract:** The first part of the Auger Engineering Radio Array, presently consisting of 21 radio-detection stations, has been installed in the surface detector array of the Pierre Auger Observatory and in close proximity to the fluorescence detectors Coihueco and HEAT. In the coming years, the number of detection stations will grow to 160 units covering an area of almost 20 square kilometers. AERA is sensitive to radio emission from air showers with an energy threshold of $10^{17}$ eV and is being used to study in detail the mechanisms responsible for radio emission in the VHF band. The design of AERA is based on results obtained from prototype setups, with which we investigated triggering methods and different hardware concepts. We present the first data from AERA, including AERA's first hybrid radio / particle detector events.

**Keywords:** radio, air shower, hybrid, AERA, Pierre Auger Observatory

## 1 Introduction

The radio detection of cosmic ray air showers offers a scalable, high-duty-cycle approach for the next generation of air shower arrays. Broadband radio pulses from air showers are coherent in the VHF band (10-100 MHz) and thus have a signal power which scales with the square of the cosmic ray primary energy. The pulse characteristics can be used to probe the electromagnetic shower development and primary particle composition [1, 2]. Recent experiments such as LOPES [3] and CODALEMA [4] have employed antennas triggered by particle detectors to demonstrate the method and verify the dominant emission mechanism — the acceleration of shower particles in the Earth's magnetic field. At the same time, significant progress has been made in the theoretical understanding of the radio signals [5], with current work focused on the more subtle sub-dominant emission mechanisms.

A technical challenge has been to develop a large-scale, autonomous antenna array which triggers directly on the radio pulses (a "self-triggered" array). At the same time, more measurements are needed to fully understand the radio signal polarization and lateral distribution, and thus to continue to refine theoretical models. The deployment of the Auger Engineering Radio Array (AERA) has started in 2010 at the Pierre Auger Observatory in Argentina, in order to address these technical and scientific issues. Stable physics data-taking has begun in March 2011, and the first hybrid radio/particle detector events have been recorded in April 2011.

## 2 AERA

AERA is a radio extension of the Pierre Auger Observatory. The observatory is a 3000-$\mathrm{km}^2$ hybrid cosmic ray air shower detector in Argentina, with an array of 1660 water-Cherenkov detectors and 27 fluorescence telescopes at four locations on the periphery. The area near the Coihueco fluorescence detector contains a number of low-energy enhancements, including the AMIGA infill array [6] and the HEAT fluorescence telescopes [7]. AERA is co-located with the infill water-Cherenkov detectors and within the field of view of HEAT, allowing for the calibration of the radio signal using "super-hybrid" air shower measurements.

Stage 1 of AERA consists of 21 radio-detection stations (RDS) arranged on a triangular grid with 150 m spacing. Stages 2 and 3, with larger detector spacings of 250 m and 375 m, will increase the total area covered to nearly 20 $\mathrm{km}^2$ (see Fig. 1).

AERA has an energy threshold of approximately $10^{17}$ eV, with its fundamental noise limit set by the radio emission from the Galactic plane. The scientific goals of the experiment are as follows:

- calibration of the radio emission from the air showers, including sub-dominant emission mechanisms, by using super-hybrid air shower measurements;
- demonstration of the physics capabilities of the radio technique, *e.g.* energy, angular, and primary mass resolution; and

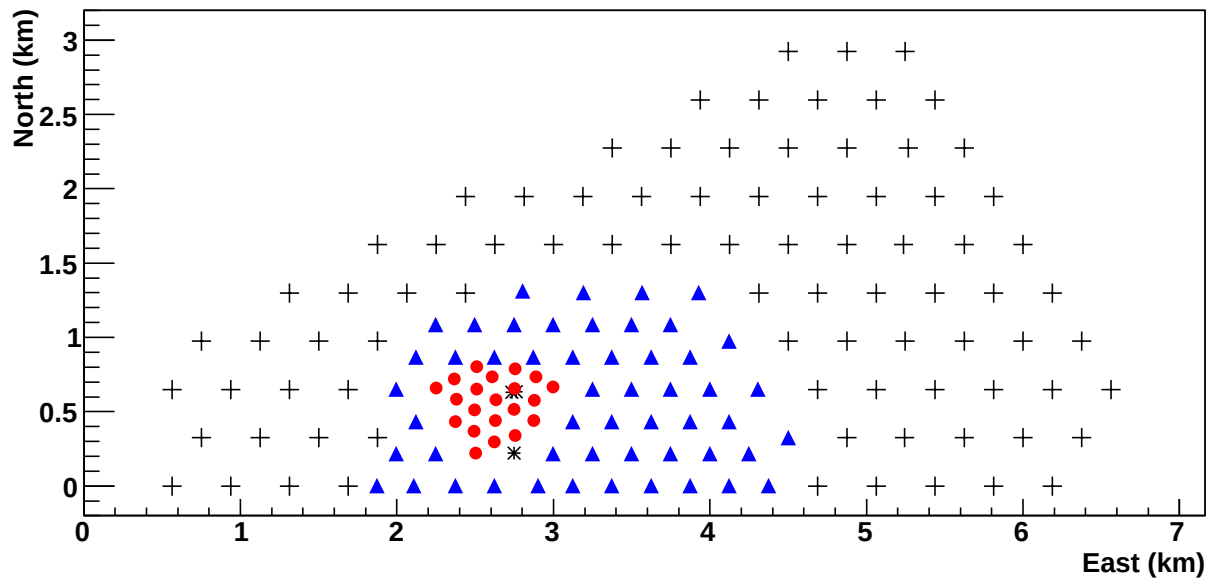

Figure 1: Layout of AERA radio-detection stations. The Stage 1 core stations (solid circles) were deployed in 2010. Planned expansions (solid triangles, crosses) will increase the instrumented area to nearly 20 $\mathrm{km}^2$.

- measurement of the cosmic ray composition in the ankle region, from 0.3 to 5 EeV, with the goal of elucidating the transition from Galactic to extra-Galactic cosmic rays.

### 2.1 Radio-Detection Station

The AERA Stage 1 radio-detection stations (RDS) consist of an antenna, associated analog and digital readout electronics, an autonomous power system, and a communications link to a central data acquisition system. The RDS design has been informed by the results of several prototype setups at the Auger site, with progress in antenna design, noise filtering, and self-triggering by radio detection stations at the Balloon Launching Station [8, 9] and the Central Laser Facility [10, 11].

The RDS antenna is a dual-polarization log-periodic dipole antenna, oriented to point north-south and east-west. The antenna is sensitive between 27 and 84 MHz, chosen as the relatively radio-quiet region between the shortwave and FM bands. The north-south and east-west antenna signals are separately amplified with two low-noise amplifiers (LNAs), each with a gain of 20 dB. The LNAs incorporate a bandpass filter (23-79 MHz) to eliminate any issues with intermodulation from the FM band.

The amplified antenna signals are routed into a custom aluminum housing containing the bulk of the station electronics. The housing includes a radio-frequency-tight chamber to mitigate issues with self-generated noise. Inside the housing, the signals are further amplified and filtered with a custom bandpass filter (30-78 MHz). Both a low-gain (+10 dB) and high-gain (+19 dB) version of each polarization is generated, to increase the dynamic range of the station.

The four analog output channels of the filter-amplifier are then digitized with a custom 12-bit 200 MHz digitizer. The digitizer employs an FPGA with specially designed filtering and triggering logic using the time-domain radio signals. Several digital notch filters can be configured to remove remaining narrowband interference before triggering. The triggering logic employs multiple voltage thresholds in order to reject signals with afterpulsing. Once a trigger is recorded, up to 2048 voltage samples of each channel are stored for transfer to the central data acquisition system (DAQ), corresponding to 10.2 $\mu$s of data. Timestamping of the triggers is performed using an on-board GPS receiver.

The station is powered using a photovoltaic system with two storage batteries. For stage 1, communication to the central DAQ is via an optical fiber network. A high-speed, low-power wireless communications system is currently under development for the next stages.

### 2.2 Data Acquisition

The central data acquisition system (DAQ) is located in a central container, the Central Radio Station (CRS). The CRS is powered by its own photovoltaic array and is equipped with a weather station to monitor the local electric field (thunderstorm conditions amplify the air shower radio signal; see *e.g.* Ref. [12]).

All timestamps from single RDS triggers are sent to the central DAQ. A multi-station trigger is formed when neighboring stations trigger within a time window consistent with a signal moving at the speed of light across the array. Once a coincidence is formed, the full radio signals are read from the stations and stored to disk for analysis.

Trigger rates are highly dependent on the RDS thresholds and the local noise conditions, but on average, the RDS trigger rates are $\sim$ 100 Hz, while the multi-station trigger rate is $\sim$ 10 Hz. The vast majority of these triggers can be localized using the RDS data to sources on the horizon — that is, man-made radio-frequency interference (RFI). A number of techniques are incorporated into the DAQ to reject RFI sources at trigger level, based on direction and periodicity.

### 2.3 Beacon

A multi-frequency narrowband transmitter (“beacon”) installed at the Coihueco fluorescence detector site transmits signals to AERA for timing calibration. Using the beacon frequencies as a phase reference, relative offsets in the individual RDS timestamps can be corrected, improving the timing resolution [13]. The transmitter signals are visible in the dynamic spectrum shown in Fig. 2.

## 3 Calibration, Analysis, and Detector Performance

Calibration of the antenna and station electronics is necessary in order to translate the recorded voltages into a measurement of the time-varying electric field vector. The gain and group delay of the analog components is measured for all channels on each RDS. The digitizers employ a self-calibration routine using a generated DC voltage, and that reference voltage is separately calibrated for each station. Frequency-dependent corrections to the digitizer DC gain have also been measured.

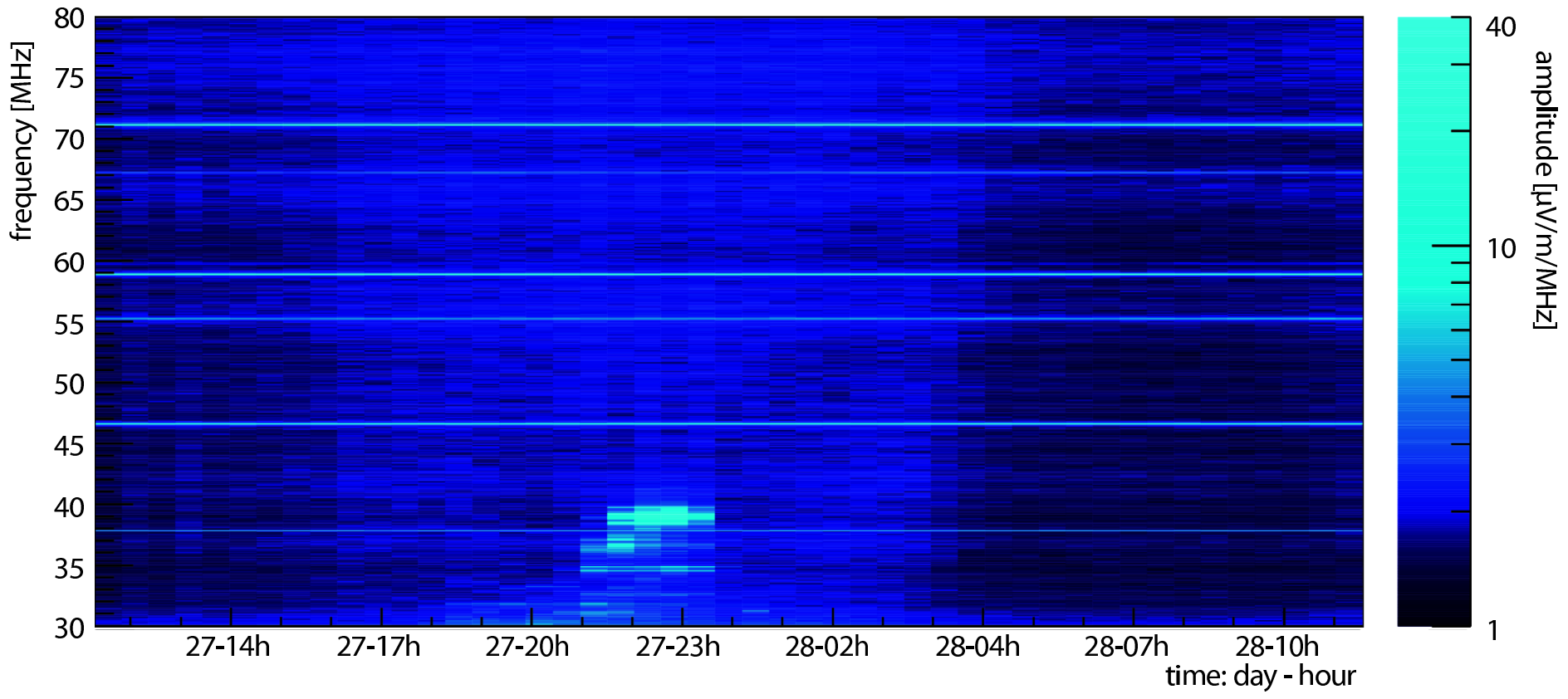

Figure 2: A daily dynamic spectrum from October 2010 — electric field strength versus frequency and time (AERA station #1, north-south polarization). The slow daily modulation of the noise level coincides with the rise and set of the Galactic Center. The calibration beacon can be seen as narrowband horizontal lines. The low-frequency noise outbursts correspond to human activity and are digitally filtered before triggering.

| | Before correction (ns) | After correction (ns) |
|---|---|---|
| mean | $2.3 \pm 0.2$ | $0.5 \pm 0.1$ |
| worst | 2.6 | 0.8 |

Table 1: RDS relative timing resolution measured with the beacon.

The direction- and frequency-dependent gain characteristics of the antenna are initially provided by detailed simulations. The simulation values are cross-checked using a calibrated reference transmitter that is flown over the AERA site using a tethered balloon. A full end-to-end cross-check of the calibration is planned using the same technique. Resistivity measurements of the ground conditions at AERA, which affect the antenna gain pattern, are in progress and will be incorporated into the analysis.

The OffLine analysis framework employs the station calibration data in order to reconstruct the time-varying electric field vector received at the RDS [14]. Both gain and group-delay-induced dispersion effects are corrected in the radio signals. In order to account for the directional dependence of the antenna gain, this process is iterative, with an initial reconstruction used to determine the estimated antenna gain for a specific event. The reconstruction can then be repeated to obtain more accurate electric field values.

An analysis of the beacon data has provided information on the relative stability of the GPS timing, as well as the precision possible after beacon phase correction. For this analysis, one RDS is used as a reference, and changes in the relative phase of the beacon frequencies in other stations are tracked and then corrected. The timing resolution achieved is better than 1 ns (see Table 1).

Reconstructed azimuth directions from approximately 14 hours of unfiltered data are shown in Fig. 3. The peaks correspond to RFI sources on the horizon and can easily be filtered out. These raw data can also be used to estimate the angular resolution of the first-guess reconstruction al-gorithm, a plane wave fit. Assuming the dominant sources are point sources on the horizon, the median space angle difference to the closest of these sources for each event is $4.2°$. As we will show in the following section, this is larger than AERA's angular resolution for more distant cosmic ray events.

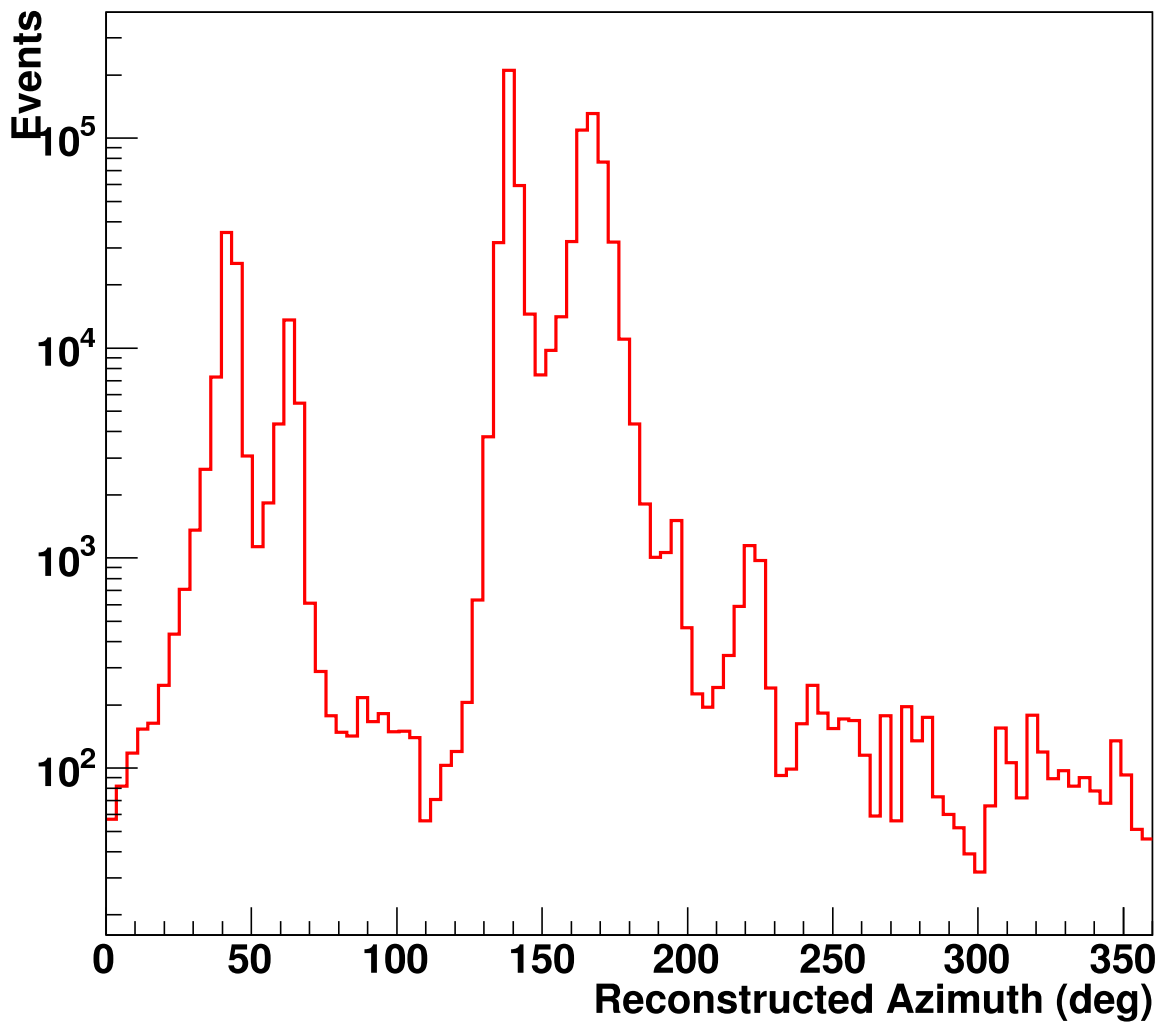

Figure 3: Reconstructed azimuth direction of 14 hours of unfiltered AERA data ($8.7 \times 10^5$ events). The peaks correspond to noise sources on the horizon.

A number of reconstruction methods are under development which will further improve the angular resolution. These include:

- spherical and/or conical wave reconstructions;
- lateral distribution fits taking into account polarization effects;
- reconstructions after beacon correction; and
- localization with coherent beamforming.

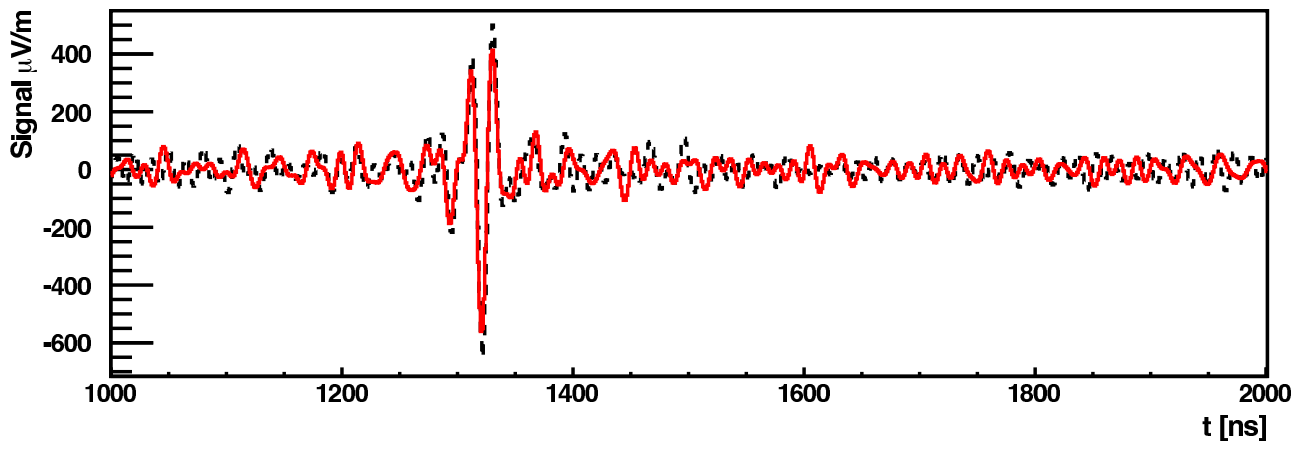

Figure 4: Calibrated radio pulse recorded for a 5.7 EeV cosmic ray event. Both north-south (solid) and east-west (dashed) polarizations are shown. The distance to the reconstructed shower axis is 114 m.

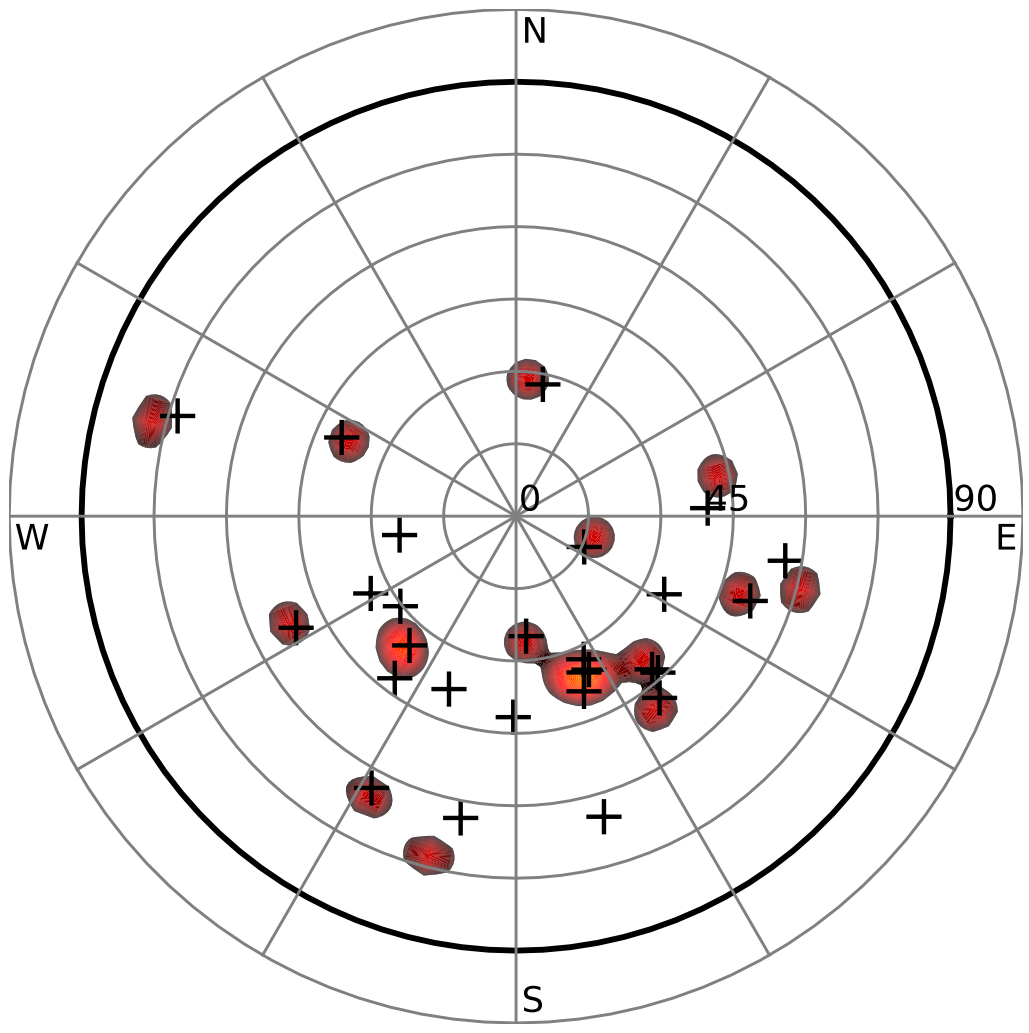

Figure 5: Polar skyplot showing reconstructed arrival directions of hybrid radio / surface detector events (the center is vertically overhead). SD directions are indicated by crosses, and for events that triggered three or more radio stations, AERA directions are shown with contours (Gaussian smearing of $3°$).

## 4 First Hybrid and Super-Hybrid Events

The AERA Stage 1 array has been stably taking data since March 2011. Following optimization of threshold and trigger settings, the first coincident hybrid radio-SD (surface detector) events were recorded in April 2011. The rate of coincident events since then is approximately $0.3$ to $0.9\ \mathrm{day}^{-1}$, depending on trigger settings and the number of stations required in coincidence. An example radio signal from a self-triggered cosmic ray event in shown in Fig. 4. A polar skyplot showing the reconstructed directions from both radio and SD is shown in Fig. 5. The median angular distance between the radio and SD reconstructions, which includes the error from the SD reconstruction, is $2.8°$. The north-south asymmetry shows the dominantly geomagnetic nature of the emission, as these data were taken primarily using a trigger on the east-west polarization. This asymmetry can be reduced by also triggering on the north-south polarization. The first super-hybrid event, including fluorescence, surface detector, and radio data, was recorded on April 30, 2011, and analysis is in progress.

## 5 Conclusions and Outlook

Building upon the previous successes of experiments such as LOPES, CODALEMA, and prototype setups at the Auger site, AERA has demonstrated the technical feasibility of a self-triggered radio air shower array. Analysis of the first hybrid events is ongoing and will provide the necessary measurements to refine our understanding of air shower emission in the VHF band. Super-hybrid events, including data from the fluorescence detectors, will provide an accurate energy cross-calibration for both AERA and the SD infill array. Advanced reconstruction methods using the polarization and frequency characteristics of the radio pulses will soon allow tests of the energy resolution and mass determination capabilities of the radio method.

Development of hardware for Stages 2 and 3 of AERA is well underway. Selected improvements to the Stage 1 design include: a high-speed wireless communications system; large storage buffers in the RDS for sub-threshold event readout; and lower power consumption, which significantly lowers the cost per station. Within the next few years, the complete 20-$\mathrm{km}^2$ array will provide the most detailed insights yet into both radio air shower physics and the cosmic ray composition at the ankle.

# Autonomous detection and analysis of radio emission from air showers at the Pierre Auger Observatory

BENOÎT REVENU[1] FOR THE PIERRE AUGER[2] COLLABORATION
[1]*SUBATECH, 4 rue Alfred Kastler, BP20722, 44307 Nantes, CEDEX 03, Université de Nantes, École des Mines de Nantes, CNRS/IN2P3, France*
[2]*Observatorio Pierre Auger, Av. San Martín Norte 304, 5613 Malargüe, Argentina*
*(Full author list: http://www.auger.org/archive/authors_2011_05.html)*
*auger_spokespersons@fnal.gov*

**Abstract:** Radio detectors acting as prototypes of the AERA project at the Pierre Auger Observatory have been used to observe air showers. We present results from a first radio setup consisting of three detection stations running in a fully-autonomous mode and with an independent triggering detection technique. Different stations of a second radio setup have been used to study the emission mechanism. The resulting data sets from both setups confirm the dominant role of the geomagnetic field in the emission mechanism. We present the results of these analyses and discuss how the polarization information is used to disentangle the geomagnetic induced emission from the emission caused by the charge excess in the shower.

**Keywords:** cosmic rays, radio detection, Pierre Auger Observatory, polarization, radio emission, AERA

## 1 Introduction

The secondary particles of extensive air showers (EAS) induced by cosmic rays emit a detectable electric field in the MHz radio frequency range. The radio-detection technique is expected to be very promising because it has a duty cycle close to 100% and could give access to the main characteristics of the primary cosmic ray: following recent simulations [1, 2], it appears that the signal is highly correlated to the longitudinal development of the shower, which would be of prime interest to estimate the primary composition of high energy cosmic rays. The results obtained by the CODALEMA [3] and the LOPES [4] radio detection experiments encouraged us to install 2 initial experiments at the Pierre Auger Observatory [5, 6] in the scope of the AERA (Auger Engineering Radio Array) experiment, described in another contribution at this conference [7]. The energy threshold of the Auger Surface Detector (SD) is $3 \times 10^{17}$ eV. The event (cosmic rays) rate is of the order of 1.6 km$^{-2}$.day$^{-1}$. In order to increase the rate of detected events at places where the AERA prototypes are installed, additional surface detectors have been installed at the center of two elementary triangular cells, leading to a local energy threshold below $10^{17}$ eV.

We present in this paper the main results obtained with these two prototypes for AERA. We first present the results of the upgraded version of the setup described in [8, 9], which was used to register the first fully autonomous cosmic-ray radio detection and gave a strong indication on the geomagnetic origin of the electric-field emission mechanism. Then, we present the polarization analysis of the setup [10] giving evidence for another electric field emission mechanism due to charge excess in the shower front.

## 2 Autonomous radio detection of air showers

One of the AERA prototype setups is made of 3 fully autonomous radio detection stations, whose principle is described in [8, 9]. The main characteristic of these stations is that they are independently triggered by the radio signal itself and do not rely on any external particle detector trigger. In the upgraded version of this system considered here, the station sensitivity has been greatly improved by using a butterfly antenna which simultaneously measures the incoming electric field in both east-west (EW) and north-south (NS) polarizations [11]. The 3 stations are located close to the center of the SD array and form an equilateral triangle. The distance between the three stations is $140$ m. The waveform is recorded in the full frequency band $0.1 - 150$ MHz for both polarizations, but the trigger decision is made when the signal exceeds a pre-defined threshold in the $45 - 55$ MHz band, using an analog trigger board. The choice of the trigger frequency band excludes the AM and FM radio emissions. The recorded signal is sampled at a frequency of 1 GHz during 2.56 $\mu$s. Each of the 3 stations sends its own data by WiFi to a central data acquisition PC and the search for coincidences with the SD is done offline by comparing the SD events timestamps with those of the registered radio events. These stations were installed on May 10, 2010 and the first coincidence with the SD has been observed on May 13, 2010. Up to March 25, 2011, a total of 40 events have been detected in coincidence with the SD array, on the basis of one event each 4 days, taking into account the effective observation time of the radio stations. Fig. 1 shows the measured time difference between the SD and the radio-detection stations as a function of the expected time difference given the shower geometry. The correlation is very clear and the slope is compatible with unity as expected. The arrival di-rection of these 40 events are not uniform in azimuth, in contrast with the arrival direction of all the events detected by the SD in the same time period. The skymap of these events in local coordinates is presented Fig. 2: 70% of the events are coming from south. This strong southern ex-cess is in good agreement with what was already observed by the previous version of this setup [9] and also by the CODALEMA [3] experiment in the northern hemisphere, with an excess of events coming from the north. This observation corresponds to an emission mechanism compatible to first order with $\vec{n} \times \vec{B}$, where $\vec{n}$ is the direction of the shower axis and $\vec{B}$ is the local geomagnetic field vector. With the reasonable hypothesis that the probability to detect the electric field is proportional to its amplitude, a computation based on this simple model gives an expected excess of $68\%$ of events coming from the south, in very good agreement with the observation.

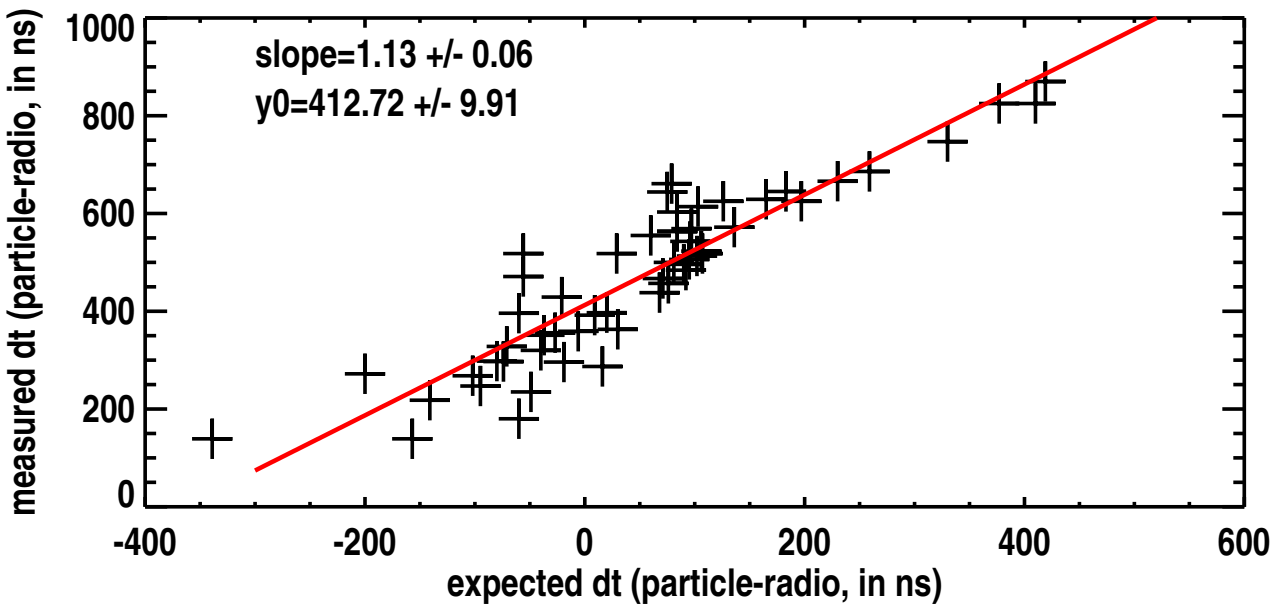

Figure 1: Correlation between the measured time difference between a coincident radio station and the SD and the corresponding time difference expected from the shower geometry.

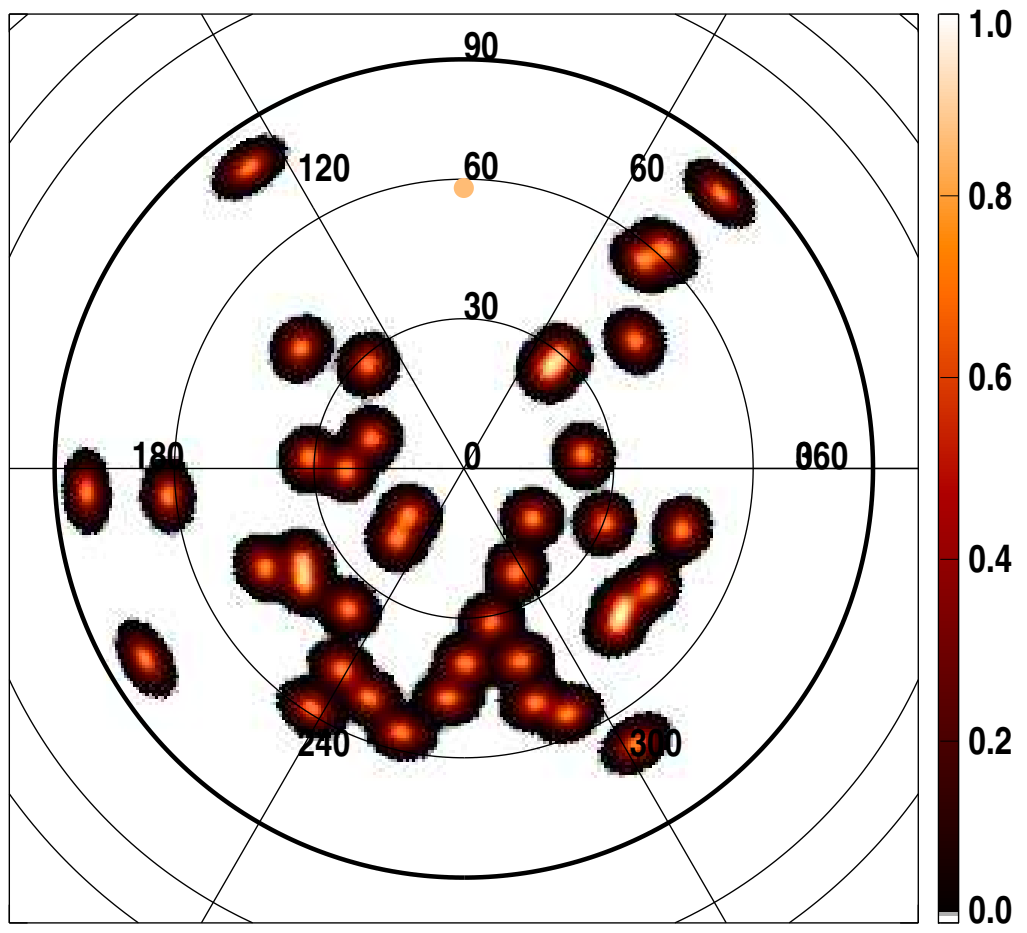

Figure 2: Skymap in local coordinates (zenith angle, azimuth) of the 40 self triggered events in coincidence with the SD, smoothed by a $5^\circ$ Gaussian beam. The orange dot stands for the geomagnetic field in Malargüe. The zenith is at the center, the north at the top and the east on the right.

Another important milestone in the radio detection technique has been the observation for the first time (on January 13, 2011) of a shower detected by 3 detection methods: 10 surface detectors, 3 telescopes located in the 3 fluorescence sites and 1 radio station detected the same event. Analysis is underway to determine the correlation between the recorded electric field and the longitudinal profile of the shower.

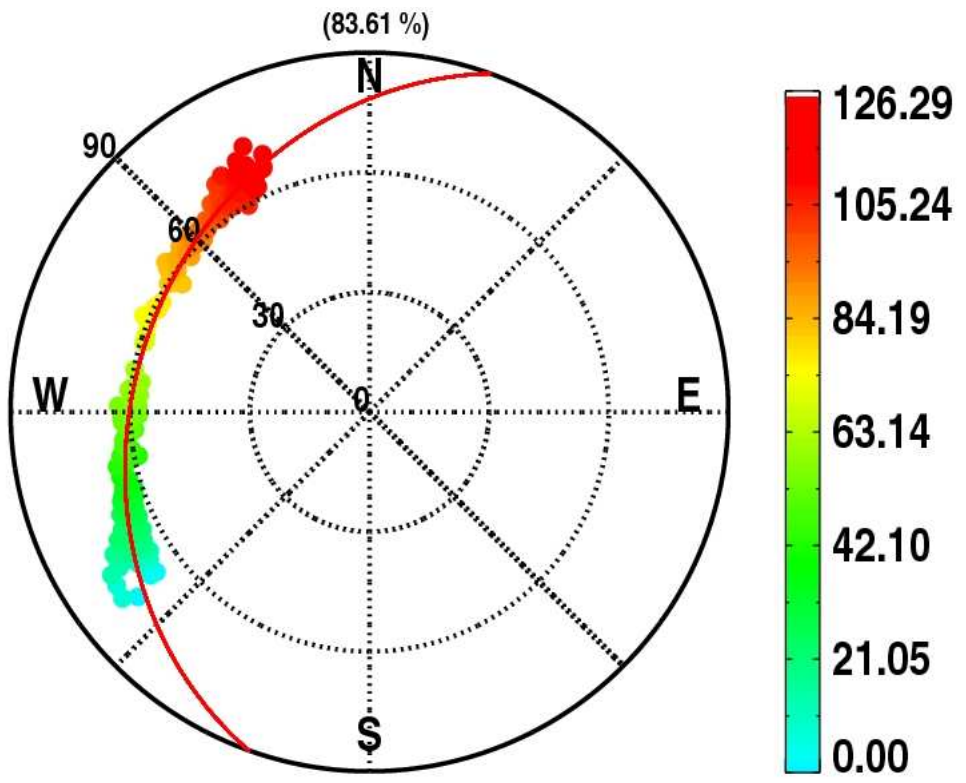

Figure 3: Detection of an airplane. The detected trajectory is indicated by the dots and the reconstructed trajectory is represented by the solid line. The angular resolution for this radio setup of three stations separated by 140 m is around $0.7^\circ$. The bar on the right side indicates the time elapsed in seconds since the first event when the airplane was detected (from south to north).

This upgraded setup was also able to detect airplane transits in the vicinity of the Pierre Auger Observatory. Fig. 3 shows an example of such a trajectory with the corresponding fit, assuming that the airplane altitude and velocity are constant and that the trajectory is linear. This allows to estimate an angular resolution around $0.7^\circ$. Airplane signal data will be used to inter-calibrate the stations.

## 3 Polarization studies

The second radio setup [12] for AERA was located in the western part of the SD array and consisted of 3 antennas positioned on an equilateral triangle separated by a distance of 100 m. These stations were triggered by an external scintillator as it was found that the radio background in this site is larger than at that of the first radio setup. We used the data from two logarithmic-periodic dipole antennas (LPDAs) [13], the third antenna was of a different type and was used for test purposes. The LPDAs are measuring the electric field in the two NS and EW polarizations. After a particle trigger the data is recorded over 10 $\mu$s at a sampling rate $f_{\rm samp} = 400$ MHz. We registered 494 events in coincidence with the SD between May 2007 and May 2008. The SD reconstruction provides the geometry of the showers. Many of these events do not present a clear radio counterpart so that we select the events having an average total power[1] within a given time window around the expected position of the maximum (knowing the shower geometry) larger than five times the average total power computed in a noise window of the time series. We ignore events detected during thunderstorms. There are 37 radio signals that pass these cuts.

The analysis of these signals confirms the conclusion obtained from the data collected with the prototype previously described: the dominant emission mechanism for the electric field is the geomagnetic radiation with a macroscopic electric field of the form $\vec{E} \propto \vec{n} \times \vec{B}$. The polarization is therefore expected to be linear with an angle $\phi_{\rm mag} = \arctan((\vec{n} \times \vec{B})_{\rm NS}/(\vec{n} \times \vec{B})_{\rm EW})$ with the EW axis. The measured polarization angle is given by $\phi(t_i) = \arctan(U(t_i)/Q(t_i))/2$ where $U$ and $Q$ are the linear Stokes parameters of the wave, given by $U(t_i) = 2{\rm Re}(E_{\rm EW}(t_i)E^*_{\rm NS}(t_i))$ and $Q(t_i) = |E_{\rm EW}(t_i)|^2 - |E_{\rm NS}(t_i)|^2$. The average polarization angle $\bar{\phi}$ is estimated as the average of the polarization angles in the time window in which the signal is present. The uncertainty on $\bar{\phi}$ is given by $\Delta\bar{\phi} = \frac{K}{N}\sqrt{\sum_{i=1}^{N}(\phi(t_i) - \bar{\phi})^2}$ where $K = f_{\rm samp}/\Delta f$, $\Delta f$ being the effective bandwidth of the measurement and $N$ the number of samples in the signal window. The correlation between the expected $\phi_{\rm mag}$ and the measured value is presented in Fig. 4. The correlation is clear and confirms the geomagnetic origin of the electric field as main contribution. Nevertheless, there should be another source of the electric field: the variation of the charge excess during the development of the air shower in the atmosphere.

## 4 The charge-excess contribution

Both electric field contributions differ by their specific polarization patterns. The geomagnetic contribution is aligned along $\phi_{\rm mag}$. The electric field from the charge excess is radial in the plane transverse to the shower axis so that the polarization pattern depends on the angle $\phi_{\rm obs}$ between the EW axis and the axis defined by the shower core and the antenna position. We can construct an observable characterizing the deviation from a pure geomagnetic electric field by first applying a rotation to the $x$=EW and $y$=NS axis to make the new $x'$ axis aligned along the $\phi_{\rm mag}$ direction and $y'$ perpendicular to $x'$. Then we can define in this rotated coordinate system a new angle $\phi'_{\rm obs} = \phi_{\rm obs} - \phi_{\rm mag}$.

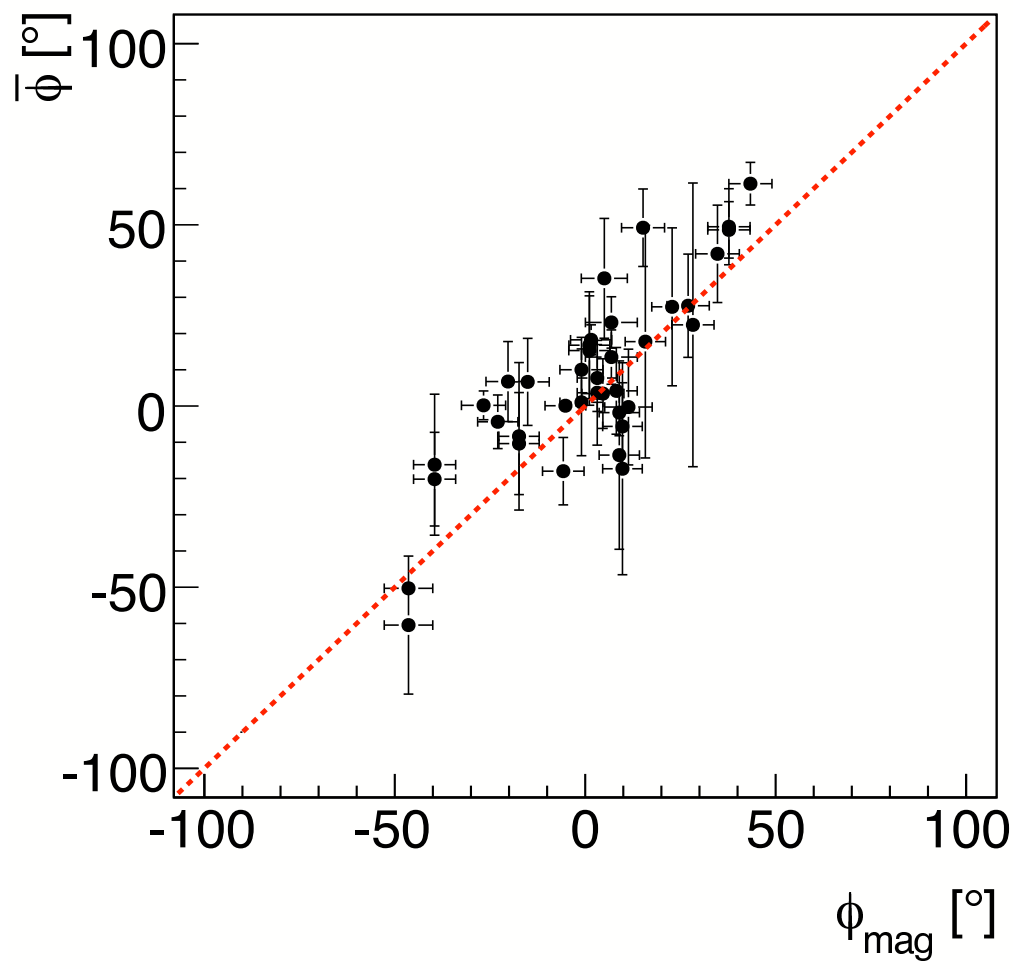

Figure 4: Measured polarization angle for the 37 signals from events in coincidence with the SD as a function of the expected polarization angle in case of a pure geomagnetic origin of the electric field. The dotted line represents full correlation.

We define $R$ by:

$$R = \sum_{i=1}^{N} E_{x'}(t_i)\, E_{y'}(t_i) / \sum_{i=1}^{N} (E^2_{x'}(t_i) + E^2_{y'}(t_i)). \quad (1)$$

For a pure geomagnetic electric field, the value of $R$ will be equal to 0 by construction. In this expression, the summation is done as before on all the samples in the signal window. To calculate $R$ from the data, we take into account the noise levels with a contribution $n_{E_{x'}}$ and $n_{E_{y'}}$ computed in the same way but outside of the signal window and using a larger number of samples to get a better estimation. For the data, the denominator in Eq. 1 takes the form $\sum_{i=1}^{N}(E^2_{x'}(t_i) - n^2_{E_{x'}} + E^2_{y'}(t_i) - n^2_{E_{y'}})$. We compare the values of $R$ for the 494 measurements of the electric field of coincident events with the SD to the values obtained for axis distance below 300 m with simulated events with REAS3 [14] and MGMR [15] having the same geometry than the observed events. From the charge-excess component, we expect a modulation in the value of $R$ with a period of 360°. This modulation appears clearly for both simulations as shown in Fig. 5. The values of $R_{\rm data}$ are calculated for the 37 signals; their uncertainties are obtained from the noise level determined on event-by-event basis. Concerning $R_{\rm sim}$, for each of the 37 signals, we generated a set of simulated showers to account for uncertainties on its reconstructed characteristics (axis direction, core position, energy). The central value and error bar of $R_{\rm sim}$ for each event is taken as the average and rms of $R_{\rm sim}$ of the set of corresponding simulated events. The set of simulated events contains 25 showers for REAS3 and 100 showers for MGMR. The correlation between $R_{\rm data}$ and $R_{\rm sim}$ is presented in Fig. 6. The correlation can be quantified using

1. Defined as the sum of the squares of the time series measurements envelope for both polarizations.

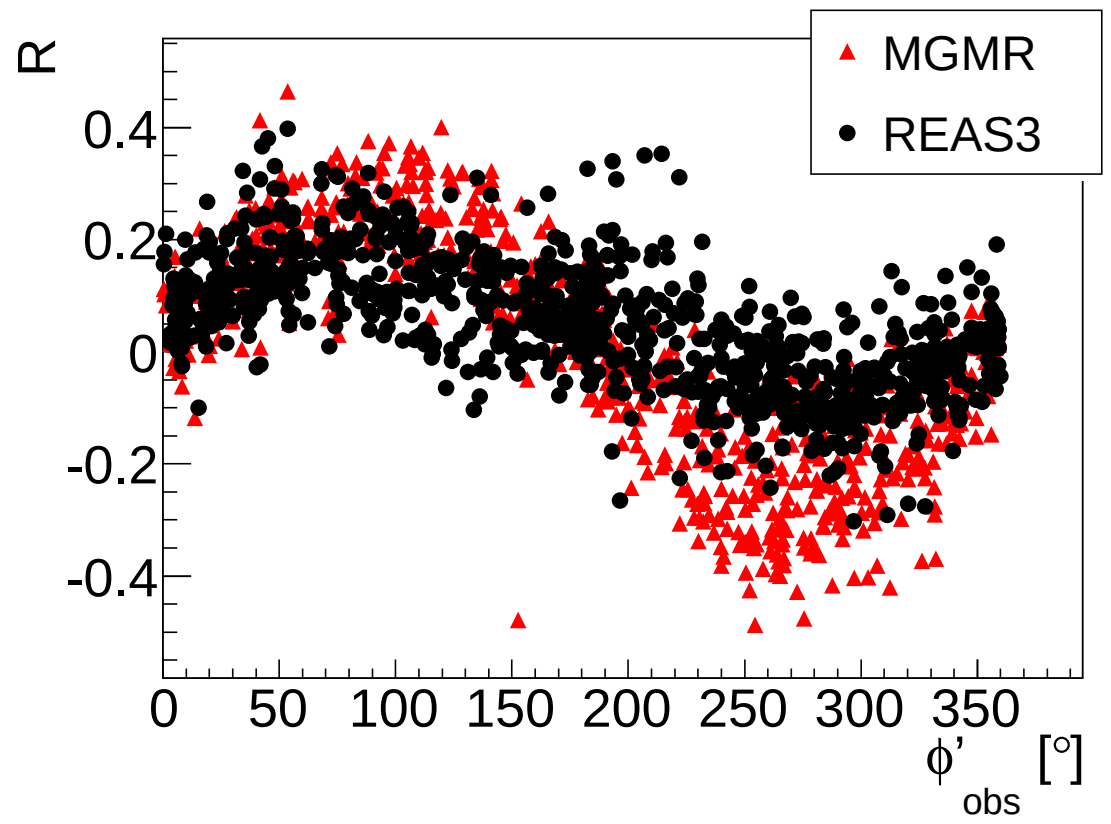

Figure 5: Values of $R$ as a function of $\phi'_{\mathrm{obs}}$. Red triangles and black circles correspond to $R$ values obtained with MGMR and REAS3 respectively, for axis distance below 300 m.

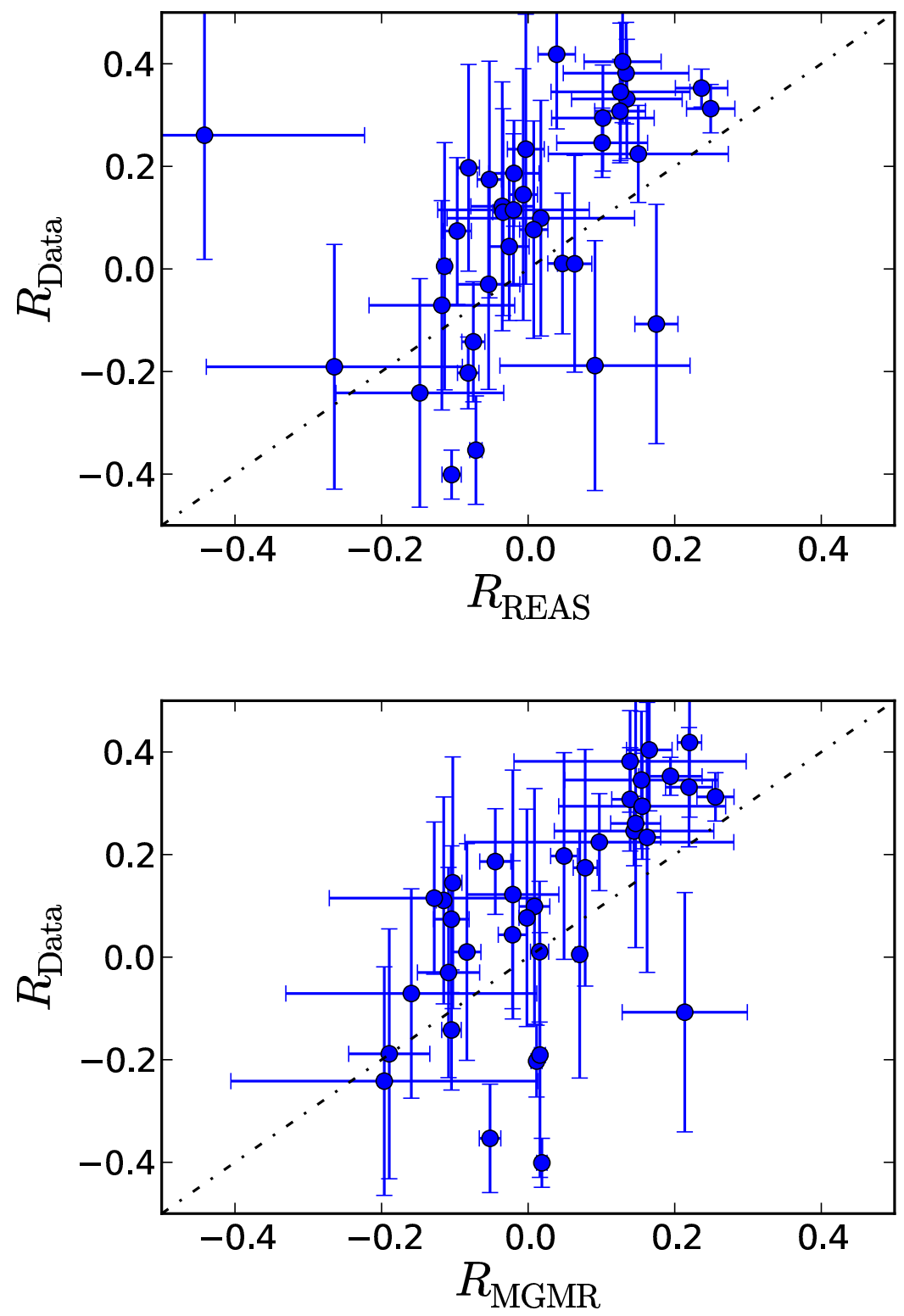

Figure 6: Correlation between $R_{\mathrm{data}}$ and $R_{\mathrm{sim}}$ for the 37 selected signals. The dotted lines represent full correlation.

$\chi^2 = \sum_{i=1}^{n}(R_{\mathrm{data},i} - R_{\mathrm{sim},i})^2/(\sigma^2_{\mathrm{data},i} + \sigma^2_{\mathrm{sim},i})$ where $n = 37$ is the number of points and $\sigma$ are the uncertainties on the corresponding axes. To compare the data with simulations without the charge-excess component, we run modified REAS3 and MGMR simulations leading to the corresponding $\chi^2_0 = \sum_{i=1}^{n} R^2_{\mathrm{data},i}/(\sigma^2_{\mathrm{data},i} + \sigma^2_{\mathrm{sim},i})$ because in this case $R_{\mathrm{sim}} = 0$. Accounting for the charge excess component reduces the $\chi^2$ for both REAS3 and MGMR: the $\chi^2/n$ goes from 6.28 to 2.64 and from 6.08 to 3.53 respectively.

## 5 Conclusion

The data from 2 prototypes of radio setups have been used to observe the dominant role of the geomagnetic field in the emission of the electric field by the secondary particles in air showers. The possibility of detecting air showers in a fully autonomous way has been demonstrated. In the analysis of the polarization data of events detected in coincidence with the SD, the inclusion of the contribution of the electric field component due to the charge excess in the showers leads to a better agreement with the data than considering only a purely geomagnetic contribution. These first very encouraging results will be checked in more details with AERA at the Pierre Auger Observatory.

# New technologies for the Pierre Auger Observatory – research and development in southeastern Colorado

FRED SARAZIN[1], FOR THE PIERRE AUGER COLLABORATION[2] AND S. COLLONGES[3], B. COURTY[3], M. DAUBENSPECK[1], A. GAUMÉ[1], B. GÉNOLINI[4], L. GUGLIELMI[3], M.A. HEVINGA[5], J. HODGSON[6], J. HOLLENBECK[6], K. KUHN[1], M. MALINOWSKI[1], M. MARTON[7], M. MASKEVICS[6], K. NELSON[6], E. RAULY[4], S. ROBINSON[1], A. ROKOS[6], C.F. SPEELMAN[5], J. THOMPSON[1], T. NGUYEN TRUNG[4], O. WOLF[1], J. YADON[6]

[1] *Physics Department, Colorado School of Mines, Golden CO, USA*
[2] *Observatorio Pierre Auger, (5613) Malargüe, Argentina*
[3] *Laboratoire Astroparticules et Cosmologie (APC), Université Paris 7, CNRS-IN2P3, Paris, France*
[4] *Institut de Physique Nucléaire d'Orsay (IPNO), Université Paris 11, CNRS-IN2P3, Orsay, France*
[5] *Kernfysisch Versneller Instituut (KVI), University of Groningen, Groningen, Netherlands*
[6] *Department of Electrical and Computer Engineering, Michigan Technological University, Houghton MI, USA*
[7] *Laboratoire de Physique Subatomique et de Cosmologie (LPSC), Université Joseph Fourier, INPG, CNRS-IN2P3, Grenoble, France*
*(Full author list: http://www.auger.org/archive/authors_2011_05.html)*
*auger_spokespersons@fnal.gov*

**Abstract:** The Pierre Auger Research and Development Array is conceived as a test bed aiming at validating an improved and more cost-effective one-photomultiplier surface detector design and a new communication system. The array of ten surface detectors and ten communication-only stations is currently being deployed in southeastern Colorado and will be operated at least until mid-2012. It is configured in such a way to allow testing of the peer-to-peer communication protocol and the new surface detector electronics that features an enhanced dynamic range aimed at reducing the distance from the shower core where saturation is observed. Atmospheric monitoring developments are also ongoing at the site and are presented in a separate paper. These developments are expected to improve the performances of the southern site of the Pierre Auger Observatory and enable future enhancements.

**Keywords:** Pierre Auger Observatory, Ultra-high energy cosmic rays, R&D for a future next-generation ground observatory

## 1 Introduction

The Pierre Auger Observatory [1, 2] aims at measuring the properties of the highest-energy cosmic rays using a large array of surface detectors combined with air fluorescence telescopes. From its original inception, the Pierre Auger Observatory was meant to achieve full sky coverage through the operation of two sites, one in each hemisphere. The southern site of the Observatory (Auger South), officially completed in November 2008, consists of an array of over 1660 surface detectors and 27 fluorescence telescopes (including the 3 elevated field-of-view HEAT telescopes [3]) deployed at 4 sites and overlooking the ground array.

The Pierre Auger Research and Development Array (RDA) in southeastern Colorado (USA) was originally conceived as a test bed for validating improved detector, communications and atmospheric monitoring techniques for the proposed northern site of the Observatory (Auger North) in Colorado. Fiscal constraints in the US have delayed Auger North until the indefinite future. Much of the development resulting from the RDA, however, was also expected to extend the physics potential of Auger South, both by improving the performance of the current detector and by enabling future enhancements. Given the substantial investment by nearly all countries of the Auger Collaboration, the scope and goals of the RDA were slightly revised to be more aligned with specific needs of Auger South. In this context, the RDA has two major goals:

*Validate a new surface detector design.* A new surface detector is being developed that is more cost-effective to build and to operate. It is based on one central photomultiplier (PMT) design coupled to electronics with updated components yielding better resolution and dynamic range. Surface detectors built to the new design can be used for in-fill arrays that will provide more detailed cosmic ray measurements at energies below the ankle of the spectrum, provid-ing a basis for understanding the transition from galactic to extragalactic sources.

*Benchmark a new communication system.* A new, versatile peer-to-peer radio-communication system will be tested at the R&D site. Originally designed to address the particular topology of the northern site, this new communication system can be used at Auger South as replacement or complement of the existing aging system in enhancements such as in-fills. Adding capacity to the communication system is critical to all future upgrades of the southern site, including, for example, large area radio arrays, if shown to be a viable alternative to air fluorescence for measuring the longitudinal profiles of cosmic-ray showers.

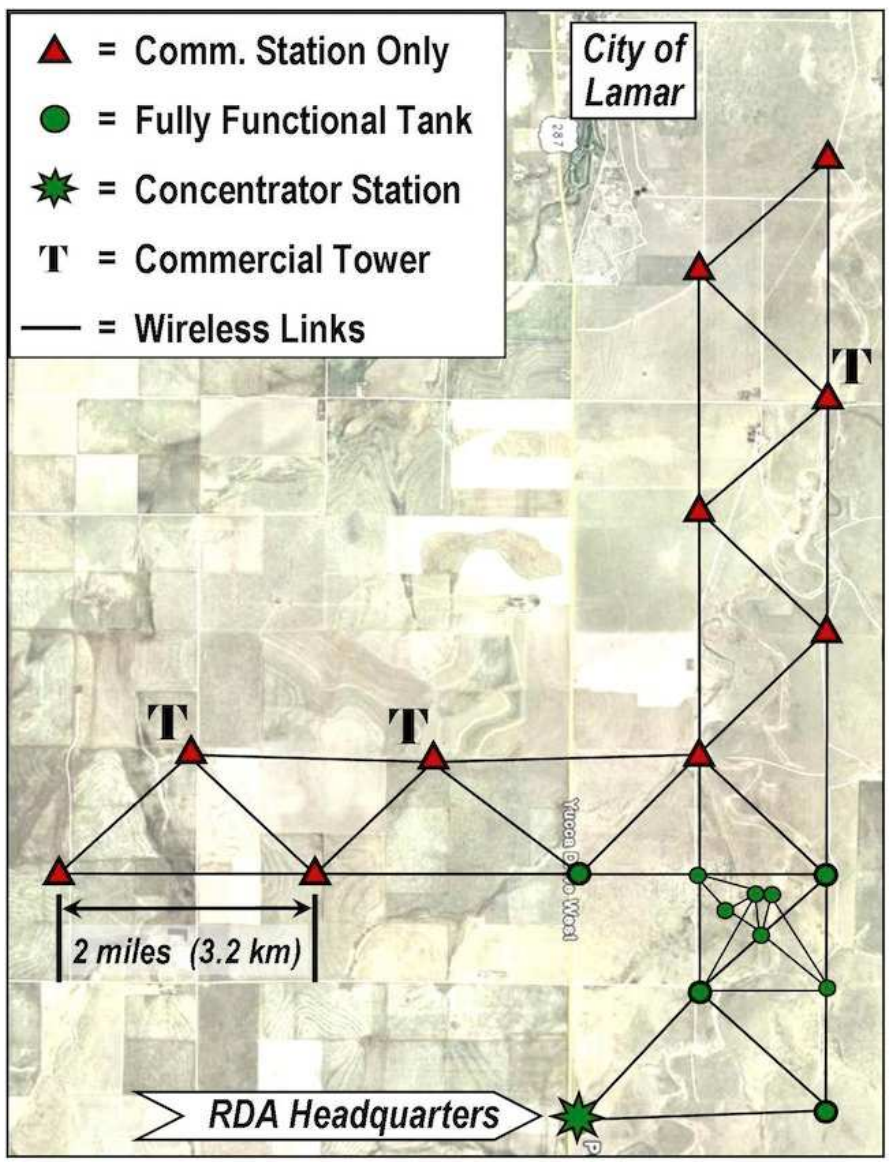

Figure 1: Layout of the R&D array in southeastern Colorado near the town of Lamar (see text for details).

Figure 1 shows the layout of the RDA located a few miles south of the city of Lamar, CO. Ten surface detectors (green circles) are concentrated in the southern part of the array within a triangle covering about 2 square miles with the detectors placed on the corners of a square-mile grid and additional in-fill positions comprising a doublet (two detectors 12 m apart) and two more detectors forming with the doublet a 575 m triangle. This configuration allows for the collection of a wide energy range of cosmic-ray showers to test the response of the new one-PMT surface detector. Ten communication-only stations (red triangles) are added to this array to specifically test the peer-to-peer communication system. The role of these stations is to simulate the broadcast of data from a large number of surface detectors upstream of the concentrator station located at the RDA headquarters (green star). The standard communication station is a standalone mast about 6 m high (the same height as that on the detectors). At three locations across the array (identified with an extra ”T”), commercial towers about 12 m high are used to achieve line-of-sight with the neighboring positions.

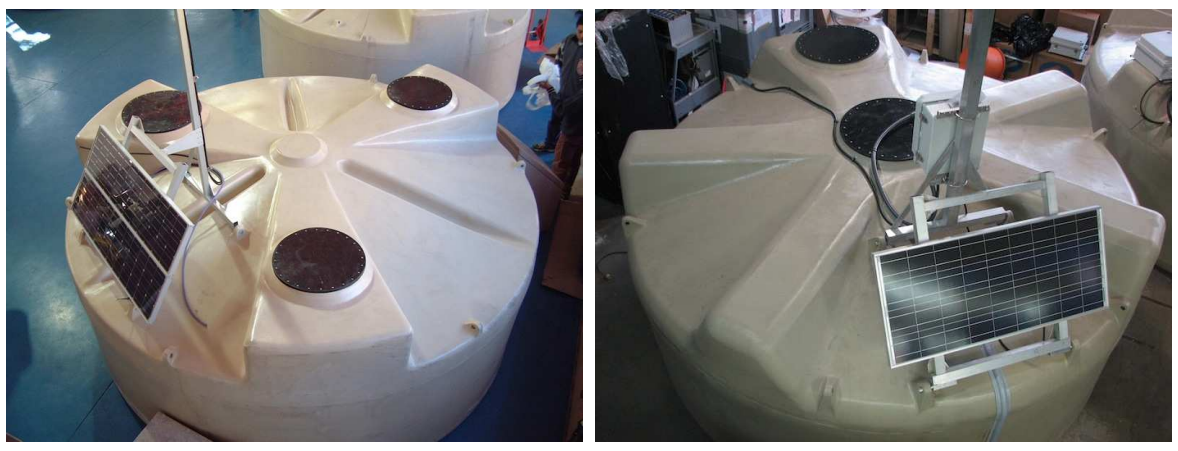

Figure 2: (Left) The original 3-PMT surface detector used in Auger South. (Right) The central 1-PMT surface detector as assembled for the RDA.

In parallel, a vigorous R&D program in atmospheric monitoring is also proceeding at the site and is presented in a separate contribution to this conference [4].

## 2 The 1-PMT Surface Detector

### 2.1 Mechanical design

A surface detector station at the southern site [1] includes a rotationally molded polyethylene tank 3.6 m in diameter with a water height of 1.2 m and water inventory of 12,000 liters. An example of the assembled station is shown in figure 2 (left). The tank being developed for possible use at the northern site, shown in figure 2 (right), uses the same molding technology but was altered to have one PMT rather than the original three. It is constructed with more gradual slopes and more rounded corners on the top structure of the tank. The features on the tank top are required to maintain high stiffness to resist creep failure, especially with the heavier mast, antenna, and electronics structures supported on the tank top. The rounding and gentle slopes were incorporated to allow rotomolding of an extra foam insulation layer on the interior, a feature necessary to prevent the water from freezing solid in a particularly harsh cold snap sometimes observed in Colorado. A research program to develop this rotationally molded foam insulation layer was underway, when the objectives of the RDA were revised to address the needs in Auger South. It was then decided to include a simpler, low-cost insulation system of foam polyethylene sheets screwed to the interior walls of the tank which should be capable of preventing freezing damage for the short operating time required.

The water inside the tank is contained in an opaque, closed polyethylene liner with a Tyvek® inner surface. The liners used were surplus from the southern site and were modified to have a single PMT port instead of three. Screw caps on several small ports allow for water filling and the installation of LED flashers. A modified solar power system has been installed for the RDA using a single solar panel (80 Wp, compared to two 53 Wp panels at Auger South) and a single 105 Ah valve regulated lead acid absorbed glass mat (VRLA-AGM) lead-acid battery. The lower power consumption (possibly as little as 5 W when fully developed, compared to 10 W in Auger South) an-ticipated for the RDA electronics allowed the reduced size of the power system, and the opportunity to compare performance of the AGM-type battery with the selectively-permeable-membrane flooded lead acid batteries used at the southern site is most welcome.

The single PMT is bonded to a thin, transparent polyethylene window with clear silicone rubber optical couplant, as at the southern site. The enclosure, however, is a new design that seals the PMT and its base away from exposure to the air inside the tank. Eventually, with production electronics of smaller size than the prototype electronics of the RDA, the electronics will be included inside this hermetically-sealed enclosure. This will result in more stable temperatures and more controlled humidity to reduce the possibility of corrosion. Rubber glands form seals where cables enter the sealed enclosures. For the RDA, the main electronics is enclosed in a separate sealed box inside the tank. The communication electronics and the Tank Power Control Board, used to monitor and provide some measure of control over the solar power system, are mounted in an enclosure outside the tank, clamped to the antenna mast behind the solar panel.

### 2.2 Surface Detector Electronics

The signal in a surface detector varies dramatically both with time and with distance from the shower core, so a wide dynamic range is required. The system must accommodate signals from the photoelectron level for small electromagnetic signals far from the core to large currents due to the passage of peak particle intensity near the shower core. The design of the electronics for the new surface detector is based on the successful design used in Auger South [5]. At the southern site, two overlapping 10-bit flash ADCs are used per PMT, to digitize signals derived from the anode and the amplified last dynode to obtain a 15-bit ($3\times10^4$) dynamic range. Nevertheless, saturated signals are observed near the core of high-energy events. The new surface detector electronics design is more highly integrated due to advances in field of programmable gate arrays (also called PLDs) and extends the dynamic range from 15 bits to 22 bits ($4\times10^6$), thereby decreasing the distance from the core where saturated signals will be observed (from 500 m to 100 m for $10^{20}$ eV showers). The dynamic range extension is achieved by using signals derived from the anode and from a deep ($5^{th}$ out of 8) dynode. The high voltage supply to the tube is designed so that space charge saturation in the last few dynodes does not feed back to the beginning of the dynode chain. This extended dynamic range also provides a more precise determination of the lateral distribution function for the highest energy events. The power consumption of the system will also be about half due to advances in low voltage CMOS components and the reduction of the number of photomultipliers from three to one.

As in Auger South, a hierarchical trigger is implemented with the Local Station Controller (LSC). The lowest level (level 1) trigger is formed by the trigger PLD, which continuously monitors the PMT signals for shower-like signature. A local low power microprocessor applies additional constraints to form level 2 triggers, which are passed on to the observatory campus through the communication system for higher level trigger formation. The LSC is based on an ATMEL AT91RM9200 microcontroller, with 128 Megabytes of RAM, several USB interfaces (one used as a 4 Gigabytes file system), Ethernet 10/100, GPS Receiver M12M Timing used to time-stamp the events, slow control ADC AD7490 (12 bits, 16 channels), slow control DAC TI DAC7554 (12 bits, 4 channels), 3 dual channels syFlash Adc LTC2280 (10 bits, 100 MHz), and an Altera Cyclone III (canbus interface for connection to the radio system, Time Tagging and Trigger firmware). The front-end analog board performs 50-MHz lowpass filtering and signal duplication. The Anode, Anode x 30, Anode x 0.1 and Dynode signals are digitized and then routed to the Cyclone FPGA. The LSC runs a Linux Debian 2.6.27 Operating System, with I-Pipe and Xenomai 2.4.10 RealTime extensions. The software-based level 2 (T2) trigger reduces the T1 event rate from 100 Hz to 20 Hz. Only the time-stamps of the T2 and a rough evaluation of the energy received in the detector are sent to the Central Data Acquisition System (CDAS) every second. All the T1 events are stored locally in DRAM and the data sent to CDAS upon request when the CDAS has detected a multi station coincidence. A maximum of 2000 events are stored, corresponding to about 20 seconds of data taking. A continuous calibration of the detector is made locally using the muon data available from the front end, and the result of the calibration sent to CDAS along with the higher-level T3 data. In addition a LED flasher can also be used for that purpose.

The various electronics boards, namely the LSC and the LED flasher driver inside the tank, and the Tank Power Control Board and the communication electronics in the enclosure on the mast, are interconnected via a ISO standard Controller Area Network (CAN) bus, which provides power, distribution of GPS timing pulses, and transfer of event and monitoring data. This scheme is an improvement on the Auger South design, with both fewer cables and more robust data transmission.

## 3 The peer-to-peer communication system

### 3.1 Concept

At the southern site, each station communicates directly with one of four tower-mounted concentrator stations located on the periphery of the array. From there, data is forwarded via commercial microwave links to a fifth tower at the Observatory campus. This is possible because the terrain is remarkably flat, but surrounded by peripheral hills upon which to place the towers, affording a clear line-of-sight between each station and a tower. However, at the northern site, direct station-to-tower routing is infeasible due to the much larger surface area of the planned array, the absence of convenient peripheral hills, and the presence of internal hills, ridges and gullies that impede station-to-tower line-of-sight. Therefore, it was decided to configure the Auger North surface detector communication system as a Peer-to-Peer Wireless Sensor Net (P2P-WSN). In a P2P-WSN, the transfer of data between a station and the concentrator is accomplished via multi-hop relaying of data between neighboring stations, rather than a single hop directly to the concentrator. The greatest disadvantage to P2P routing is reliability, because the failure of one radio will disconnect not only itself, but all stations upstream of that point (away from the concentrator). Thus, redundant routing and fault-tolerance must be present in the WSN. The particular WSN paradigm chosen for the northern site is the Wireless Architecture for Hard Real-Time Embedded Networks (WAHREN) [6]. WAHREN routes all messages via a second order power chain topology, in which each station, or node, communicates not only with its nearest, or first order, neighbors but also with its second-nearest, or second order, neighbors, providing the required redundancy.

The Medium Access Control (MAC) layer of WAHREN is a Spatial Reuse Time Division Multiple Access (SR-TDMA) protocol. SR-TDMA divides time into constant-length transmission slots. A fixed number of slots are then grouped into a TDMA Window. Within each window, each node is assigned one slot in which it is allowed to transmit. If two nodes are outside of interference range from each other, then they can be assigned to the same slot (hence the term Spatial Reuse). WAHREN employs a systolic broadcast scheduling protocol, shown in figure 3.

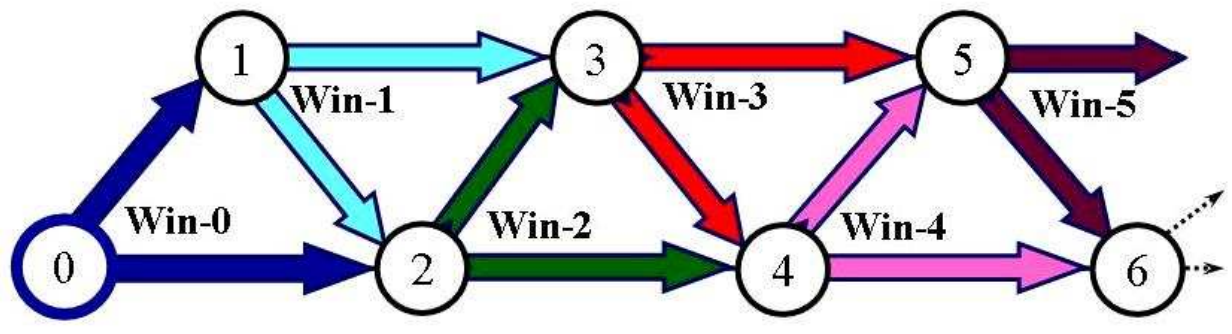

Figure 3: Schematic of single-source systolic broadcast along a second-order power chain.

In this example, node 0 transmits its own message in window 0. Then, in each subsequent window, k, node k forwards the message received during the previous window to two downstream neighbors (k+1 and k+2). Except for the initial transmission from node 0 to node 1, all nodes receive two copies of the message from two different sources. In the single-source example above, each node transmits in only one window. This observation makes it possible for every node in the system to transmit its own original message simultaneously (each within its assigned slot of window 0). Then, in each subsequent window, all messages are forwarded at the rate of one node per window.

Power chains can turn a sharp corner in a manner that is transparent to the graph topology. Such a structure is called a Möbius Fold. This allows the surface detector array to potentially be partitioned into several sectors, in each of which, several side chains intersect a backbone chain at a right angle. The RDA layout shown in figure 1 features such a structure with a backbone running north-to-south and a side chain running west-to-east. The WAHREN paradigm has been verified by a variety of formal verification methods and testbed systems. However, the RDA will permit extensive end-to-end testing to be done in-situ on a full-scale grid with the hardware. RDA testing will include: (1) interference, signal strength, and bit and packet error rate studies; (2) fault injection experiments wherein nodes are purposely put into a failure mode; (3) tests of trigger packet transmission and read out of data from stations; (4) traffic studies wherein fake traffic is introduced to study system behavior at saturation.

### 3.2 Hardware implementation

To avoid the excessive overhead and RF band restrictions imposed by commercial standards such as IEEE 802.11 (WiFi), custom circuit boards were designed to directly implement the WAHREN protocol over four licensed, dedicated channels in the 4.6 GHz band. The surface detector station radio is split into a baseband board and an RF daughter card, both of which can be customized to other applications. Primary control of the radio is provided by an Altera Cyclone III FPGA. Its configuration includes modulator/demodulator circuitry, as well as dual NIOS II processor cores, one serving as the main CPU and the other as an input/output processor (IOP). The interface to the RF daughter card uses a Maxim 19707 ADC/DAC chip. The final component on the baseband board is a TMS470 microcontroller, serving as a channel guardian. It independently monitors transmitter output and system timing.

## 4 Projected timeline

The RDA is expected to be operational in the Fall of 2011 and will collect data until at least June 2012.

# Microwave detection of cosmic ray showers at the Pierre Auger Observatory

PATRICK S. ALLISON[1], FOR THE PIERRE AUGER COLLABORATION[2]
[1]*Ohio State University, Columbus, OH 43210, United States*
[2]*Observatorio Pierre Auger, Av. San Martín Norte 304, 5613 Malargüe, Argentina*
*(Full author list: http://www.auger.org/archive/authors_2011_05.html)*
*auger_spokespersons@fnal.gov*

**Abstract:** Microwave emission from the electromagnetic cascade induced in the atmosphere by ultra-high energy cosmic rays (UHECR) may allow for a novel detection technique, which combines the advantages of the well-established fluorescence technique - the reconstruction of the shower profile - with a 100% duty cycle, minimal atmospheric attenuation and the use of low-cost commercial equipment. Two complementary techniques are currently being pursued at the Pierre Auger Observatory. AMBER (Air-shower Microwave Bremsstrahlung Experimental Radiometer), MIDAS (Microwave Detection of Air Showers) and FDWave are prototypes for a large imaging dish antenna. In EASIER (Extensive Air Shower Identification using Electron Radiometer), the microwave emission is detected by antenna horns located on each surface detector of the Auger Observatory. MIDAS is a self-triggering system while AMBER, FDWave, and EASIER use the trigger from the Auger detectors to record the emission. The coincident detection of UHECR by the microwave prototype detectors and the fluorescence and surface detectors will prove the viability of this novel technique. The status of microwave R&D activities at the Pierre Auger Observatory will be reported.

**Keywords:** Pierre Auger Observatory, microwave, radio, detectors

## 1 Introduction

The observation of microwave emission from electromagnetic cascades at accelerator experiments in 2003 and 2004, along with results from a simple prototype detector [1] suggest that the ultra-high energy cosmic rays (UHECRs) may be detectable from microwave emission alone. Since microwave attenuation in the atmosphere is minimal (less than $0.05\,\mathrm{dB\,km^{-1}}$ [2]) and the noise temperature of the sky is relatively small (below $10\,\mathrm{K}$ for zenith angles less than $60\,^\circ$), a UHECR microwave detector could benefit from the longitudinal observation of the air shower profile with 100% duty cycle and minimal atmospheric loss concerns.

To demonstrate whether or not microwave detection is feasible, several detectors have been developed to operate in conjunction with the Pierre Auger Observatory.

## 2 AMBER

The AMBER detector is an upgrade from the Gorham et al. prototype [1], consisting of a $2.4\,\mathrm{m}$ low-emissivity off-axis parabolic dish with a look angle of 30 degrees, viewed by 4 dual-polarization, dual-band (C: $3.4-4.2\,\mathrm{GHz}$ and Ku: $10.95-14.5\,\mathrm{GHz}$) and 12 single-polarization C-band antenna horns. The signal from the antenna horns is passed through a low-noise block (LNB) which amplifies and downconverts the signal to below $2\,\mathrm{GHz}$, where it is put into a logarithmic (progressive-compression) power detector which produces a signal roughly proportional to the observed radio frequency (RF) power (in dBm). The installed AMBER detector is shown in Fig. 1.

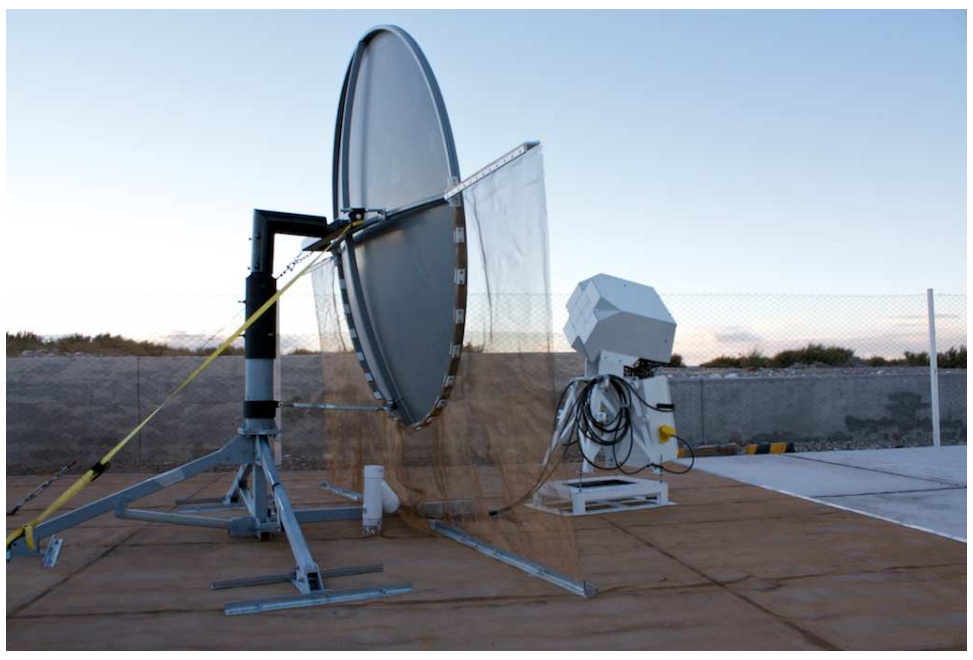

Figure 1: The upgraded AMBER antenna array installed at the HEAT enclosure at the Pierre Auger Observatory.

The goal of the upgraded AMBER detector is to observe cosmic ray showers in coincidence with surface detectors of the Pierre Auger Observatory without self-triggering. The Surface Detector (SD) trigger latency is very long - nominally $5\,\mathrm{s}$, however, a modified array trigger (T3) algorithm was developed which outputs approximately 70% of all T3s within 3 s. The signals from the power detector are digitized at 100 MHz and placed into a large circular buffer (32 GB for all input channels). Time within the circular buffer is tracked using a GPS pulse-per-second (PPS) output and a 100 MHz clock for in-second tracking. The accuracy of this procedure was measured by placing the GPS PPS signal into one input channel, and requesting data around the second boundary. From this, an observed time resolution of 11 ns was obtained.

When the SD registers a possible cosmic ray via a T3 trigger, an approximate time at ground, core location, and incident direction will be derived from the trigger times and tank positions. From the time, location, and direction provided from this reconstruction, the time that the shower crosses the AMBER field of view (FOV), if at all, is then calculated, and the data corresponding to that time will be read out. From a comparison with fully-reconstructed Auger events, the approximation is valid to within $10\,^{\circ}$ degrees in solid angle and 500 m in core position. For all cosmic rays within the AMBER FOV, this accuracy is sufficient for a reasonable readout window (100 $\mu$s) to contain a possible cosmic ray track.

The calibration of the AMBER instrument was done by first injecting a signal of varying power from a network analyzer into the power detector modules to convert the signal seen at the digitizer to power observed at the power detector input. Next, the front of the feed array was placed into a liquid nitrogen bath in an anechoic chamber to perform a Y-factor measurement to calibrate the gain and noise figure of the LNB to obtain the power at the input of the feed array. Finally, the dish itself was calibrated separately using a Y-factor measurement using RF absorber foam and a calibrated LNB. The combined system noise temperature ranges from $\sim 45$ K for the outer single-band antennas, and $\sim 65$ K for the inner antennas in C-band. Ku-band system temperatures were significantly higher ($\sim 100$ K) due to the LNBs.

The upgraded AMBER system was in operation at the University of Hawaii from January to June 2010. A search for cosmic ray candidates was performed using a separate self-trigger board. However, the RF environment was significantly worse than during the operation of the original prototype, and no candidates were found. During the operation period, several observations of Sun transits were performed to validate the expected optical performance, as seen in Fig. 2. The full-width at half-max (FWHM) of the Sun was $2.4\,^{\circ}$, and the expected FWHM based on the dish characteristics was $2.3\,^{\circ}$, giving a total field of view of $\approx 7^{\circ} \times 7^{\circ}$.

Following the operation in 2010, the AMBER detector was packaged and shipped to Argentina. The detector was reassembled and integrated at the Coihueco Fluorescence Detector (FD) site, alongside the HEAT [3] fluorescence telescopes (which have a similar viewing angle), and overlooking the SD infill array. Data taking in coincidence with the Auger SD is currently underway.

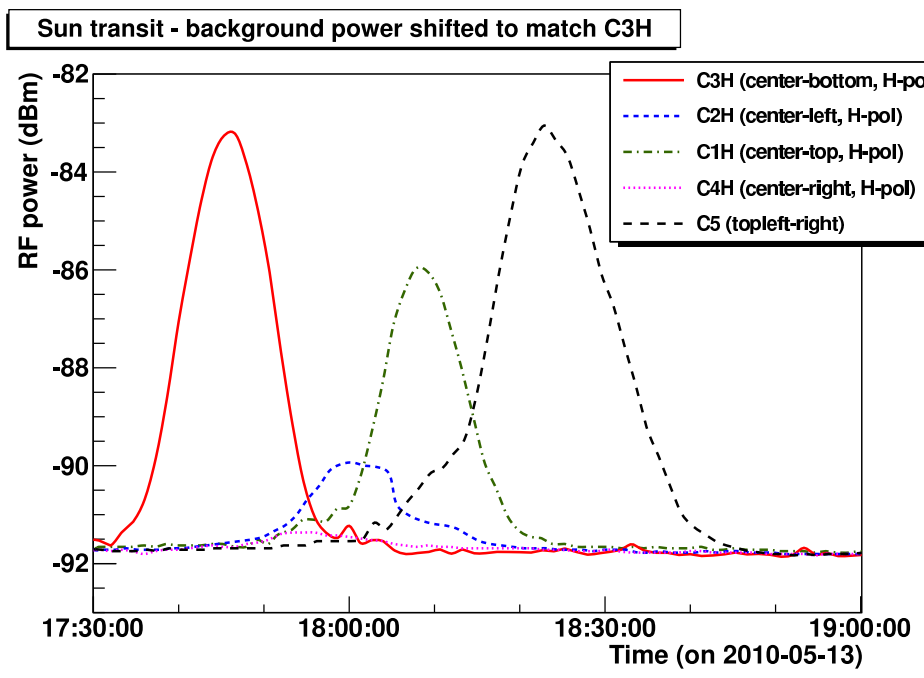

Figure 2: RF power as a function of time for a Sun transit observed by AMBER. The Sun's path was directly through the C3H/C5 pixels, partially through the C1H/C2H pixels, and not directly in C4H at all. Observed FWHM for the Sun was $2.4\,^{\circ}$, compared to an expected $2.3\,^{\circ}$ for the dish geometry.

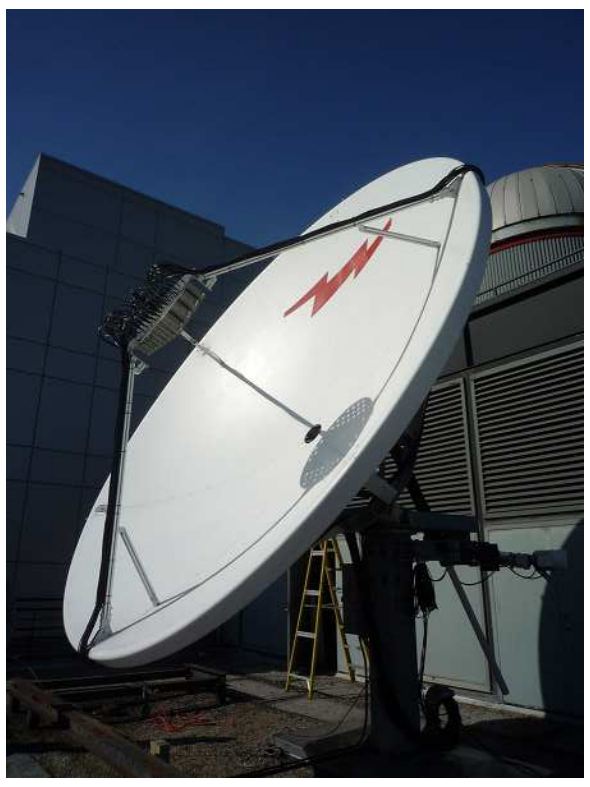

Figure 3: The MIDAS prototype at the University of Chicago.

## 3 MIDAS

The MIDAS detector (shown in Fig. 3) consists of a 4.5 m parabolic reflector with a 53-pixel camera with 7 rows of 7 or 8 pixels, arranged in a staggered layout to maximize coverage. The pixel's field of view ($\approx 1.3\,^{\circ} \times 1.3\,^{\circ}$ at center) depends on its position on the camera due to aberrations, for a total camera FOV of $\approx 20\,^{\circ} \times 10\,^{\circ}$. Pixels are integrated with commercial satellite television LNBF (LNB with feedhorn), and the down-converted RF signal is passed into a logarithmic power detector.

The 53 analog channels are organized in groups of 8 in rack-mount electronics enclosures, where the output signal of the power detector is digitized by a 20 MHz 14-bit FADC on a VME board developed by the Electronics Design Group at the Enrico Fermi Institute at University of Chicago. An on-board FPGA is used for digital signal processing and trigger.

The MIDAS trigger, implemented in the FPGAs of the FADC boards, selects candidate events by pixel topology and time coincidence requirements. For each pixel, the signal is continuously digitized and the running ADC sum of 10 consecutive time bins is calculated. Whenever this sum falls below a preset threshold, a First Level Trigger (FLT) is issued for the pixel and a $10\,\mu\mathrm{s}$ gate is opened. This threshold is adjusted every second to keep the FLT rate close to $100\,\mathrm{Hz}$. The Second Level Trigger (SLT) performs a search for pre-defined patterns of FLT triggers whose gates overlap in time. Valid patterns correspond to the expected topology of a cosmic ray shower (straight tracks across the camera). When a SLT is issued, a stream of $100\,\mu\mathrm{s}$ of ADC data (including 500 pre-trigger samples) is stored in memory for each of the 53 channels.

Several sources of GHz RF interference were observed to be present in the urban environment of the University of Chicago campus, including cellular phone towers, various motors, and most notably the navigation system of airplanes overflying the MIDAS antenna on their route to Midway Airport. The RF interference may increase suddenly, generating bursts of events during several seconds. Similar interference was observed at the AMBER installation at University of Hawaii. Significant improvement of the background conditions were observed with the installation of bandpass filters designed specifically to cut out the airplane transmission frequency.

Calibration of the MIDAS prototype was performed both during its comissioning and periodically during data taking. A relative calibration is performed using a log-periodic antenna positioned at the center of the reflector, excited by an RF pulse generator. The antenna illuminates the whole camera with a $4\,\mathrm{GHz}$ RF pulse of a few $\mu\mathrm{s}$ pulse width, with pulse power varied from -60 to $0\,\mathrm{dBm}$ in $5\,\mathrm{dBm}$ steps. The antenna will be used to monitor the stability of the system during data taking by firing a set of 10 pulses with fixed power and duration every 15 minutes.

An absolute calibration of MIDAS was performed using several astronomical sources, including the Sun, the Moon ($\approx 0.01\,F_{\odot}$), and the Crab Nebula ($\approx 10^{-3}\,F_{\odot}$), giving an effective system temperature of $\simeq 120\,\mathrm{K}$.

Based on the measurements from Gorham et al. [1], using quadratic energy scaling and the measured system temperature, a MIDAS pixel viewing the shower maximum of a $5\,\mathrm{EeV}$ shower at $10\,\mathrm{km}$ distance would measure $\approx 2000\,\mathrm{ADC\,counts}$ under the baseline. With linear scaling, a $10\,\mathrm{EeV}$ shower at the same distance would yield $\approx 200\,\mathrm{ADC\,counts}$, as compared to the measured pixel baseline fluctuation of about $70\,\mathrm{ADC\,counts}$. Thus, the MIDAS prototype has a good sensitivity for UHECR detection.

In addition, an end-to-end Monte Carlo simulation of the MIDAS prototype was developed, including the camera beam patterns and the absolute calibration, which is being used for a realistic estimate of the event rate, and for the characterization of the expected events in the same format as the data. An example of a simulated event is shown in Fig. 4.

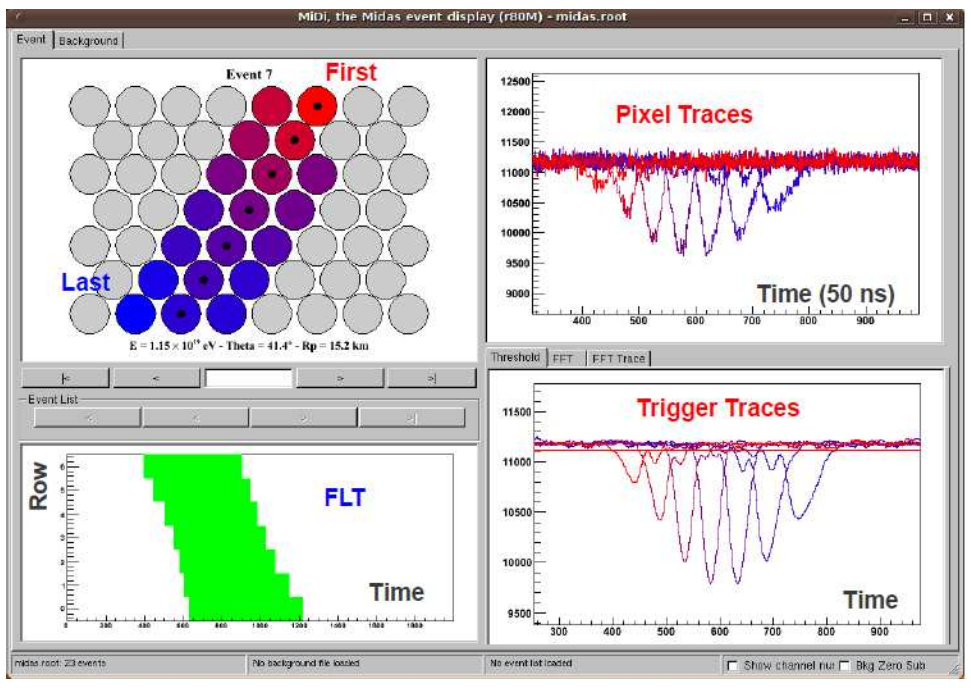

Figure 4: Event display of a simulated MIDAS event. Top left: FLT pixels with color indicating increasing trigger time from light (red) to dark (blue). Top right: FADC traces for selected pixels. Bottom left: FLT trigger gate integrated over all pixels in a row versus time. Bottom right: ADC running sum for selected pixels.

The MIDAS prototype has been in stable operation for 6 months at the University of Chicago, and is in the process of being relocated to the Los Leones FD site.

## 4 FDWave

The aim of the FDWave project is to develop a microwave telescope equipped with a matrix of radio receivers taking advantage of the existing infrastructure in the Pierre Auger Observatory. The integration of microwave detectors in the Auger Observatory will be performed by installing GHz antennas in installed pixels without photomultipliers (PMTs) at the Los Leones FD site (264 total). The firmware of the FD trigger boards will be modified in order to exclude the microwave signal from the trigger decision. In coincidence of a shower detected by the FD telescopes, the microwave signal will be acquired and registered in the standard data format used by the Auger Collaboration. Extrapolating the reconstructed longitudinal profile within the FOV of the antennas and comparing the energy deposit with the microwave signal, it will be possible to obtain important information on the emission yield and then on the feasibility of this new detection technique.

A careful study of the FD optics [5] and of its use at the GHz frequencies has been performed. The spherical mirrors (3.4 m curvature radius) of Los Leones are of the aluminum kind and therefore conveniently reflecting. The spherical focal surface has a radius of $1.743\,\mathrm{m}$ and the detectors are placed in the holes of an aluminium camera body. The camera geometry puts a constraint on the lowest detectable frequency - feeds below $9\,\mathrm{GHz}$ are too big and cannot be used with the present camera geometry. This minimum frequency is very close to the Ku-band. Staying within this band has the important advantage of the low cost of the instrumentation due to their large commercial diffusion. The optimal radio receiver fitting the camera geom-etry constraints is the ”RED Classic” straight-feed $40\,\mathrm{mm}$ LNB manufactured by Inverto.

The simulation of the FD optics including the diaphragm aperture and camera shadow shows that at $11\,\mathrm{GHz}$, the pixel field of view is $\sim\ 0.7\,^{\circ}$. The telescope gain is $\sim\ 44\,\mathrm{dBi}$ and the dish effective area is $\sim\ 1.35\,\mathrm{m}^2$. With those parameters the telescope should provide evidence of the microwave emission for showers above $3\,\mathrm{EeV}$ with quadratic signal scaling. The sensitivity can be significantly improved averaging the FADC traces over many shower profiles, an operation that can be successfully and easily performed using the shower parameters reconstructed by the FD photomultipliers.

## 5 EASIER

The aim of the EASIER project is to observe radio emission in both the GHz and MHz regime from air showers by antennas installed on each water Cherenkov tank of the SD. Each microwave antenna covers a large field of view of $\approx 60\,^{\circ}$ angle around zenith. The RF signal from the antenna is fed into a logarithmic power detector, and then digitized at $40\,\mathrm{MHz}$ using the existing SD electronics. In addition, EASIER will benefit from the existing power distribution and communication at the SD station, which greatly simplifies its integration into SD data taking. Since the signal of the EASIER antenna is digitized by the same FADC as the PMTs at each SD station, timing is automatically provided and EASIER data are saved whenever an SD station triggers. As with AMBER and FDWave, the external triggering approach imposes no requirements on the signal to noise ratio of the radio emission, which, under realistic noise conditions, gives a gain of a factor of 3 to 5 relative to self-triggering.

While the effective area of the EASIER antennas is much smaller than the other detectors, the antennas will be significantly closer to the shower axis and within $\approx 3\,\mathrm{km}$ from shower maximum, as opposed to $\approx\ 10\,\mathrm{km}$ for dish-type antennas. In addition, due to the short distance from the shower axis, the radio signal is compressed in time typically by a factor of 10. In principle, these factors will compensate for the smaller effective area of the EASIER antenna.

EASIER prototype antennas have been installed in two hexagons of SD stations, one with GHz antennas (seen in Fig. 5) and one with MHz antennas. The MHz detectors consist of a fat active dipole antenna (as used in CODALEMA [4]) with a $36\,\mathrm{dB}$ LNA, with additional filters restricting the frequency range to $30-70\,\mathrm{MHz}$. The GHz detectors consist of a DMX241 LNB feed, at the top of a $3\,\mathrm{m}$ mast looking vertically, and passed through a power detector. The signals are then adapted to the SD FADC input range, resulting in a measurement range of -20 to +30 dB relative to background for the MHz detector, and -20 to +20 dB relative to background for the GHz detector. An event seen by the SD and EASIER MHz antennas is shown in Fig. 6.

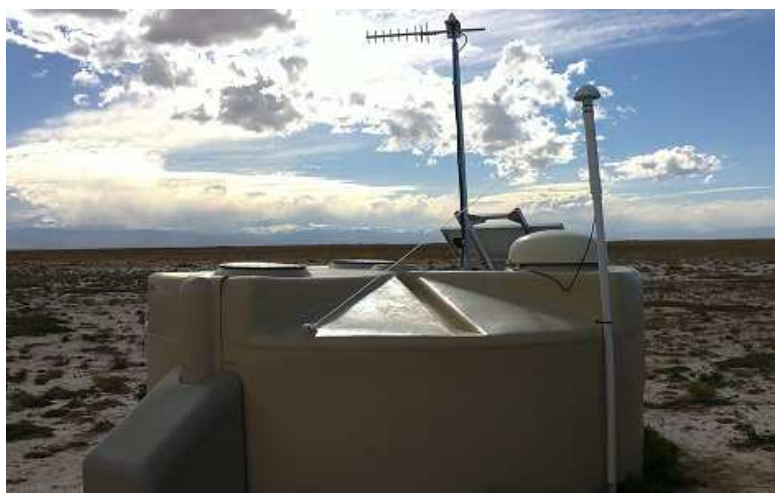

Figure 5: One of 7 Auger SD stations with a GHz EASIER antenna (right mast).

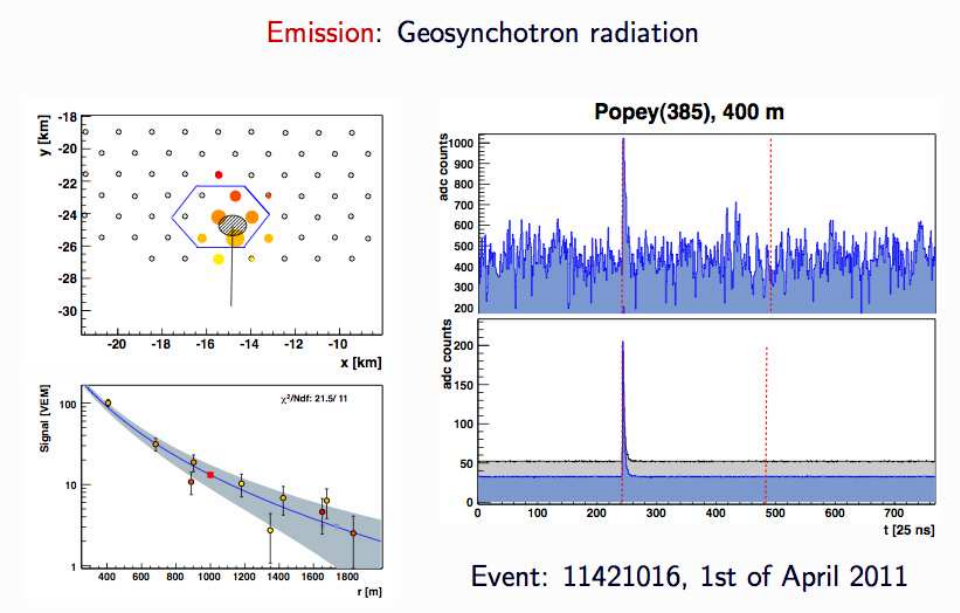

Figure 6: Event display of a cosmic ray seen in the SD and in an EASIER MHz antenna. The hexagon in the top left indicates the tanks with EASIER MHz antennas. Signals in the MHz antenna and SD station are visible in the top and bottom right, respectively.

## 6 Conclusion

Multiple detector prototypes are currently under construction at the Pierre Auger Observatory to attempt to observe microwave emission from cosmic ray showers. Using the unique resources of the world’s largest cosmic ray observatory, data regarding the microwave emission from UHECRs will soon determine whether or not this detection technique is viable.

# Acknowledgments

The successful installation, commissioning and operation of the Pierre Auger Observatory would not have been possible without the strong commitment and effort from the technical and administrative staff in Malargüe.

We are very grateful to the following agencies and organizations for financial support:

Comisión Nacional de Energía Atómica, Fundación Antorchas, Gobierno De La Provincia de Mendoza, Municipalidad de Malargüe, NDM Holdings and Valle Las Leñas, in gratitude for their continuing cooperation over land access, Argentina; the Australian Research Council; Conselho Nacional de Desenvolvimento Científico e Tecnológico (CNPq), Financiadora de Estudos e Projetos (FINEP), Fundação de Amparo à Pesquisa do Estado de Rio de Janeiro (FAPERJ), Fundação de Amparo à Pesquisa do Estado de São Paulo (FAPESP), Ministério de Ciência e Tecnologia (MCT), Brazil; AVCR AV0Z10100502 and AV0Z10100522, GAAV KJB100100904, MSMT-CR LA08016, LC527, 1M06002, and MSM0021620859, Czech Republic; Centre de Calcul IN2P3/CNRS, Centre National de la Recherche Scientifique (CNRS), Conseil Régional Ile-de-France, Département Physique Nucléaire et Corpusculaire (PNC-IN2P3/CNRS), Département Sciences de l'Univers (SDU-INSU/CNRS), France; Bundesministerium für Bildung und Forschung (BMBF), Deutsche Forschungsgemeinschaft (DFG), Finanzministerium Baden-Württemberg, Helmholtz-Gemeinschaft Deutscher Forschungszentren (HGF), Ministerium für Wissenschaft und Forschung, Nordrhein-Westfalen, Ministerium für Wissenschaft, Forschung und Kunst, Baden-Württemberg, Germany; Istituto Nazionale di Fisica Nucleare (INFN), Ministero dell'Istruzione, dell'Università e della Ricerca (MIUR), Italy; Consejo Nacional de Ciencia y Tecnología (CONACYT), Mexico; Ministerie van Onderwijs, Cultuur en Wetenschap, Nederlandse Organisatie voor Wetenschappelijk Onderzoek (NWO), Stichting voor Fundamenteel Onderzoek der Materie (FOM), Netherlands; Ministry of Science and Higher Education, Grant Nos. 1 P03 D 014 30, N202 090 31/0623, and PAP/218/2006, Poland; Fundação para a Ciência e a Tecnologia, Portugal; Ministry for Higher Education, Science, and Technology, Slovenian Research Agency, Slovenia; Comunidad de Madrid, Consejería de Educación de la Comunidad de Castilla La Mancha, FEDER funds, Ministerio de Ciencia e Innovación and Consolider-Ingenio 2010 (CPAN), Xunta de Galicia, Spain; Science and Technology Facilities Council, United Kingdom; Department of Energy, Contract Nos. DE-AC02-07CH11359, DE-FR02-04ER41300, National Science Foundation, Grant No. 0450696, The Grainger Foundation USA; ALFA-EC / HELEN, European Union 6th Framework Program, Grant No. MEIF-CT-2005-025057, European Union 7th Framework Program, Grant No. PIEF-GA-2008-220240, and UNESCO.